\documentclass[apj]{emulateapj}

\slugcomment{Submitted to AJ}

\begin{document}

\title{MULTIWAVELENGTH VARIABILITY OF THE BLAZARS MRK 421 AND 3C 454.3 IN THE HIGH-STATE}

\author{Haritma Gaur\altaffilmark{1,2}, Alok C.\ Gupta\altaffilmark{1,2}, Paul J. Wiita\altaffilmark{3}}

\altaffiltext{1}{Aryabhatta Research Institute of Observational Sciences (ARIES),
Manora Peak, Nainital - 263129, India}
\altaffiltext{2}{Department of Physics, DDU Gorakhpur University, Gorakhpur - 273009, India}
\altaffiltext{3}{Department of Physics, The College of New Jersey, P.O.\ Box 7718, Ewing, NJ 08628, USA}

\email{haritma@aries.res.in}

\begin{abstract}
We report the results of photometric observations of the blazars Mrk 421 and 3C 454.3 designed to search
for intraday variability (IDV) and short-term variability (STV). Optical photometric observations were spread 
over eighteen nights for Mrk 421 and seven nights for 3C 454.3 during our observing run in 2009-2010
at the 1.04 m telescope at ARIES, India. Genuine IDV is found for the source 3C 454.3 but not
for Mrk 421.  Genuine STV is found for both sources. Mrk 421 was revealed by the MAXI X-ray detector 
on the International Space Station to be in an exceptionally high
flux state in 2010 January - February.  We performed a correlation between the X-ray and optical bands 
to search for time delays and found a weak correlation with higher frequencies leading 
the lower frequencies by about ten days.
The blazar 3C 454.3 was found to be in high flux state in November-December 2009. We performed correlations in 
optical observations made at three telescopes, along with X-ray data from the MAXI satellite and
public release $\gamma$-ray data from the Fermi space telescope. We found strong correlations between the $\gamma$-ray and optical bands
at a time lag of about four days but the X-ray flux is not correlated with either. We briefly discuss the possible reasons for the time delays between these bands within the framework of  existing models for  X-ray and $\gamma$-ray emission mechanisms.
\end{abstract}

;

\section{\bf Introduction}

Blazars are the radio$-$loud Active Galactic Nuclei (AGN) that are classified as BL
Lacertae objects (BL Lacs) if they have largely featureless optical spectra
or as flat spectrum radio quasars (FSRQs) if they have prominent emission lines.
All blazars are characterized by broadband 
non-thermal emission extending over the complete electromagnetic spectrum, strong 
polarization from radio to optical wavelengths,  
and displays of violent variability on timescales that can extend from a fraction of an hour to many years. 
In the usual orientation-based unified model of radio$-$loud AGN,  blazar jets 
make an angle of $\leq$ 10$^\circ$ from the line of sight and the emission from these jets is thus Doppler
boosted and dominates what we observe (Urry \& Padovani 1995).

Blazar variability timescales have often been arbitrarily divided into three classes: timescales 
from a few tens of minutes to less than a day are called intraday variability (IDV) (Wagner \& 
Witzel 1995) or micro-variability or intranight variability, those from several days to a few
months are short timescale variability (STV), while long timescale variability (LTV) covers 
changes from several months to many years (Gupta et al.\ 2004).  The spectral energy distribution (SED) of blazars have two 
peaks (e.g., Giommi, Ansari \& Micol 1995; Fossati et al.\ 1998).  The locations of those
peaks can be used to classify blazars into LBLs (Low Energy Peaked 
Blazars)  with the first hump in the near infrared (NIR) or optical band and the second hump usually peaking
at GeV $\gamma$-ray energies, while HBLs (High Energy Peaked Blazars) are those with first peak in the 
UV or X-ray band and the second  peak located at  up to TeV energies (e.g., Padovani \& Giommi 1995). 
The high polarization of the radio to optical emission suggests that the lower energy peak is 
produced via the synchrotron process but the high energy emission mechanism in blazars is not
yet fully understood, though it is probably due to the inverse Compton (IC) mechanism. 

\subsection{\bf Mrk 421}

With its redshift $z=0.031$, Mrk 421 (B2 1101$+$384) ($\alpha_{2000.0}$=11h 04m 27.2s and 
$\delta_{2000.0}$=$+38^{\circ} 12^{'} 32^{\prime \prime}$) is among the closest blazars, at a distance of 
134 Mpc (H$_{0} =$ 71 km s$^{-1}$ Mpc$^{-1}$, $\Omega_{m} =$ 0.27, $\Omega_{\lambda} =$ 0.73).  It 
is classified as an HBL because the energy of its synchrotron peak is higher than 0.1 keV.
It is the brightest TeV $\gamma-$ ray emitting blazar in northern hemisphere.  Mrk 421 was first noted 
to be an object with a blue excess which turned out to be an elliptical galaxy with a bright point 
like nucleus (Ulrich et al.\ 1975). The object showed optical polarization and the spectrum of the 
nucleus was seen to be featureless and so it was classified as a BL Lac.   Mrk 421 was detected 
in the GeV band by the EGRET instrument on the Compton Gamma$-$ray Observatory (CGRO) (Lin et al. 1992; 
Michelson et al. 1992). It was the first known
extragalactic TeV $\gamma-$ray emitter (Punch et al.\ 1992), and has been repeatedly
confirmed as a TeV source by  ground-based $\gamma$-ray telescopes (Aleksic et al. 2011; 
Acciari et al. 2011 and references therein).  

Mrk 421 has been extensively observed at all wavelengths and some noticeable studies in low-energy bands 
include an exhaustive compilation of radio data at 22 and 37 GHz over about 25 years (Ter{\"a}sranta et al.\ 2004, 2005).
The source is  characterized by strong variability in the optical region (e.g., Miller 1975; Liu et al.\ 1997) including LTV  of $\sim$ 4.6 mag  (Stein et al.\ 1976) and extreme rapid optical 
variability  exemplified by a 1.4 magnitude brightness change
in only 2.5 hours (Xie et al.\ 1988).  NIR data extending over about three decades were given by Fan \& 
Lin (1999) and Gupta et al.\ (2004) reported detection of IDV and STV in the blazar in the IR.  Weakly correlated X-ray and 
NIR variability was reported 
in the source by Gupta et al.\ (2008). 

It is worth noting that Mrk 421 is almost always very active at high energies. 
In spring/summer 2006, Mrk 421 reached its highest X-ray flux recorded 
until that time with a peak flux $\sim$85 mCrab in the 2.0--10.0 keV band, with the corresponding first peak of 
the SED often occurring at $>$10 keV (Tramacere et al.\ 2009).   A hard X-ray flare from Mrk 
421 was detected by Super-AGILE on 2008 June 10 (Costa et al.\ 2008) which was followed by detection in 
$\gamma-$rays (Pittori et al. 2008) by the AGILE/GRID (Gamma-ray Imaging Detector).   Ushio et al.\ (2009) 
presented the observations of X-ray variability of Mrk 421 with Suzaku.  Recently, strong X-ray flares were 
detected from the source in 2010 January and February with the Monitor of All-sky X$-$ray Image (MAXI) 
instrument on the International Space Station (ISS).  The February 2010 flare reached 164 mCrab and 
is the strongest among those reported from the object (Isobe et al.\ 2010). 

The time-average energy spectrum of Mrk 421 
during the flaring stage has been measured at high energies with HESS using large-zenith-angle observations 
(Aharonian et al.\ 2005) and with MAGIC (Albert et al.\ 2007). X$-$ray and TeV flares were observed around May 16, 
1994 (Takahashi et al.\ 1994; Kerrick et al.\ 1995) and around April 25, 1995 (Takahashi et al.\ 1995). The observed 
correlated variability between X-rays and TeV $\gamma-$rays (Maraschi et al.\ 1999; Fossati et al.\ 2008) can be
explained in the synchrotron-self Compton (SSC) framework (Ghisellini et al.\ 1998), whereas 
the external Compton (EC) scenario is unlikely to  apply in HBLs due to the low density of ambient photons. 
There was a detection of a rapid variability timescale of TeV $\gamma-$ray 
emission from Mrk 421 ($\sim$10 min; Gaidos et al.\ 1996) that may require a very large Doppler factor 
$\delta \geq$ 50 (Ghisellini \& Tavecchio 2008).   Mrk 421 has been a target of several simultaneous multi-wavelength 
monitoring campaigns (Takahashi et al.\ 2000; Rebillot et al.\ 2006; Fossati et al.\ 2008; Lichti et al.\ 2008).
Despite all the studies of this source, we found that there have been a very few investigations of IDV and STV of Mrk 421 in optical bands.  So we decided to pursue the present study which addresses 
both the IDV and STV of Mrk 421. Our optical observations are synchronized with the X$-$ray observations from
MAXI.  The present optical observations give simultaneous information in an additional
spectral window and make it possible to search for the correlations and time delays 
between optical and X$-$ray bands. This work could give us useful input for multi$-$wavelength 
modelling of blazars and lead to a better  understanding of the cause of their variability.

\subsection{\bf 3C 454.3}

 3C454.3 (PKS 2251+158) ($\alpha_{2000.0}$=22h 53m 57.75s
$\delta_{2000.0}$=$+16^{\circ} 08^{'} 53.56^{\prime \prime}$) is a well known flat 
spectrum radio quasar (FSRQ) at redshift z=0.859. It has displayed
pronounced variability at all wavelengths and has been extensively observed over the years
in most energy bands, from radio (e.g., Bennett 1962) through microwave (Bennett et al.\ 2003), 
optical (e.g., Sandage 1966; Raiteri et al. 1998), X$-$ray (e.g., Worrall et al.\ 1987; 
Tavecchio et al.\ 2002), low energy $\gamma-$ray (Blom et al.\ 1995;
Zhang et al.\ 2005) and high energy $\gamma-$ray (Hartmann et al.\ 1993, 1999). 

3C 454.3 entered a bright phase starting at 2000, and has
shown remarkable activity in the past decade (Fuhrmann et al.\ 2006; Villata et al.\ 2006, 2007, 
Raiteri et al.\ 2007, 2008). In spring 2005 it underwent major outbursts, reaching an R-band 
magnitude of 12.0 that was the largest apparent optical luminosity ever recorded from this blazar 
(Villata et al.\ 2006). 
A Whole Earth Blazar Telescope monitoring effort continued after the optical outburst and followed the subsequent
radio activity (Villata et al.\ 2007) and then the faint state in the 2006-2007 observing
season (Raiteri et al.\ 2007). In this last period, the relatively low contribution of the synchrotron emission from the jet meant that the ``little blue bump'', believed to arise from Fe-line emission from the broad line region, as well as the
``big blue bump'', due to the nearly thermal emission from the accretion disc were recognizable.
An increase in activity occurred at X-ray and radio wavelengths as well, with the 230 GHz radio
variations having a delay of $\sim$ 2 months with respect to the optical variability (Raiteri et
al.\ 2008).
 After the quiescent state, the quasar underwent a new stage of high optical activity 
(Raiteri et al. 2008) that continued to the end of 2008.  
 UV excesses observed at some times are likely to be a signature of the 
thermal radiation from the accretion disk feeding the central black hole.

Swift telescope observations of 3C 454.3 (Giommi et al.\ 2006) and hard X-ray observations (Pian et al.\ 2006) covering 
this active phase in 2005 detected large fluxes between 10 and 100mCrab.
Starting with the 2007 July AGILE detection above 100 MeV (Vercellone et al.\  2008), 3C454.3
has been very active in $\gamma$-rays. It was subsequently monitored by AGILE in 2007-2009 and
showed repeated flares that usually coincide with periods of intense optical and enhanced X-ray
activity (Chen et al.\ 2007, Donnarumma et al.\ 2009).
During this time span, very bright $\gamma-$ray emission was detected (Tosti et al.\ 2008), with
an excellent correlation between the  $\gamma$ ray and NIR/optical variations (Bonning
et al.\ 2009).

The source is listed as 1FGL J2253.9$+$1608 in the First FERMI-LAT active galactic nucleus (AGN)
catalog (Abdo et al.\ 2010). 3C454.3  is the first source for which daily resolved broadband
spectral energy distribution (SEDs) with GeV data have been obtained (Abdo et al.\ 2009).
Strong Ly$\alpha$ radiation has been seen from 3C454.3 (Bonnoli et al. 2010), indicating 
the presence of an external
photon source for the Compton scattering aside from torus emission (Sikora et al.\ 2009).
Modeling of SEDs has been performed by Finke \& Dermer (2010), and Pacciani et al.\ (2010).

This FSRQ showed strong activity at optical frequencies in 2008-2009.
(Villata et al.\ 2008; Sasada et al.\ 2009; Gupta et al.\ 2009).
Foschini et al.\ (2010) claim $\gamma-$ray variability from 3C454.3 on timescales as short
as a few hours from the LAT data. The continuous monitoring by the Fermi-LAT showed that the 
source activity faded considerably in early
2009 and then rose back up from June onward. It underwent an exceptional outburst in 2009 November$-$
2010 January when it became brightest $\gamma-$ray source in the sky for over a weak.
(Striani et al.\ 2009, 2010; Escande \& Tanaka 2009).

Here we present our optical observations from ARIES, India during the outburst 
flare in November-December 2009, along with data from the optical monitoring program during this period with the 
Small and Moderate Aperture Research Telescope System (SMARTS) in Chile as well as optical data from the 1.5 m 
telescope of KANATA observatory in Japan.
In order to see if there is related variability of the blazar 3C 454.3 in optical, X-ray and gamma bands, 
we correlate the above optical data with the X-ray data (in 2-10 KeV) obtained from MAXI 
 as well as with 1-300 Gev fluxes made public by the Fermi Science Support Center. 
These results should be useful for understanding the $\gamma$-ray and X-ray emission mechanisms and 
could also offer a test of the existing models (leptonic and hadronic) for the $\gamma$-ray emission.

The paper is structured as follows.  In Section 2, we give brief descriptions of observations and data 
reduction methods. In Section 3, we discuss the techniques we used to search for variability properties  
and we provide the results in Section 4.  A discussion and our conclusions are given in Section 5.

\section{\bf Observations and Data Reductions}

\subsection{\bf ARIES observations and data reduction}

Our  optical photometric observations were carried out in the B, V, R and I pass bands between November 2009 and
June 2010 using the 104-cm Sampurnanand telescope (ST) located at the Aryabhatta Research Institute of Observational 
Sciences (ARIES) in Nainital, India.  It has Ritchey$–$Chretien (RC) optics with a f/13 beam. The detector was a 
cryogenically cooled 2048$\times$2048 CCD chip mounted at the Cassegrain focus. This chip has a readout noise of 
5.3 e$^{-}$ pixel$^{-1}$ and a gain of 10 e$^{-}$ADU$^{-1}$ in the employed slow readout mode. Each pixel 
has a dimension of 24 $\mu$m$^{2}$, corresponding to 0.37 arc-sec$^{2}$ on the sky, thereby covering a total field 
of 13$^{\prime} \times$ 13$^{\prime}$. We carried out observations in a 2 $\times$ 2 binned mode to improve the 
signal-to-noise (S/N) ratio. The seeing usually ranged between $\sim 1.^{\prime \prime}5$ and $\sim 3.^{\prime \prime}0$.  The detailed 
observation logs of the blazars Mrk 421 and 3C 454.3 are given in Table 1.

Image processing was done using the standard routines in Image Reduction 
and Analysis Facility (IRAF)\footnote{IRAF is distributed by the National Optical Astronomy Observatories, 
which are operated by the Association of Universities for Research in Astronomy, Inc., under cooperative
agreement with the National Science Foundation.}.
\par


Data processing to provide the  instrumental magnitudes of the stars and the target source was
done  using the Dominion Astronomical Observatory Photometry (DAOPHOT II) software to perform
the concentric circular aperture photometric technique (Stetson 1987, 1992). 
We observed local standard stars\footnote{http://www.lsw.uni-heidelberg.de
/projects/extragalactic/charts} in the field of blazars.
For Mrk 421, we observed three local standard stars, labeled 1, 2 and 3
while we had nine local standard stars for 3C 454.3.   
Aperture photometry was carried out with four concentric  aperture radii, i.e., 
 $\sim$ 1$\times$FWHM, 2$\times$FWHM, 3$\times$FWHM and 4$\times$FWHM. On comparing 
the photometric results, we found that aperture
radii of 2$\times$FWHM almost always provided the best S/N for both blazars  
and  the standard stars, so we adopted those apertures for our final data reductions. 
The flux from the nucleus of Mrk 421 is contaminated by the emission of the host galaxy. 
To remove this constant component, we used  
the measurements of Nilsson et al.\ (2007)  to estimate the host galaxy emission in the
R-band. This flux is used to obtain the corresponding contributions for the B and V bands (Fukugita et al.\ 
1995) and we corrected for  Galactic extinction using the extinction map of Schlegel,
Finkbeiner \& Davis (1998).

Two standard stars in each blazar field, Stars 1 and 2 for Mrk 421 and Stars C and D for 3C 454.3, 
were used to check that the standard stars were mutually non-variable  and finally one 
standard star each, Star 1 and Star D, were used to calibrate the instrumental magnitudes of  
Mrk 421 and 3C 454.3, respectively.

\subsection{\bf Optical Data from KANATA}
The KANATA 1.5 m telescope performed long term monitoring of the source 3C 454.3  in the R-band with a 
cadence of one-day. This data have been reported
in Pacciani et al.\ (2010) and the details of the observations are given there.
The data were kindly provided to us by M.\ Uemura and cover the range from 
5 November 2009 to 3 January 2010.

\subsection{\bf Optical Data from SMARTS}
The SMARTS\footnote {http://www.astro.yale.edu/smarts/glast/3C454.3lc.html} photometric data and light curves
(LCs) for 3C 454.3  are publicly available on the web (Bailyn 1999).
Monitoring of the sources are carried out on the 1.3 m telescope located at Cerro Tololo Inter-American
Observatory (CTIO) with the ANDICAM instrument. ANDICAM is a dual-channel imager with 
a dichroic that feeds an optical CCD and an IR imager, which can obtain simultaneous data 
from 0.4 to 2.2 $\mu$m (Bailyn 1999).
We have taken data from the SMARTS web archive for 3C 454.3 from 5 November 2009 to 13 December 2010 in the
B, V and R bands.

\subsection{\bf X-ray archival data from MAXI}

MAXI is the first astronomical payload to be
installed in the Japanese Experiment Module$-$Exposed Facility (JEM$-$EF or Kibo$-$EF) on the 
ISS and has high sensitivity as an all-sky X-ray monitor.   It started operation in 2009 August 
(Matsuoka et al.\ 2009) and has two types of 
X-ray slit cameras with wide fields of view and two types of X-ray 
detectors.    Although MAXI has two X-ray instruments, 
the Gas Slit Camera (GSC) and the Solid State Slit Camera  (SSC; Tsunemi et al.\ 2010), we have
analyzed only GSC data because it has both higher sensitivity and larger sky coverage.  The GSC consists 
of 12 one-dimensional position sensitive proportional counters having 5350 cm$^{2}$ detection 
area in total  that are sensitive to X-ray photons with energies from 2-20 keV, while the 
SSC is composed of 32 
X-ray CCD cameras with an energy range of 0.15-12 keV. The MAXI GSC signals for the sources were 
integrated orbit by orbit within a 3$^{\circ}$ $\times$ 3$^{\circ}$ square aligned to the scan 
direction centered on the source; the background was evaluated from two squares  offset 
by $\pm$3$^{\circ}$ along the scan direction in the sky, each of the same size as that of 
the source region. The extracted counts were normalized by dividing by a total exposure (in units 
of cm$^{2}$ sec) obtained with a time integration of the collimator effective area.  The data 
from all the activated counters are summed up. 
We downloaded 1 day average X-ray fluxes from MAXI\footnote {http://maxi.riken.jp/top/} 
for both Mrk 421 and 3C 454.3. This data
cover the period November 2009 to June 2010 for Mrk 421 and the period 
November--December 2009 for 3C 454.3.

\subsection{\bf Fermi $\gamma$-ray data}

The Fermi Space Telescope's Large Area Telescope (LAT) is designed to measure the cosmic gamma-ray flux up to 
$\sim$300 GeV. It is an imaging, wide field-of-view high-energy pair conversion
telescope with energy range from $\sim$20 MeV to $\ge$300 GeV (Michelson 2007).
As a service to the community, the LAT Instrument Science Operations Center provides daily 
and weekly averaged fluxes for a number of blazars. Fluxes and 1$\sigma$ uncertainties for 
1--300 GeV band, using preliminary instrument response functions and calibrations, are made available
on-line. We obtained the FERMI-LAT data for 3C 454.3 (on daily basis synchronized with optical data). 
Daily fluxes are not available 
for the source Mrk 421 but weekly values are available.   However, this sparse data
would be insufficient to provide adequate information on any correlated variability between 
the $\gamma$-ray and optical and X-ray bands, so we do not consider it further.

\section {METHODS}

\subsection {\bf Variability Detection Criterion}
Variability of the sources Mrk 421 and 3C 454.3  was investigated by computing  the commonly
used quantity 
$C$ (Romero et al.\ 1999) that is defined as the average of C$_{1}$ and C$_{2}$: 
\begin{equation}
C_{1} = \frac {\sigma(BL-Star A)}{\sigma(Star A-Star B )}~
\&~
C_{2} = \frac {\sigma(BL-Star B)}{\sigma(Star A-Star B)}.
\end{equation}

Using aperture photometry of the source and standard stars in the field, we determined the 
differential instrumental magnitude of the blazar and standard star  A, 
blazar and standard star B and standard star A vs.\ standard star B.
 Then, we determined observational scatters $\sigma$ (BL$-$Star) and $\sigma$ (Star A$-$Star B).  
If $C > 2.57$, the nominal confidence limit of the presence of variability is  99\%;
however, $C$ is not a true statistic and this confidence level is usually too conservative (de Diego 2010).
As discussed above, we used Star 1 and Star 2 for Mrk 421 and Star C and Star D for 3C 454.3 as
Star A and Star B, respectively, in the above expression.

We also test any claims of variability using a  proper statistic that is  reasonable
to employ for differential photometry, the $F$-test (de Diego 2010).  Given two sample variances such as 
$s_Q ^{2}$ for the blazar instrumental LC measurements and $s_* ^{2}$ for those of the standard star, 
then 
\begin{equation}
F=\frac {s_Q ^{2}}{s_* ^{2}} .
\end{equation}
The number of degrees of freedom for each sample, $\nu_Q$ and $\nu_*$, will be the same and equal to
the number of measurements $N$ minus 1 ($\nu = N - 1$). The $F$ value is then compared with the
$F^{(\alpha)}_{\nu_Q,\nu_*}$ critical value, where $\alpha$ is the significance level set for the test.
The smaller the $\alpha$ value, the more improbable that the result is produced by chance. If $F$ 
is larger than the critical value, the null hypothesis (no variability) is discarded. We have performed
the $F$-test at two significance levels (0.1\% and 1\%) which correspond to 3$\sigma$ and 2.6$\sigma$
detections, respectively.

The percentage variation in the LCs is calculated by using the variability amplitude parameter
$A$, introduced by Heidt \& Wagner (1996) and  defined as
\begin{equation}
A =  \frac{100}{<A>}\times \sqrt{{(A_{max}-A_{min}})^2 - 2\sigma^2}(\%), 
\end{equation}
where $A_{max}$ and $A_{min}$ are the maximum and minimum fluxes in the calibrated LCs of
the blazar, $<A>$ is their mean, and the average measurement error of the blazar LC is $\sigma$. 

The calculated $F$ statistics, $C$ ``statistics'' and variability amplitude ($A$) 
values are listed in Tables 2 and 3. 

\subsection{\bf Structure Function}

The structure function (SF) is a technique that can provide some information on the nature of the physical
process causing any observed variability. 
The SF is free from any constant offset in the time series (Rutman 1978; Simonetti et al.\ 1985; 
Paltani et al.\ 1997). For details about the SF as we have employed it, see Gaur et al.\ (2010).

We have carried out the SF analysis of all of those LCs which satisfy the
variability detection criteria. Recently, Emmanoulopoulos et al.\ (2010) have discussed
the weaknesses of the SF method, including spurious indications of timescales and periodicities.
So, we have cross checked the SF results by the DCF method to look for any hints of periodicity.

\subsection{\bf  Discrete Correlation Function Analysis}

The Discrete Correlation Function (DCF) was first introduced by Edelson \& Krolik (1988) and 
was generalized by Hufnagel \& Bregman (1992) to include a better error estimate. 
For details about the DCF see 
Tonnikoski et al.\ (1994), Hovatta et al.\ (2007) and references therein.
For two different data trains, any strong peak in the DCF can indicate the possible time lag.

\section{\bf RESULTS}
\subsection{\bf Intra$-$Day Variability of blazars in the R-band}

{\bf Mrk 421 } We intensively observed the blazar Mrk 421 using a R filter during nine nights from 
21 November 2009 to 9 April 2010. The complete observing log of the blazar is in Table 1.  
The LC of the blazar Mrk 421 (calibrated)  and the differential instrumental magnitude (of 
Star 1-Star 2) are displayed in Fig.\ 1 for those nine nights. 
We have performed both $C$ and $F$ tests on those nine nights; however, no
genuine intra-day variability was
found during any of them. The $C$ and $F$ values are given in Table 2 and they never 
exceed  the formal significance criteria.

{\bf 3C 454.3} We observed the blazar 3C 454.3 through an R filter on seven nights from 22 November
2009 through 21 December 2009. The LC of 3C 454.3 and the differential instrumental magnitude
(StarC - StarD) are displayed in Fig.\ 2. The complete observing 
log for this blazar is given in
Table 1.
 The  $C$, $F$ and $A$ values for this IDV are listed in Table 2.
We found that the $C$ values and results of the $F$-test both show significant values for four 
nights (22 Nov, 13 Dec, 15 Dec and 20 Dec 2009) so  
it is clear that the source has shown IDV during four nights of our observations.
We have carried out the SF and DCF analysis of those four LCs
satisfying the variability detection criteria and these are shown in Fig.\ 3;  
 however, no significant variability timescale was detected in
any of those LCs. 
   
\subsection{\bf Short$-$Term Flux and Color Variability}

\subsubsection{\bf Mrk 421}

The nightly LCs of Mrk 421 (calibrated magnitude) in B, V, R, (B$-$V), (V$-$R) and (B$-$R) 
are plotted in the different panels in Fig.\ 4. Here we estimate the 99\% confidence 
detection level of short-term variability using the detection tests described in Section 3.1 
and calculate the variability amplitude  using Eq.\ (3).

{\bf B pass-band:} The short-term LC of Mrk 421 in the B-band is displayed in the
upper left panel of Fig.\ 4. The maximum variation noticed in the source is 0.70 mag (between
its brightest level at 14.26 mag on JD 2455187.52290 and the faintest level at 14.96 mag on
JD 2455296.10980). The values of the $C$ and $F$-tests support the existence of short-term variations 
in the source in $B$-band observations. We calculated short-term variability amplitude using Eq.\ (3) 
and found that the source has varied $\sim$62\%.

{\bf V pass-band:} The short-term LC of Mrk 421 in the V-band is shown in the middle left 
panel of Fig.\ 4. The maximum variation noticed in the source is 0.59 mag (between its brightest level 
at 13.59 mag on JD 2455207.59130 and the faintest level at 14.18 mag on JD 2455296.10703). The values of 
the $C$ and $F$-test also support the existence of short-term variation in the source in these V-band observations. The
short-term variability amplitude is $\sim$53\%.

{\bf R pass-band:} The corresponding LC of Mrk 421 in the R-band is in the lower left panel 
of Fig.\ 4. The maximum variation noticed in the source is 0.49 mag (between its brightest level at 12.64 mag 
on JD 2455187.50831 and the faintest level at 13.16 mag on JD 2455158.44563). Again, the  $C$ and $F$-tests 
both indicate short-term R-band variations are present with an amplitude of $\sim$44\%.

{\bf (B$-$V) color:} The short-term LC of Mrk 421 in the (B$-$V) color is shown in the
lower right panel of Fig.\ 4. The maximum variation noticed in the source is 0.17 mag (between
its color range 0.39 mag at JD 2455271.11511 and 0.56 mag at JD 2455358.12050). 
However, neither the $C$- nor $F$-test provide  support for the existence of significant (B$-$V) color 
variations in our observations.

{\bf (V$-$R) color:} The short-term LC of Mrk 421 for (V$-$R) is displayed in the
upper right panel of Fig.\ 4. The maximum variation noticed in the source is 0.13 mag between
the color range 0.25 mag at JD 2455207.59130 and 0.38 mag at JD 2455296.11602.  Again, no significant (V$-$R) color variations are seen in our observations.

\subsubsection{\bf 3C 454.3}

The nightly LCs of 3C 454.3 (calibrated magnitude) in B, V, I, (B$-$V) and (V$-$R) 
are plotted in Fig.\ 2 and that for the R band is plotted in Fig.\ 5.
Here, $C$- and $F$-tests could not be performed on the entire large data-sets because only the nominal 
calibrated magnitudes of 3C 454.3 are
available on SMARTS site  (for B, V and R bands) and without the unavailable data for
comparison stars we cannot compute those quantities.  The same is the case for the KANATA  
data (which is only for the R band). Therefore we performed C- and F-tests on the ARIES data only
and those values are quoted in Table 3. However, when computing the amplitude of variation, 
we have used the whole data-set including data from SMARTS as well as KANATA.

{\bf B pass-band:} The short-term LC of 3C 454.3 in the B-band is displayed in the
top of the middle bottom panel of Fig.\ 2. The maximum variation noticed in the source is 1.593 mag 
(between its brightest level at 14.937 mag on JD 2455179.05072 and the faintest level at 16.53 mag on
JD 2455143.58696). We performed C- and F-tests on the ARIES data and found the variations to be
highly significant. We calculated the STV amplitude of whole dataset using Eq.\ (3) and found that 
the source has varied by $\sim$ 126\%.

{\bf V pass-band:} The short-term LC of 3C 454.3 in the V-band is given in the
middle of the middle bottom panel of Fig.\ 2. The maximum variation noticed in the source is 
1.581 mag (between its brightest level at 14.331 mag on JD 2455179.05491 and the faintest level 
at 15.912 mag on JD 2455141.57627).  
Our C- and F-tests  performed on the ARIES data yielded  highly significant values.
The amplitude of the variation of whole data-set is $\sim$ 125\%. 

{\bf R pass-band:} The short-term LC of 3C 454.3 in the R-band is shown in 
 Fig.\ 5.  In this plot we have combined our data with those provided by the
SMARTS and KANATA telescopes. Again, C- and F-tests are performed only on ARIES data and values are 
much higher than 0.999\% significance.
The maximum variation noticed in the source is 1.578 mag (between
its brightest level at 13.709 mag on JD 2455167.91 and the faintest level at 15.505 mag on
JD 2455141.5774). From the figure, it is clear that there are two flares with first flare of 0.44
mag peaking near JD 2455152 and second flare with 1.31 mag peaking near JD 2455171. 
The amplitude of total variation in the STV light curve is $\sim$ 125\%. 

{\bf I pass-band:} The short-term LC of 3C 454.3 in the I-band is displayed in the
lower portion of the lower middle panel of Fig.\ 2. The C- and F-tests performed on ARIES data
still yield high significance for the variations.
The maximum variation noticed in the source is 0.656 mag (between
its brightest level at 13.105 mag on JD 2455179.04786 and the faintest level at 13.761 mag on
JD 2455186.08768).  As SMARTS data is not available in the I band, we have many fewer data points 
and so the observed STV  amplitude is only $\sim$ 59\%.

{\bf Correlated variations between color and magnitude:}
\ \ Color--magnitude plots of 3C 454.3 are displayed in bottom right panel of Fig.\ 2. The upper
 and lower sub-panels respectively show the (V$-$R) and (B$-$R) colors plotted with respect to V magnitude.  The straight 
lines shown are the best linear fit for each of the color indices, $Y$, against magnitude, $X$, for each 
of the sources: $Y=mX+c$. For (V$-$R), the fitted value for the slope of the curve, $m=-$0.08 and the
constant, $c=1.67$. Also, the linear Pearson correlation coefficient, $r =-0.81$ and the corresponding
null hypothesis probability value is minute, at $p=5.26 \times 10^{-10}$, thus indicating a very strong correlation.  
Similarly, for (B$-$V), the fitted values 
for the slope of the curve is $m=-0.05$ and that for the constants is $c=1.30$. 
The linear Pearson correlation coefficient is, $r =-0.48$ and the corresponding $p= 0.002$. 
The negative slopes imply the opposite correlation between brightness and color, so the source 
exhibits a redder when brighter behavior.

\subsection {\bf Correlated Variability }
\subsubsection{\bf Mrk 421}
Fig.\ 6 displays the X$-$ray and optical LCs of the  2009--2010 observing season.
We can see from the  X$-$ray LC that a  brightness 
increase, apparently corresponding to a modest flare, occurred around 
JD=2455197.  Meanwhile, the optical observations in R-band  show an increase in brightness  
around JD=2455177, which peaks at 12.93 mag and a decline after   
at JD=2455197 (though we don't have data for that entire interval and we probably  have
 missed the actual peak). Still we can say that the flares 
in the X$-$ray and optical bands are seen in the same general time span. 

In the X$-$ray LCs, there is an even bigger flare peaking at JD=2455243.  Unfortunately,
we could not obtain any optical data between 21 January and 13 March 2010, so we appear to have 
missed this flare in the optical. Therefore we have performed the DCF analysis only in the temporal 
region containing the first flare region from the beginning of the data train at JD=2455150 to   
JD=2455230 (shown by a vertical line in Fig.\ 6). In the X$-$ray data-set, we 
found that the source flux counts were given as negative on a few days 
of observations and such days were omitted in our analysis. The DCF between the X$-$ray 
is displayed in the top right panel of Fig.\ 6. The distribution of points 
have two significant maximas, $\sim$ 0.88, at a negative time lag of 9.5 days, and
$\sim$ 0.64 at a positive time lag of 6.2 days.
The negative time lag has a greater DCF value which apparently
implies that variations at the lower frequency lag behind  those at the higher 
frequency; however, given our sparse optical data we cannot claim it a strong correlation.

\subsubsection{\bf 3C 454.3}

For this source we are able to cross-correlate the variability across $\gamma$-ray, X-ray and optical bands. 
A prominent peak at around JD 2455167 and a short flare near JD 2455170 are seen in both 
the optical LC and the $\gamma$-ray LC (Fig.\ 7).  However, these features are 
not seen in the X-ray LCs also shown in Fig.\ 7. We performed a DCF between the $\gamma$-ray (1-300 GeV) 
flux and the LCs in the optical R-band. This DCF shows a large peak correlation amplitude of
$\sim$ 0.90 at $\tau=4.5\pm1$ days which indicates gamma-ray frequencies are leading the optical 
frequencies.
We have used Monte-Carlo simulations to test the strength of the peak correlation amplitude.
To do so we fit the original optical and $\gamma$-ray light
curves with low order polynomials to extract those trends. From the de-trended LCs, we obtained
the underlying probability distributions of the fluctuations. Using this probability distribution, we performed random
sampling and thereby generated 3000 realizations of a random LC with the underlying statistical
properties of the original LC  for both the optical
and $\gamma$-ray bands. Next, taking each combination of these randomly generated LCs, we determined their
DCFs. 
We checked the ability of DCF to find the real time lags. Since the simulated DCFs also
 gave time lags of 4.5 days, we examined the $p$-values at a lag of 4.5 days.  
The null hypothesis examined here is that the
highest correlation value at the given lag of the actual data is higher than the simulated 
light curve's DCF at the same time lag.  We found 
correlation value of 0.70 to be at a $p$-value of 0.99 at the time lag of 4.5 days. Hence, our observed
correlation amplitude of 0.90 indicates significance well above 0.99.

\subsection{X-ray Hardness$-$Ratio Analysis}
We can crudely study the spectral variability of the source through the hardness ratio. It is defined as 
either the ratio of counts ($b/a$) or the ratio of the difference and sum of the counts $(b-a)/(b+a)$, 
(Zhang et al.\ 2006).  We use the former definition. In Fig.\ 6, the top left panel shows the 
hardness ratio, 4-10 keV/2-4 keV, as a function of intensity for Mrk 421, while the middle right panel 
in Fig.\ 7 shows the same quantity for 3C 454.3. 
For Mrk 421, there is no correlation ($r =-0.17$ with $p = 0.50$) between the source hardness 
ratio and its intensity. For 3C 454.3, there appears to be a weak correlation ($r=0.40$ with $p = 0.1$) 
between source hardness ratio and its intensity.

\section {\bf Discussion And Conclusions} 

During our observation period of November 2009 to June 2010, we monitored Mrk 421 for IDV in 9
nights, but genuine IDV in any of the B, V and R pass-bands was not detected 
in any night as both $C$ and $F$ values were always less than the 99\% significance levels.
We noticed the existence of significant short-term variability in this blazar from our 
observations and the source showed a maximum variation in the B band of 0.70 magnitudes.  
The total short-term variation detected in our observations 
in the B, V, and R bands are $\sim$ 62\%, $\sim$ 53\% and $\sim$ 44\%, 
respectively. Our data don't show any truly significant color variations in the (B$-$V) 
 and (V$-$R) colors.

We monitored 3C 454.3 for IDV during 7 nights during November and December 2009.
We performed $C$ and $F$-tests and found that 4 of these 7 nights showed genuine IDV. 
To search for any variability timescale, we computed the SF and DCF, but no significant variability 
timescales were 
detected. We noticed the existence of significant short-term variability in this FSRQ blazar but could 
not perform $C$ or $F$-tests on the whole data-sets as we had only calibrated data from the Chilean and 
Japanese telescopes and not the needed data for comparison stars. So we performed $C$- and $F$-tests only 
on the ARIES data but we calculated variability amplitudes based on the full data-set. 
We found short-term variability amplitudes in the B, V, R and I pass-bands to be $\sim$126\%,
$\sim$125\%, $\sim$125\% and $\sim$59\%, respectivel; however the last is based on much less data, 
and so is
consistent with the others.   We also find that the colors of 3C 454.3 are redder at brighter levels,
supporting the findings of Villata et al.\ (2006).  \par

Several models have been developed to explain IDV and short-term variability in radio-loud
AGNs, and most of these can be classified as the shock-in-jet models and accretion-disk-based models (Wagner \& Witzel 1995;
Ulrich et al.\ 1997; Urry \& Padovani 1995; Mangalam \& Wiita 1993; Chakrabarti \& Wiita 1993 and 
references therein). For blazars in the outburst state, both IDV and short-term variability are strongly 
presumed to be completely dominated by the relativistic jet required in all models of radio-loud AGNs. 
However, IDV and short-term variability of 
blazars in the low state, as well as that of radio-quiet AGN, might be explained by  models based on instabilities or hot-spots on the accretion disk (e.g.\ Mangalam \& Wiita 1993; Chakrabarti \& Wiita 1993).

Short-term variations were seen between many nights. These flux variations can be reasonably explained 
by models involving relativistic shocks propagating outwards (e.g., Marscher \& Gear 1985; Wagner \& Witzel 1995; 
Marscher et al.\ 1996). The larger flares are expected to be produced by the emergence and motion 
of a new shock triggered by some strong variation in a physical quantity such as velocity, electron 
density or magnetic field moving into and through the relativistic jet. 
IDV reported in the flaring and high states is generally attributed to the shock moving down the 
inhomogeneous medium in the jet. Non-detection of IDV in Mrk 421 indicates that the relativistic 
shock along  the jet  has neither changed significantly in speed, nor direction with respect 
to the line of sight nor were any significant non-antisymmetric 
structures carried outward in the relativistic magnetized jets.

In the X$-$ray LC, there are two strong flares, with the first flare essentially coinciding with
the flare in the optical band. Flares from the source was reported by Isobe et al.\ (2010) in their 
2010 January-February observations from MAXI GSC. The maximum 2-10 KeV flux in the January and February 
flares were 120$\pm$10 mCrab and 164$\pm$17 mCrab, respectively, and the latter maximum is the highest 
among those reported from Mrk 421 so far. Also, the MAXI GSC spectrum around the maximum of the 
flares was found to be consistent with a spectral index indicative of synchrotron radiation. 
We found that there was
a correlation between the optical and X-ray bands ($DCF = 0.88$) at a negative 
lag of 9.5$\pm$2 days, which indicates variations at the higher frequencies lead those at  the lower frequencies.
However, because of the sparse optical data, we cannot claim this to be an absolutely convincing  correlation.

Such a time delay between bands would be naturally produced by the frequency stratification 
expected in the simplest shock-in-jet model. This effect occurs when the electrons
are energized along a surface (e.g., shock front) and then move away from it at a speed close to $c$
 as they  lose energy via both synchrotron and IC processes (Marscher \& Gear 1985). The 
highest energy electrons will suffer the most severe radiative losses so that they only maintain 
these high energies and produce high energy photons over short distances.   Hence,
the highest frequency radiation can be emitted only within a thin sheet behind the shock
front. The thickness of the sheet increases as the frequency decreases until the frequency is so low
that radiative losses are negligible across the entire shocked region.  Thus, in 
this standard model a flare caused by a
shock spreads across  multiple wavebands, but the time-scale of variability can be much shorter at
higher frequencies.  The X-ray flux in a shock-induced
flare should therefore peak first, followed by the optical and then lower frequencies, 
whereas the $\gamma$-ray flux, if produced by IC scattering  from the X-ray (or EUV) photons 
could then lead the optical, while if produced from the optical or IR photons, would lag the optical.

The blazar 3C 454.3 was also found in flaring state during November-December 2009. 
The  present observations confirmed the presence of fairly significant color variations that
support the presence of thermal emission beneath the dominant non-thermal jet. Raiteri et al.\ (2007) 
previously found good evidence for big and little ``blue bumps" in the SED of 3C 454.3 during periods 
of low emission.
Prominent peaks are seen nearly simultaneously in the optical and $\gamma$-ray bands but 
these features are not seen in the X-ray band. The DCF of the $\gamma$ and optical bands 
shows a peak correlation amplitude of $\sim$0.90 at $\tau$=4.5$\pm$1 indicating $\gamma$-rays 
leading the optical rays.   
Similar behavior was found by Bonning et al.\ (2009) for the flare in July 2008. 
The correlated optical/$\gamma$-ray variability supports the external Compton
model in which relativistic electrons in the jet radiate radio through UV synchrotron photons and inverse
Compton scatter IR/optical photons to hard X-ray and $\gamma$-ray energies (Dermer \& Schlickeiser 1993;
Sikora et al.\ 1994).  The lack of correlation seen in the DCF for X-rays with respect 
to optical/gamma bands can be reasonably understood if the X-rays are coming from 
low-energy electrons inverse Compton scattering
external UV photons, rather than higher energy electrons producing synchrotron photons.
These lower energy electrons would vary more slowly and thus plausibly give rise to the relatively 
stable X-ray emission. The modest quantity of our data, which allows for the likely identification of 
just one flare between the optical and $\gamma$-ray bands, precludes our attempting to produce more 
detailed models.

\acknowledgements
 
We gratefully acknowledge Kanata team for observations and Prof.\ M.\ Uemura for providing us published Kanata data on
 the blazar 3C 454.3. HG is thankful to Dr.\ K.\ Nilsson for a discussion about host galaxy 
contributions to observed flux.
We thank the referee for several very helpful suggestions.
This research has made use of MAXI data provided by RIXEN, 
JAXA and the MAXI team. The acquisition and analysis of the SMARTS data are supported by Fermi 
GI grants 011283 and 31155 (PI C.\ Bailyn). This research has made use of the NASA/IPAC 
Extragalactic Database (NED)
which is operated by Jet Propulsion Laboratory, California Institute of Technology,
under contract with the National Aeronautics and Space Administration.
{}

\clearpage
\begin{figure}
 \centering
\includegraphics[width=2.2in,height=2.5in]{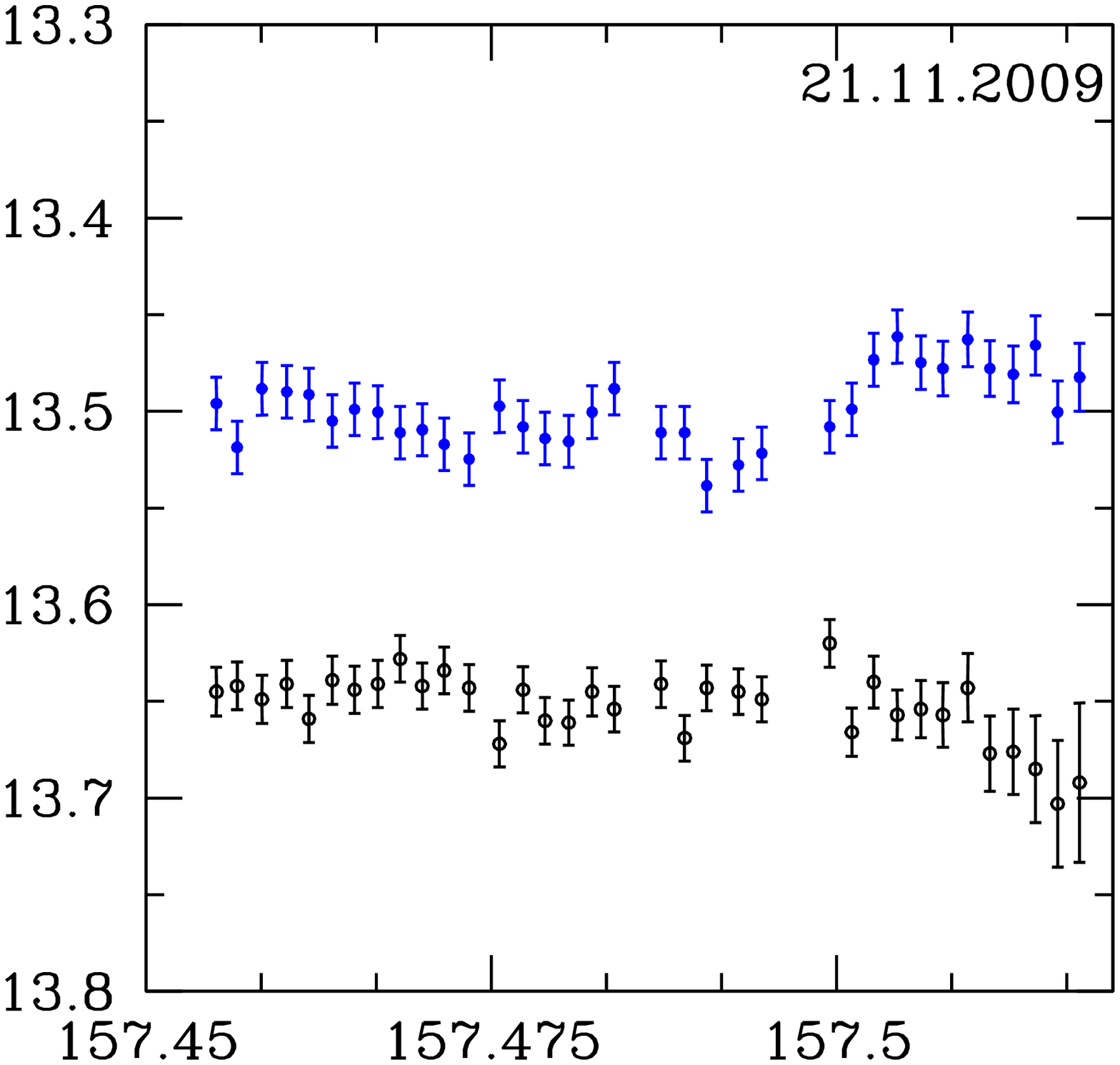}
\includegraphics[width=2.2in,height=2.5in]{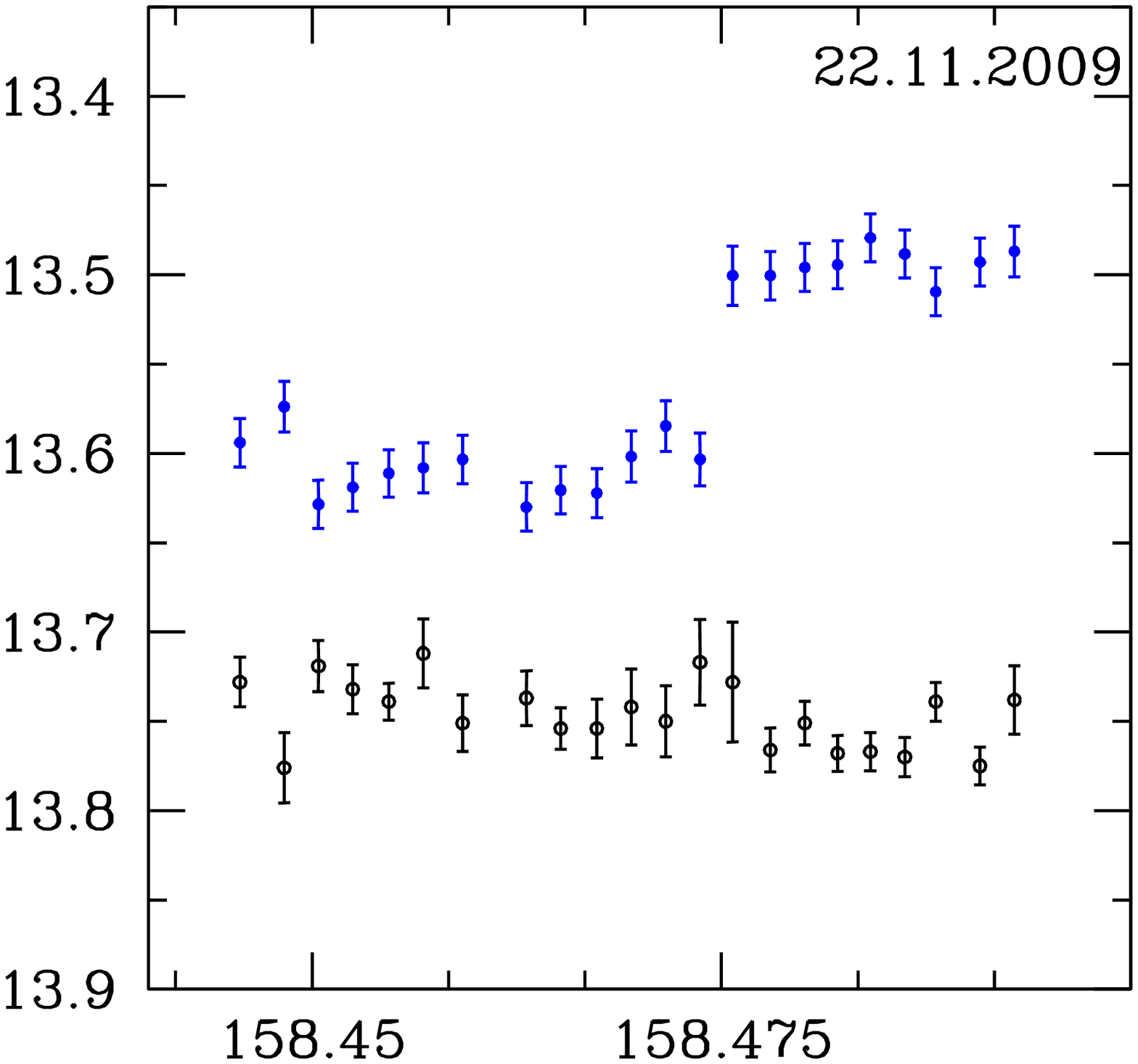}
\includegraphics[width=2.2in,height=2.5in]{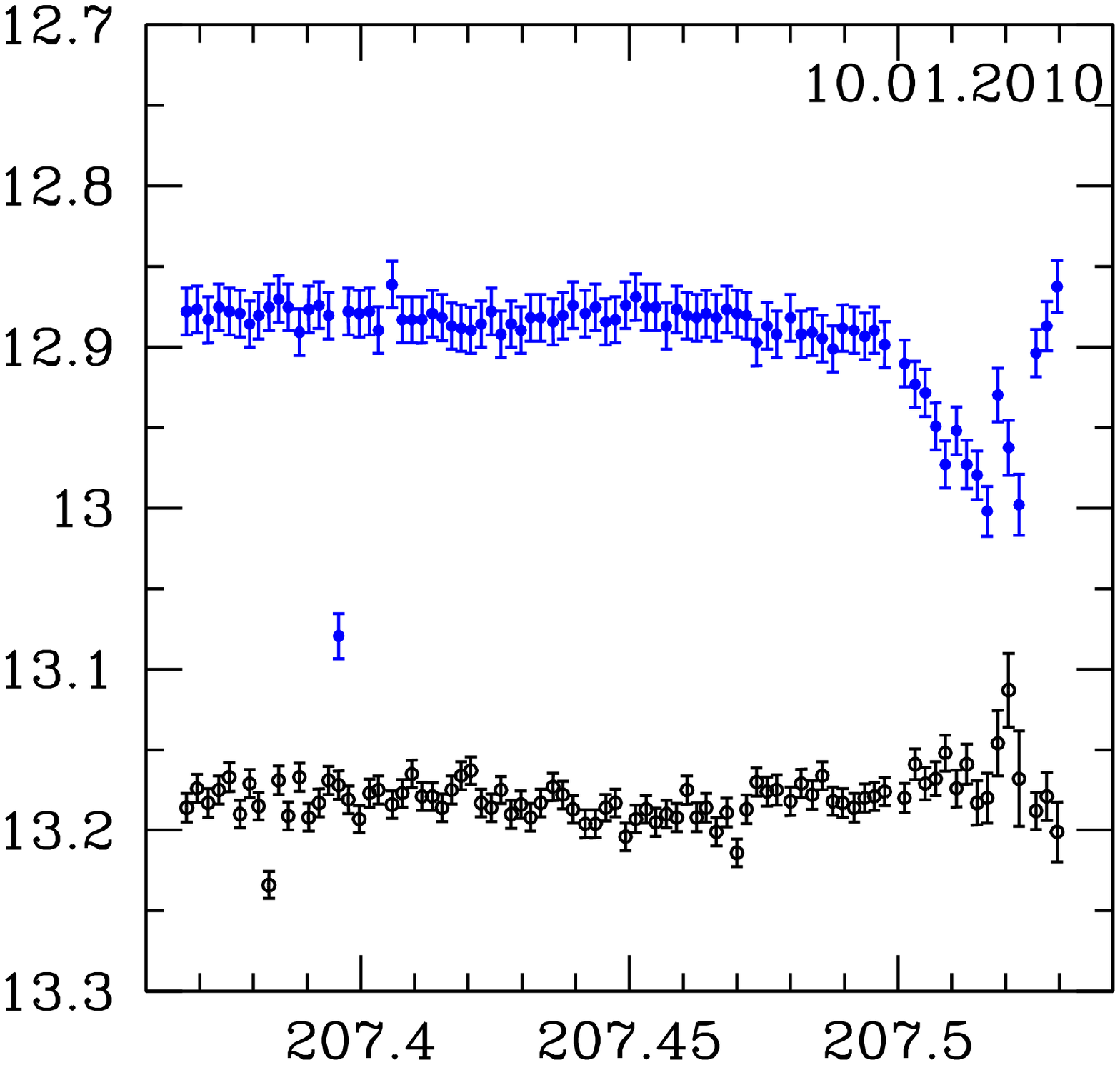}
\includegraphics[width=2.2in,height=2.5in]{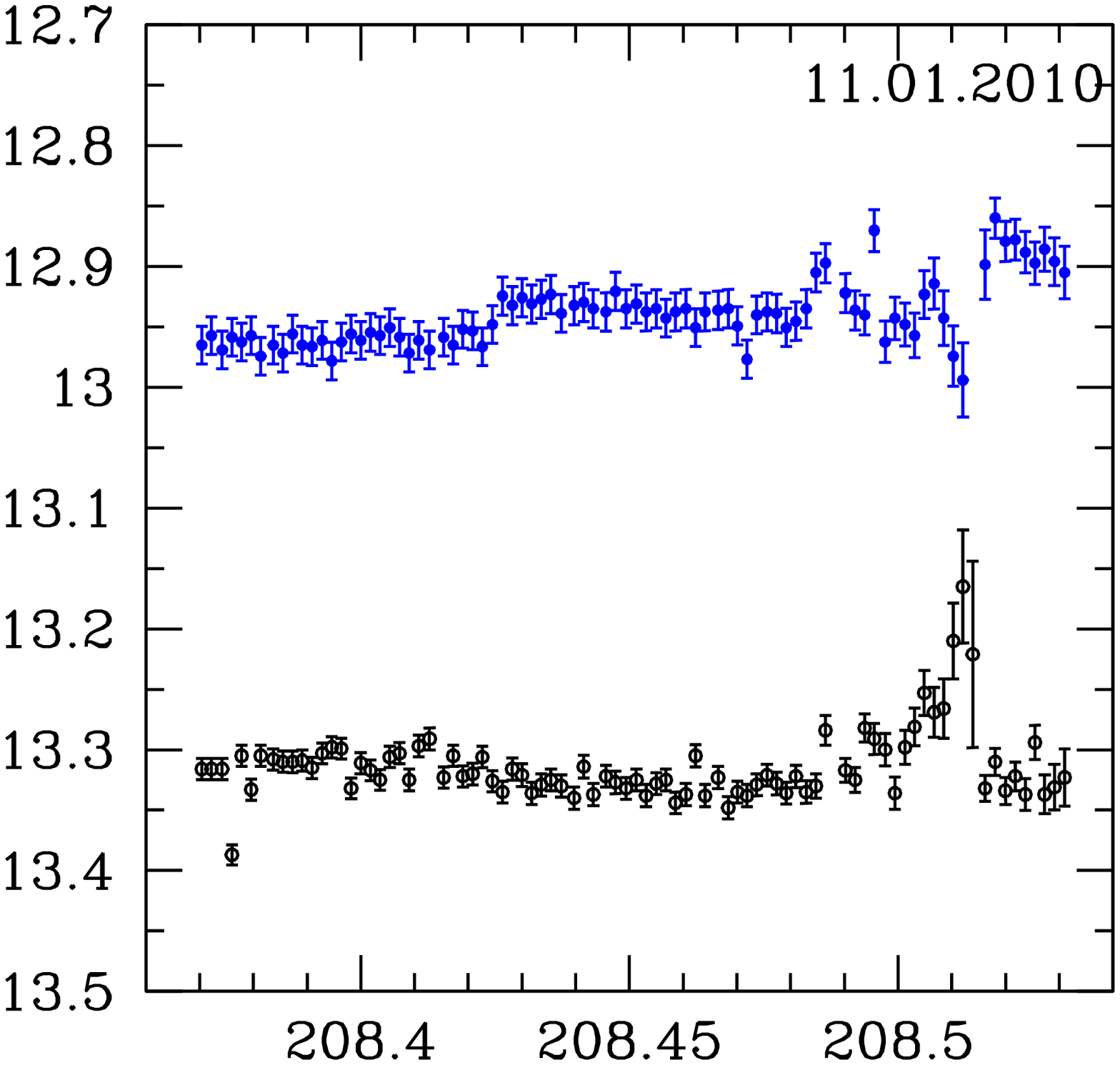}
\includegraphics[width=2.2in,height=2.5in]{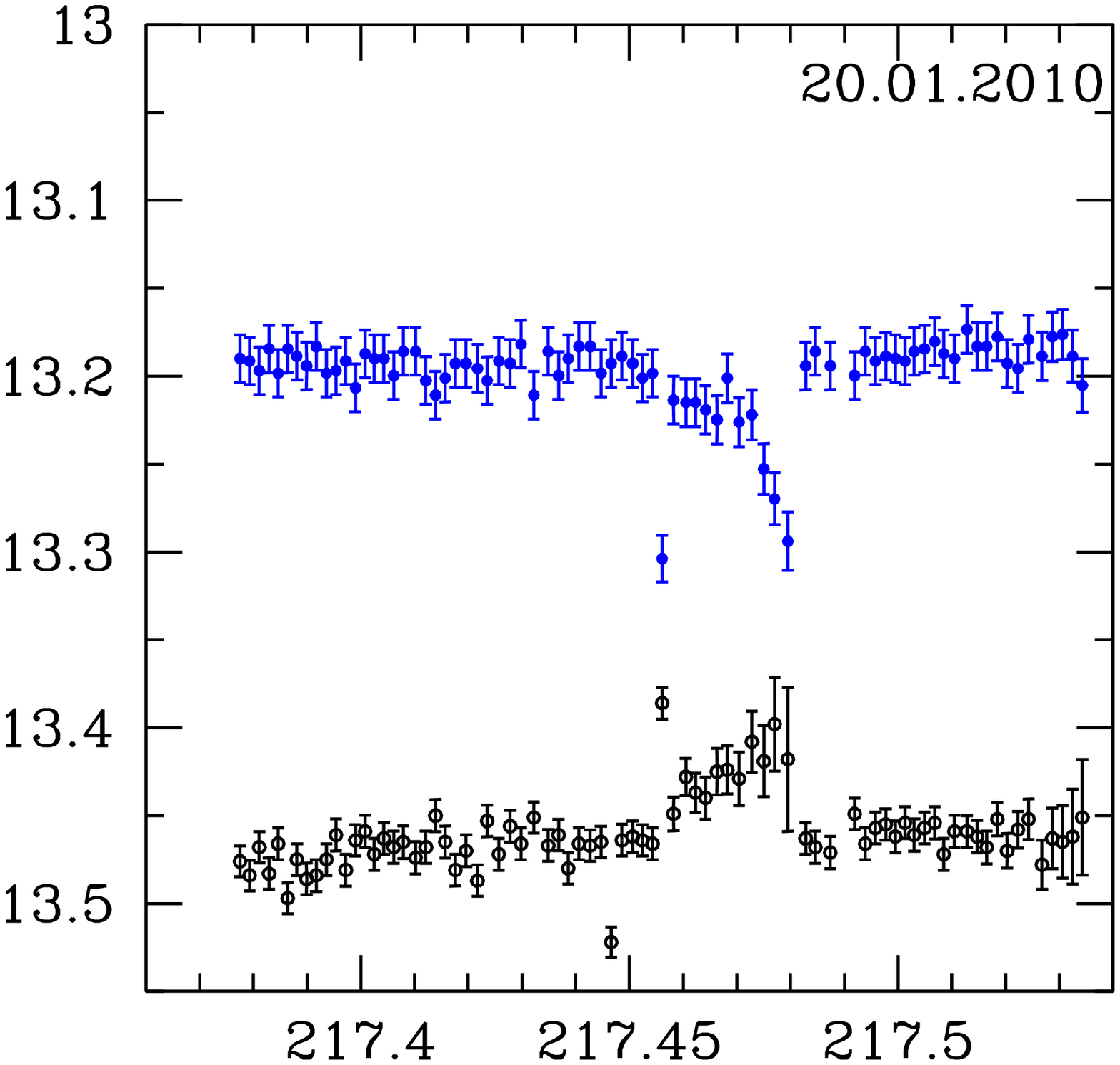}
\includegraphics[width=2.2in,height=2.5in]{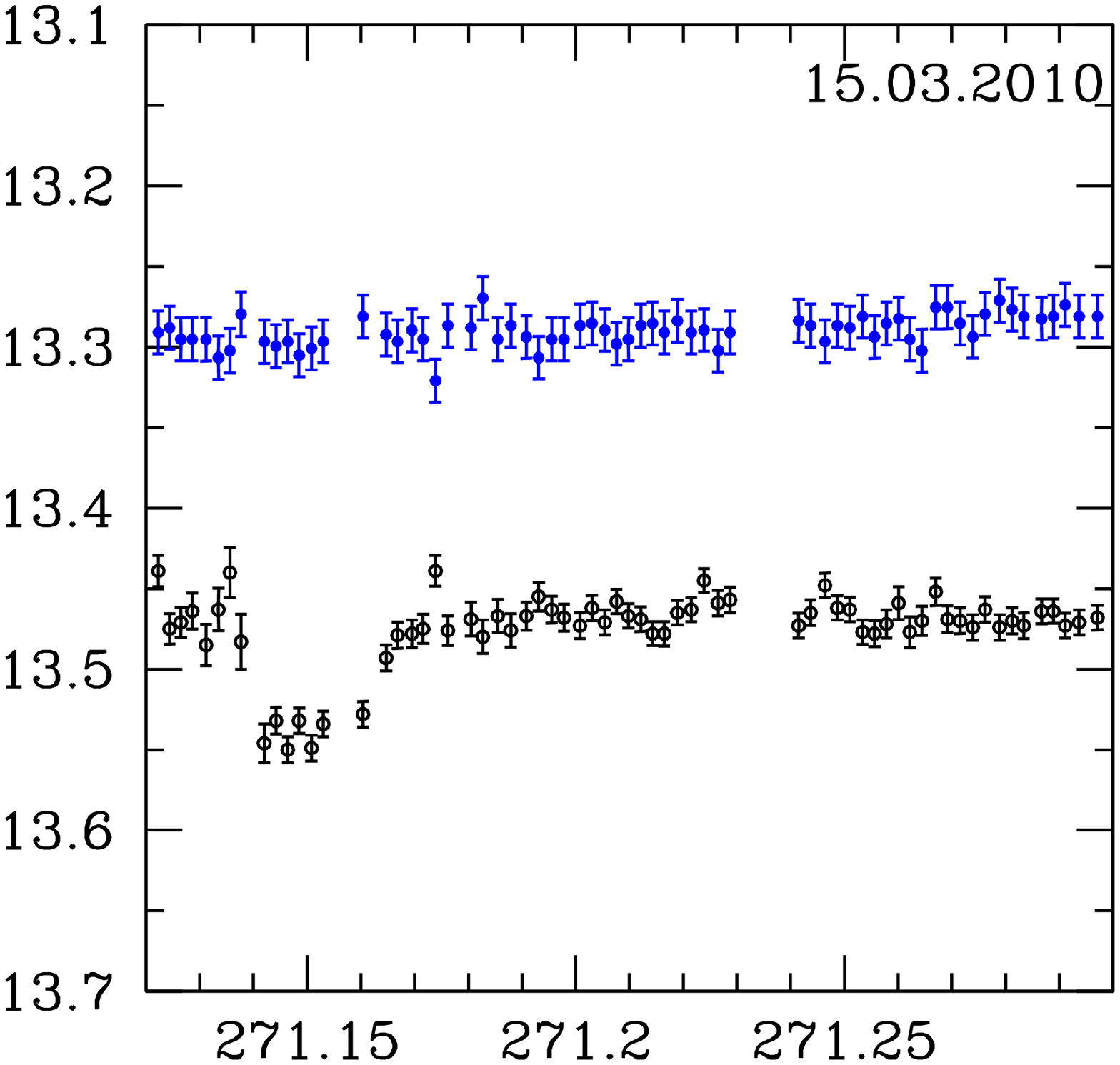}
\includegraphics[width=2.2in,height=2.5in]{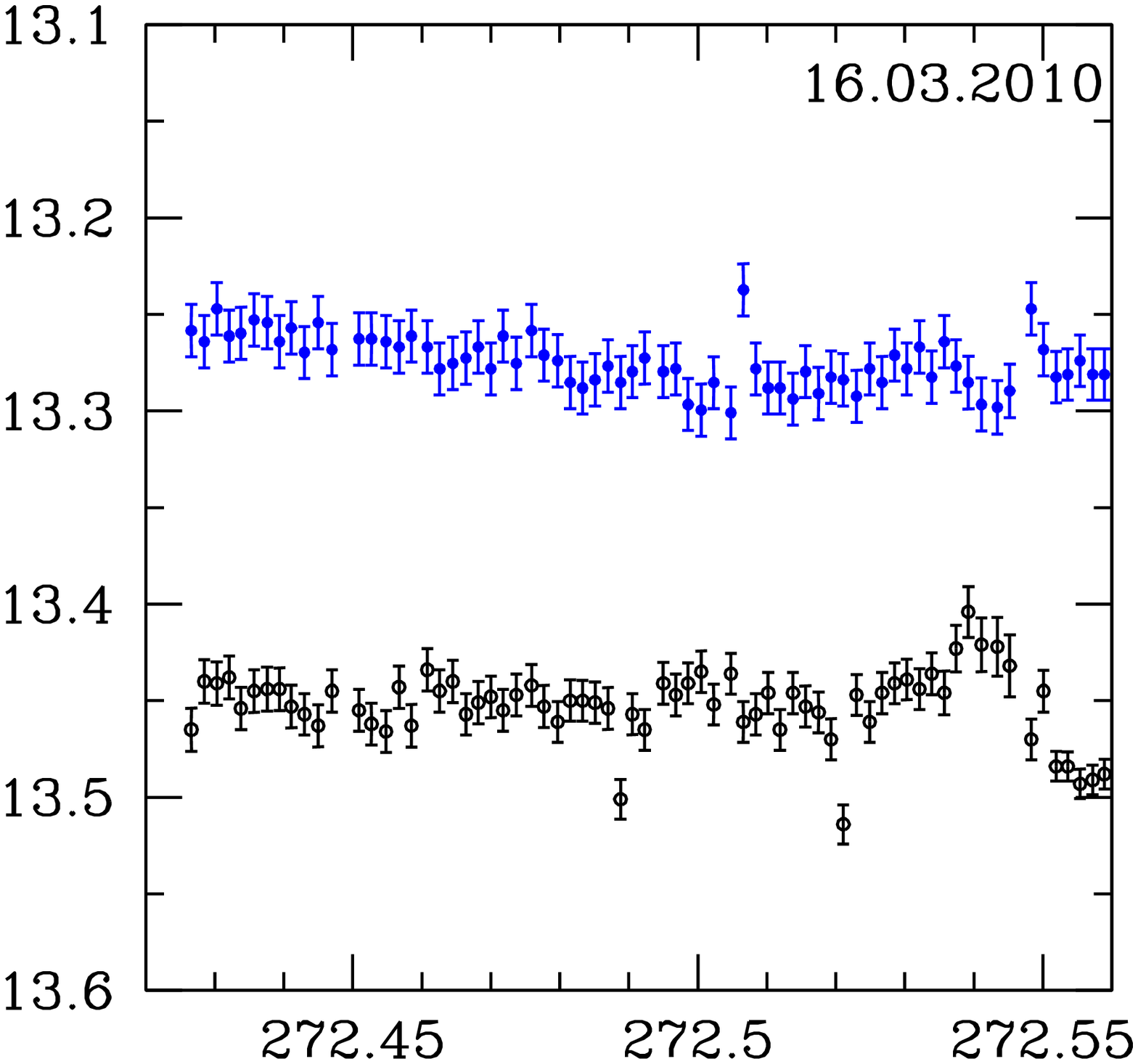}
\includegraphics[width=2.2in,height=2.5in]{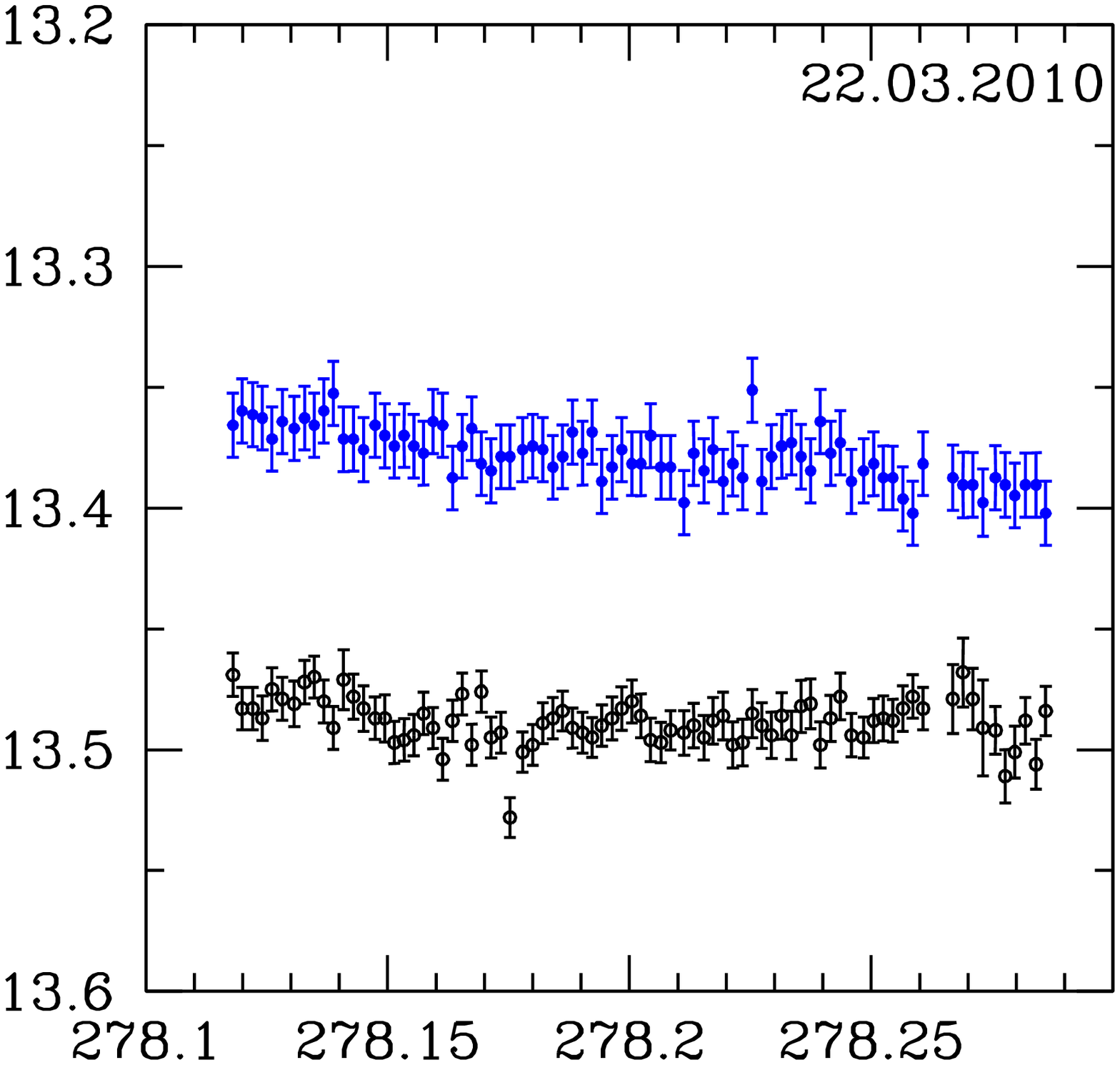}
\includegraphics[width=2.2in,height=2.5in]{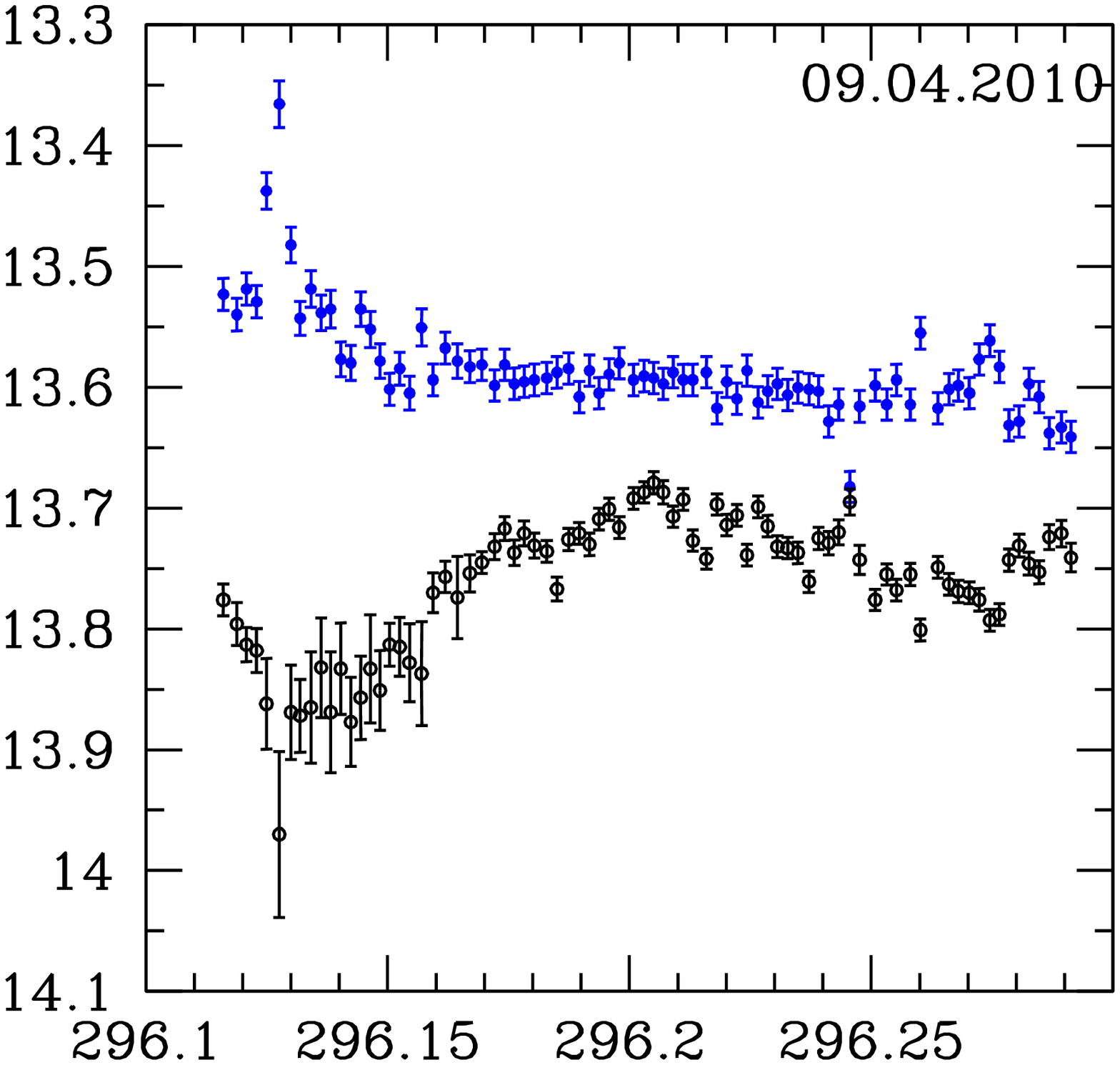}
\caption{R band light curves (LCs) of Mrk 421. Upper curves are the calibrated LCs of Mkr 421 
(w.r.t Star 1). Lower curves are the differential instrumental magnitudes of Star 1 \& Star 2 
with arbitrary offsets.}
\end{figure}

\clearpage
\begin{figure}
 \centering
\includegraphics[width=2.2in,height=2.5in]{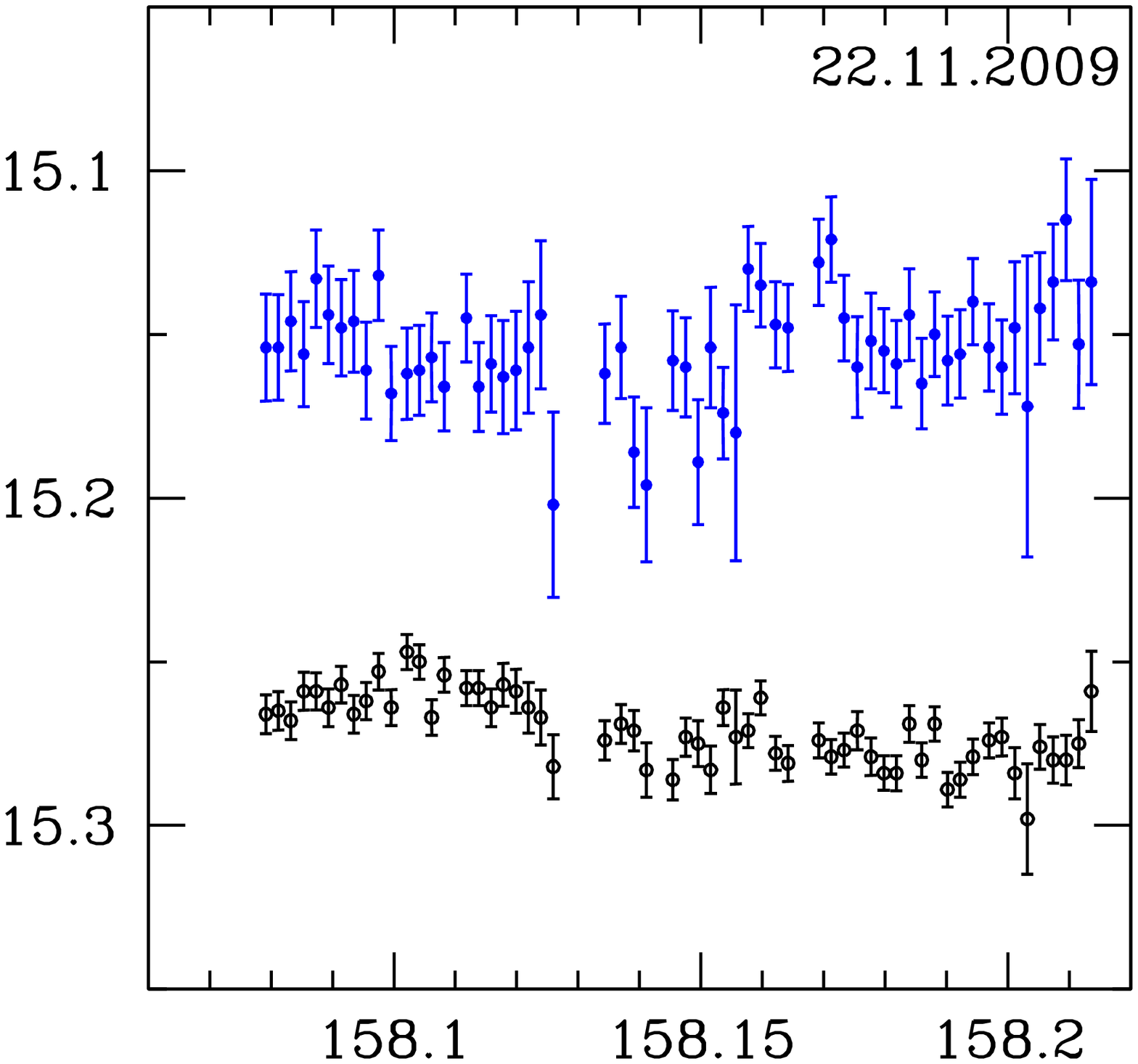}
\includegraphics[width=2.2in,height=2.5in]{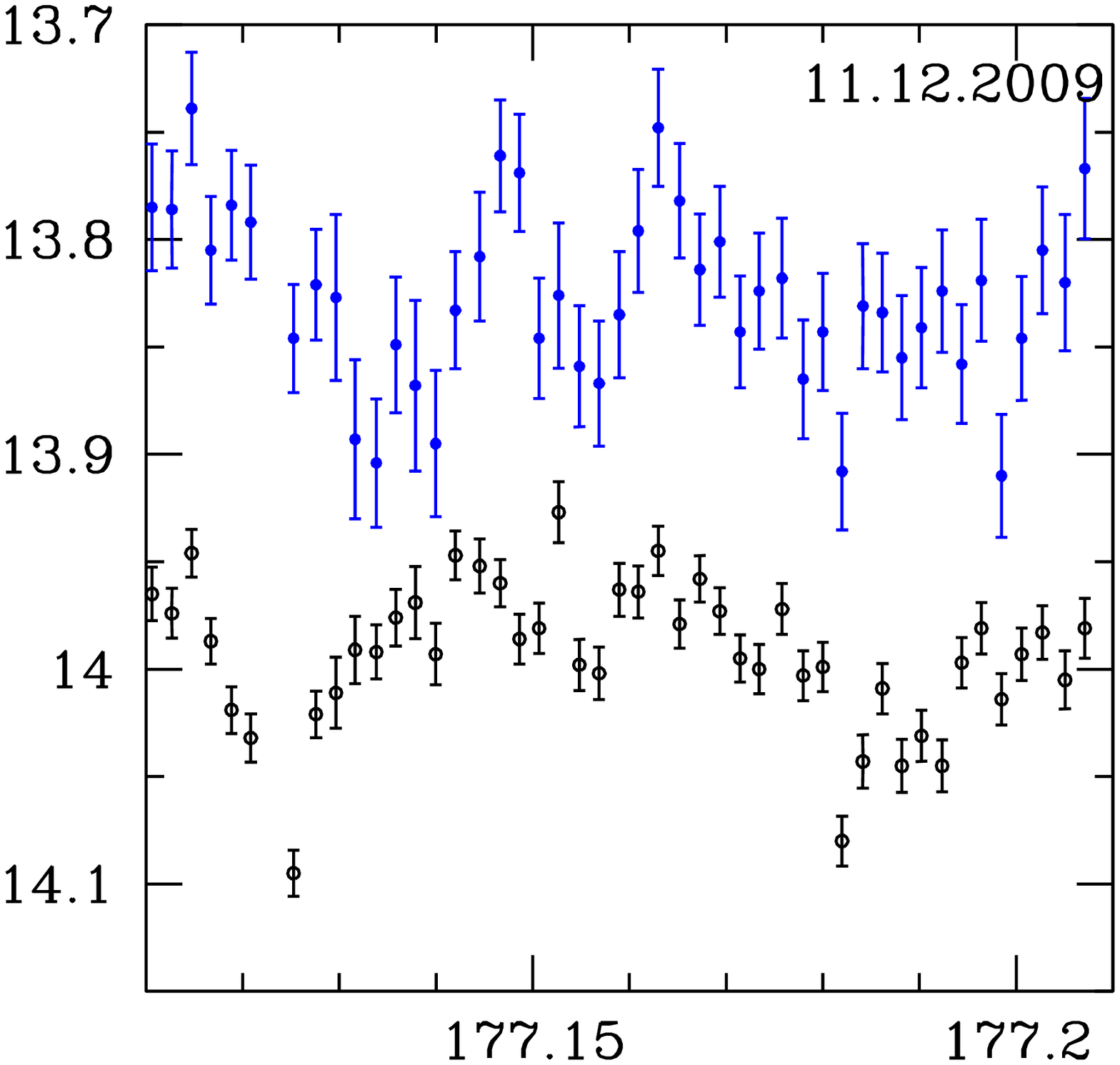}
\includegraphics[width=2.2in,height=2.5in]{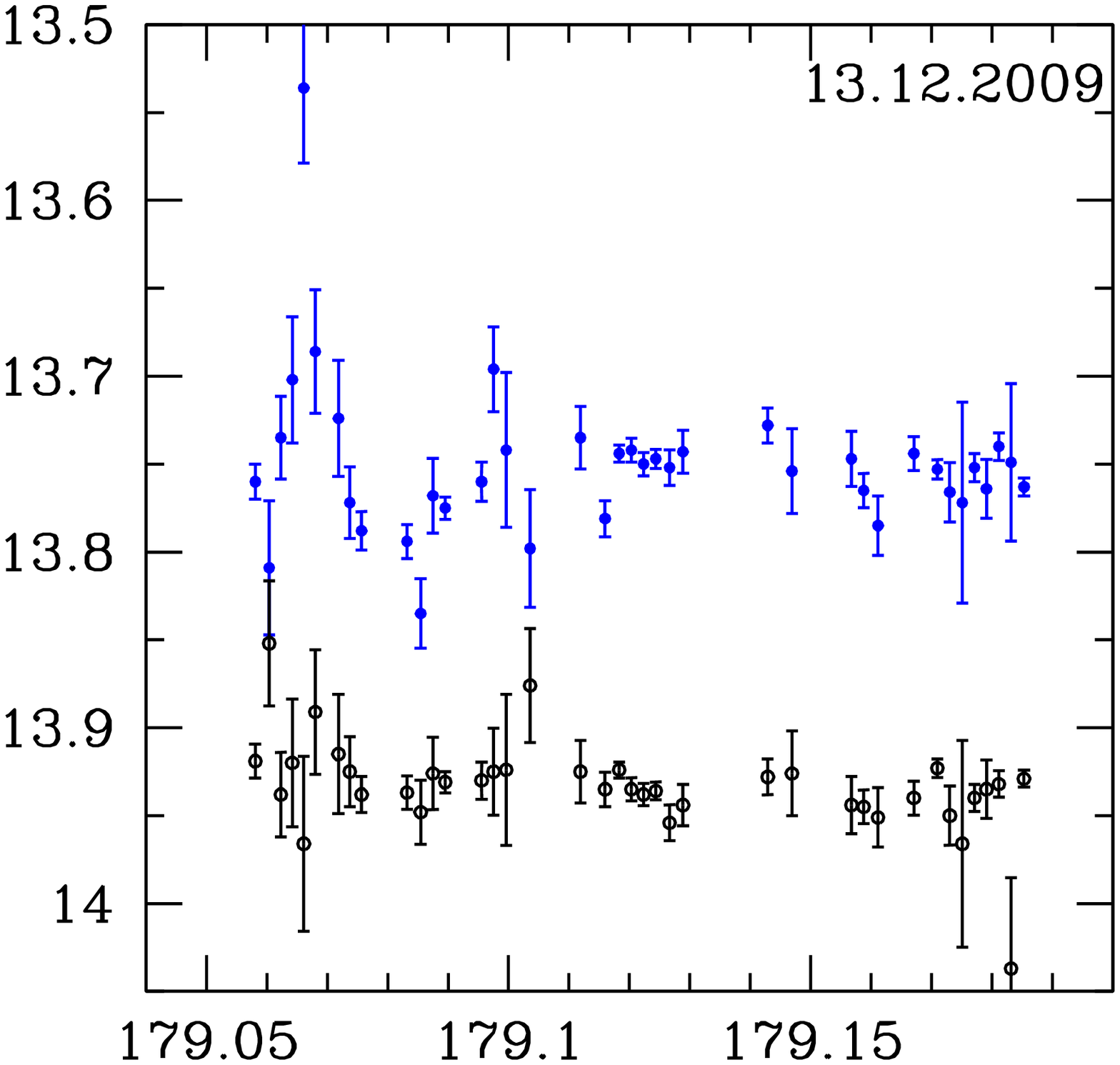}
\includegraphics[width=2.2in,height=2.5in]{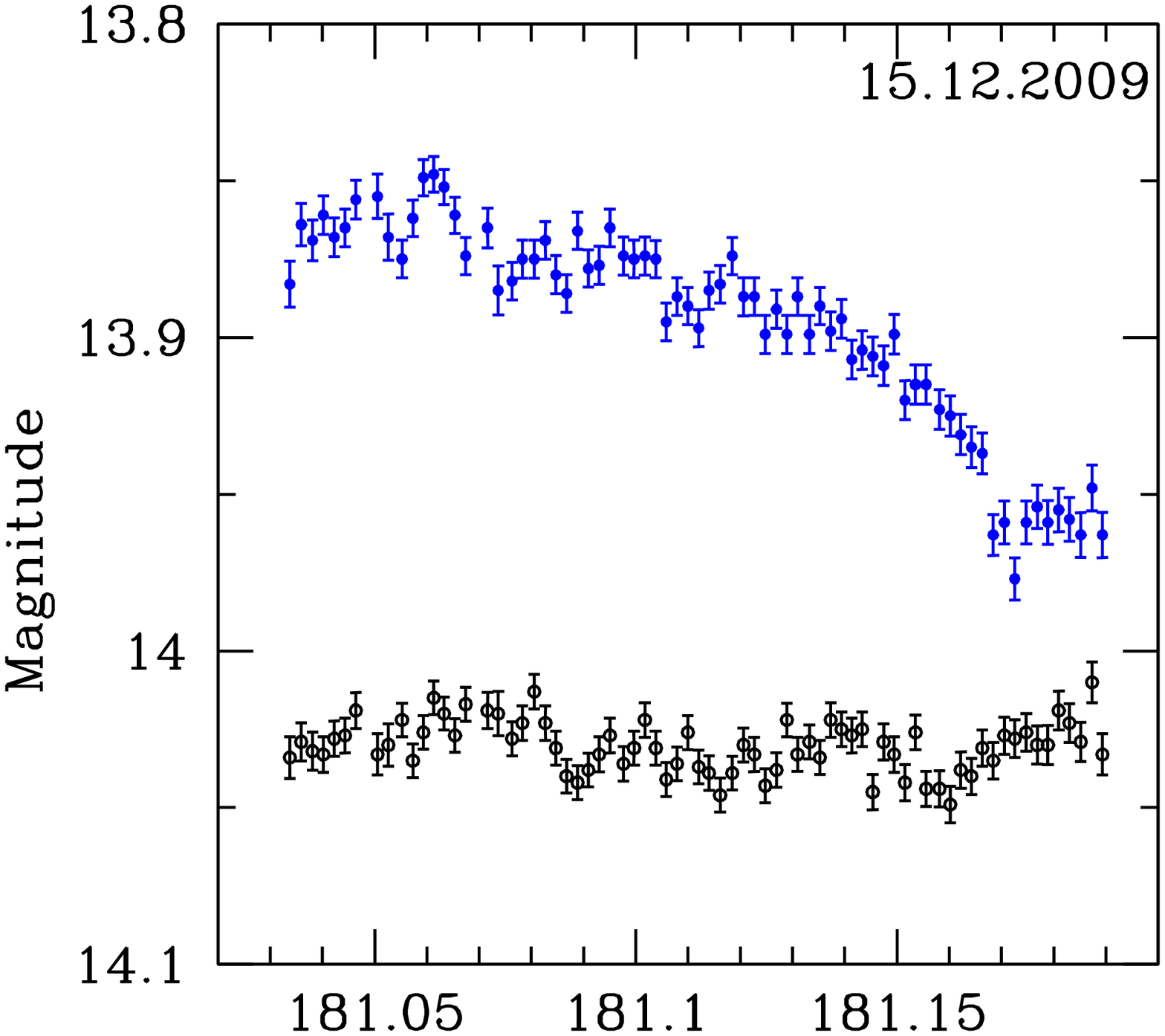}
\includegraphics[width=2.2in,height=2.5in]{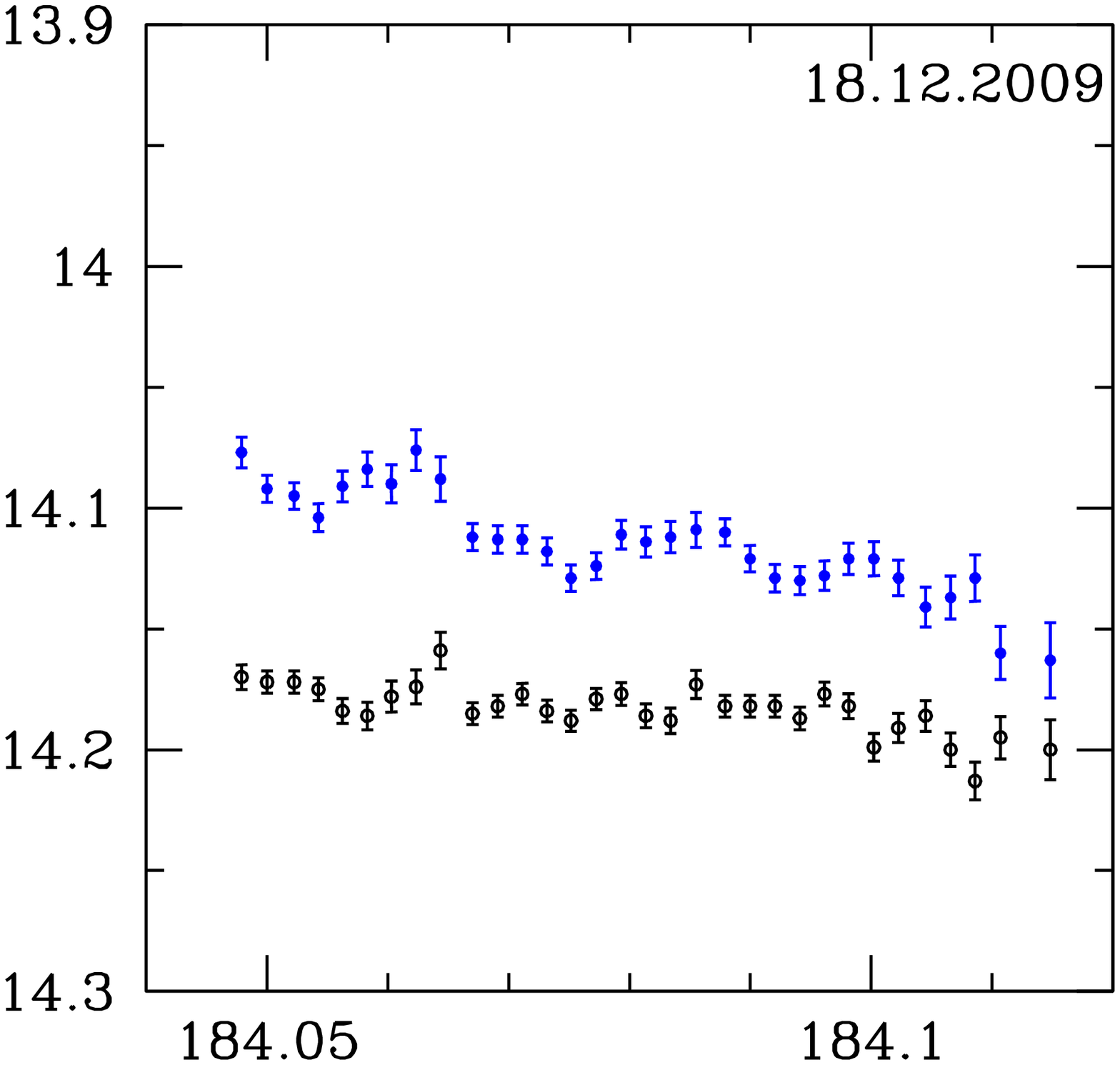}
\includegraphics[width=2.2in,height=2.5in]{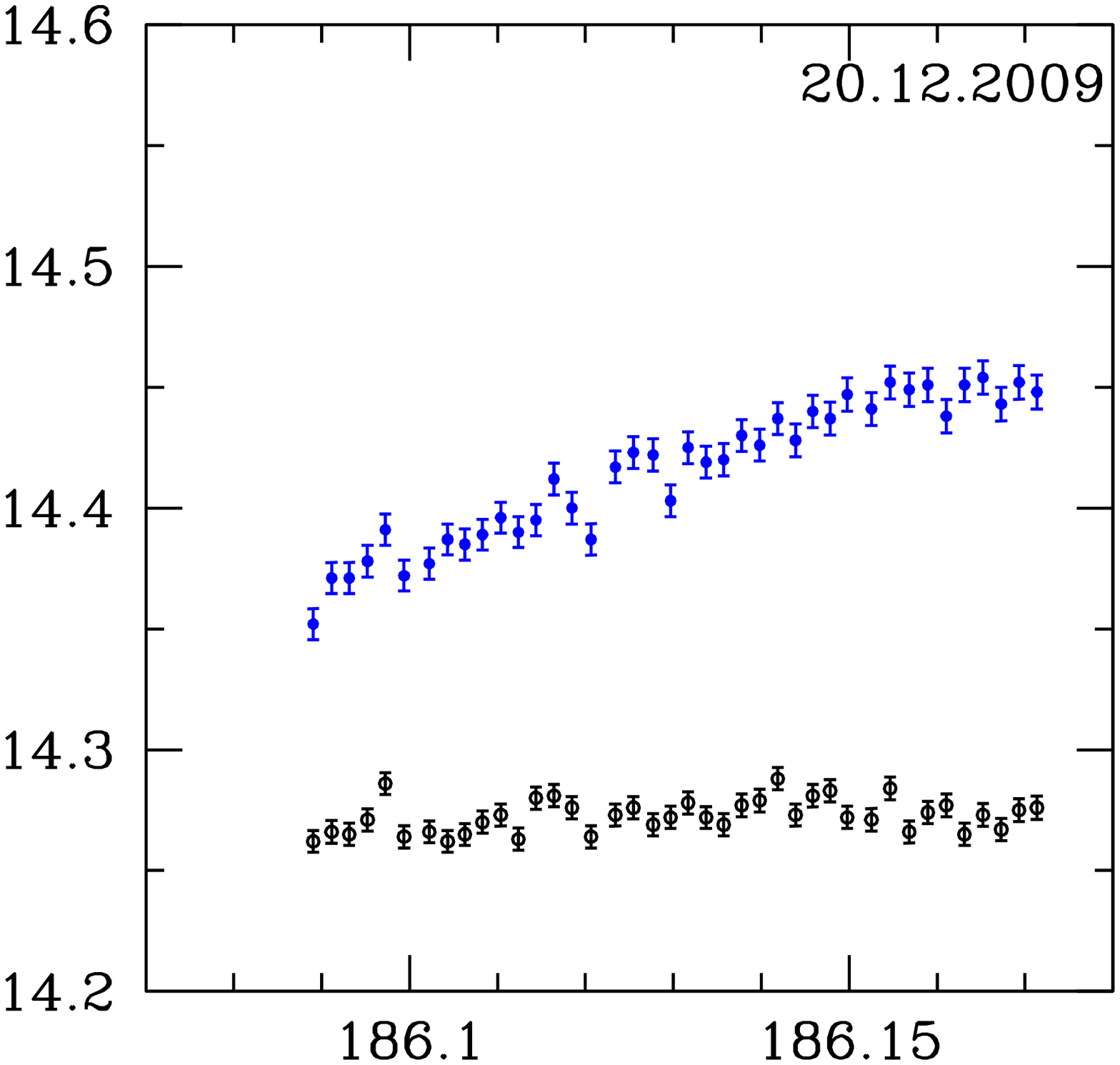}
\includegraphics[width=2.2in,height=2.5in]{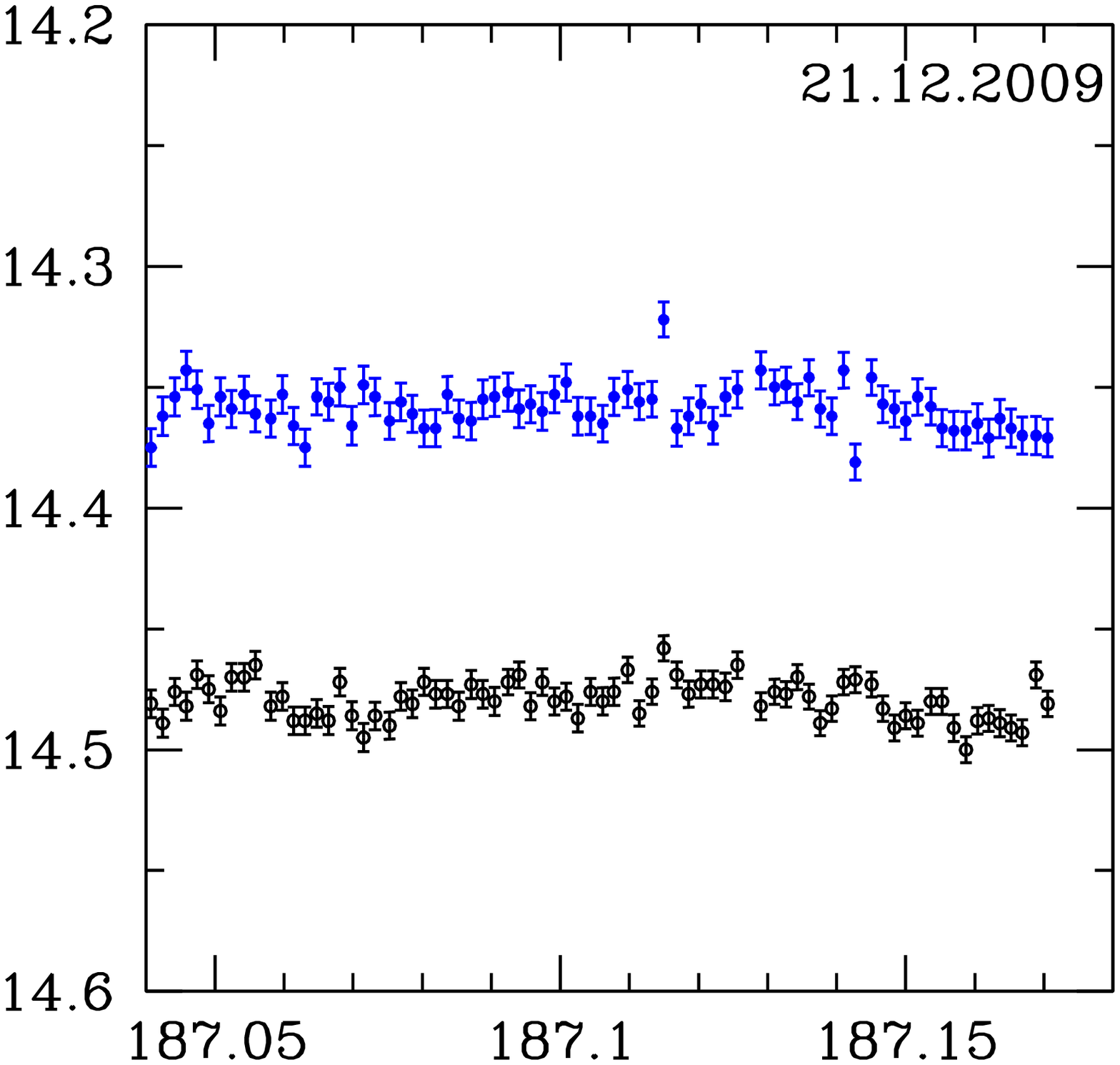}
\includegraphics[width=2.2in,height=2.5in]{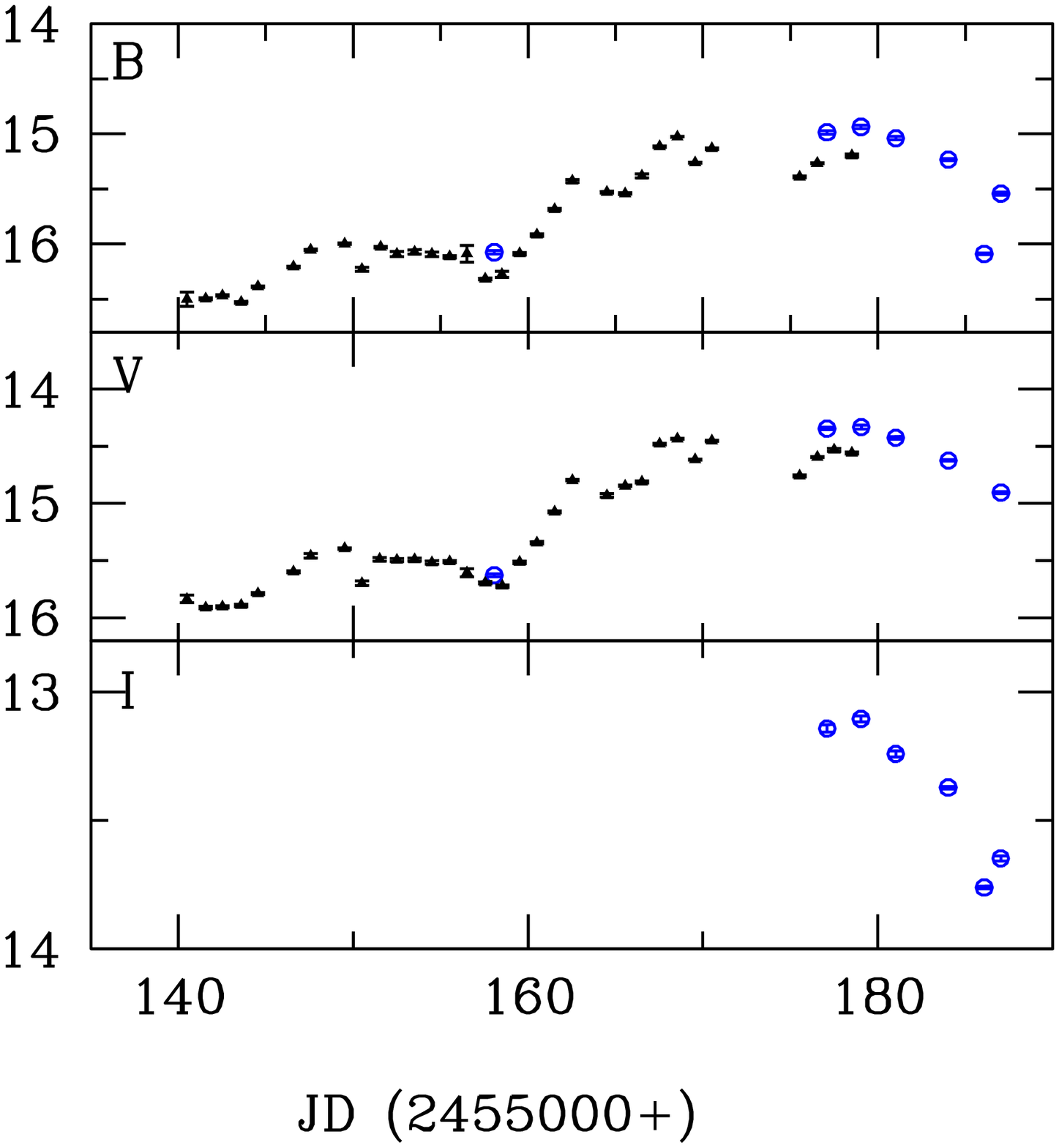}
\includegraphics[width=2.2in,height=2.5in]{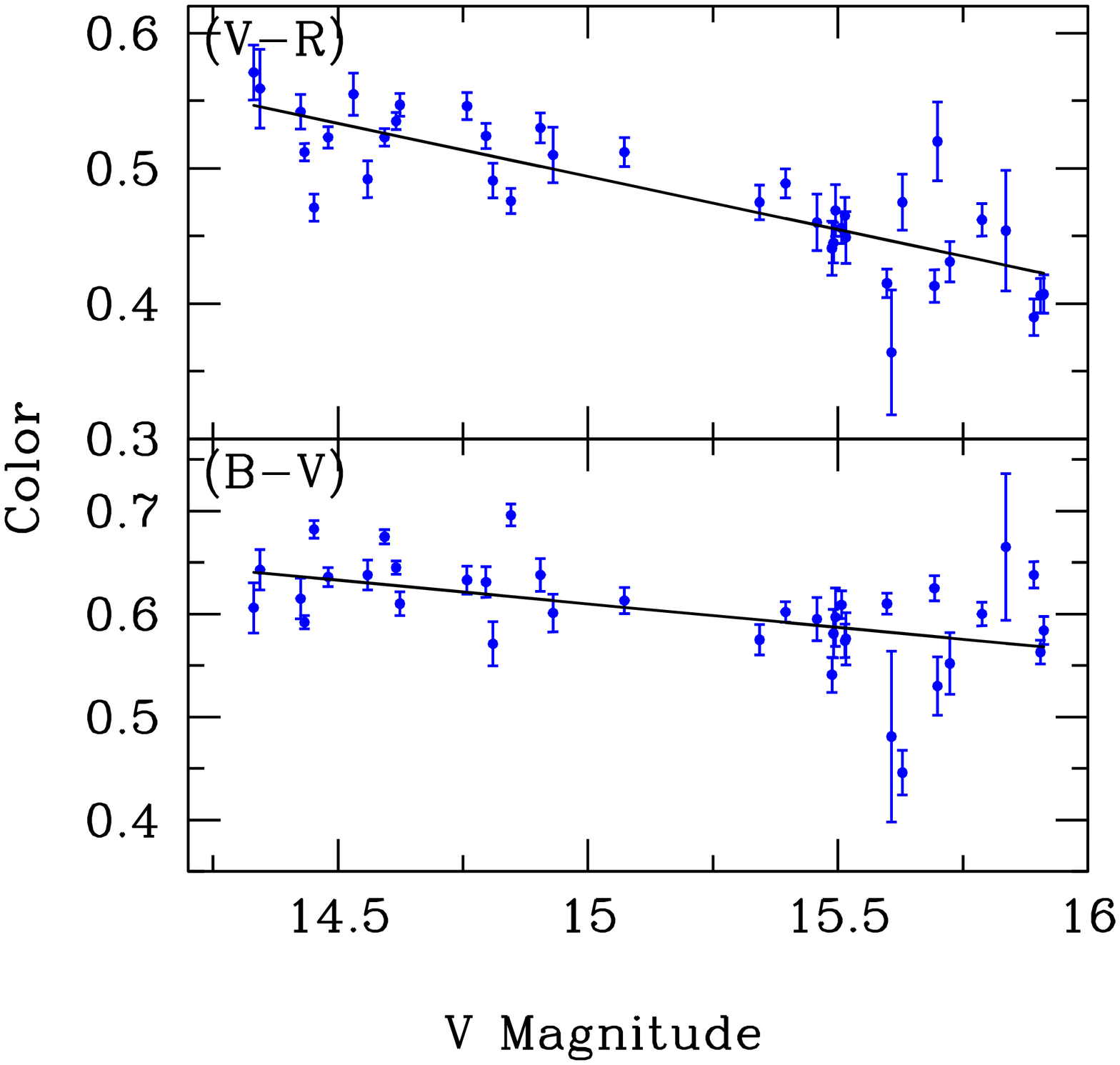}
\caption{LCs and color indices of 3C 454.3.  The first seven panels contain R-band LCs. Upper curves 
are the calibrated LCs of 3C 454.3 (w.r.t. star D).
Lower curves are the differential instrumental magnitudes of stars C and D with arbitrary 
offsets. The eighth panel shows B, V and I band light curves of 3C 454.3 (SMARTS data in starred symbols 
and ARIES data in open circles); the last panel shows the (V-R) and (B-V) colors against V magnitude.}
\end{figure}

\clearpage
\begin{figure}
 \centering
\includegraphics[width=2.2in,height=2.5in]{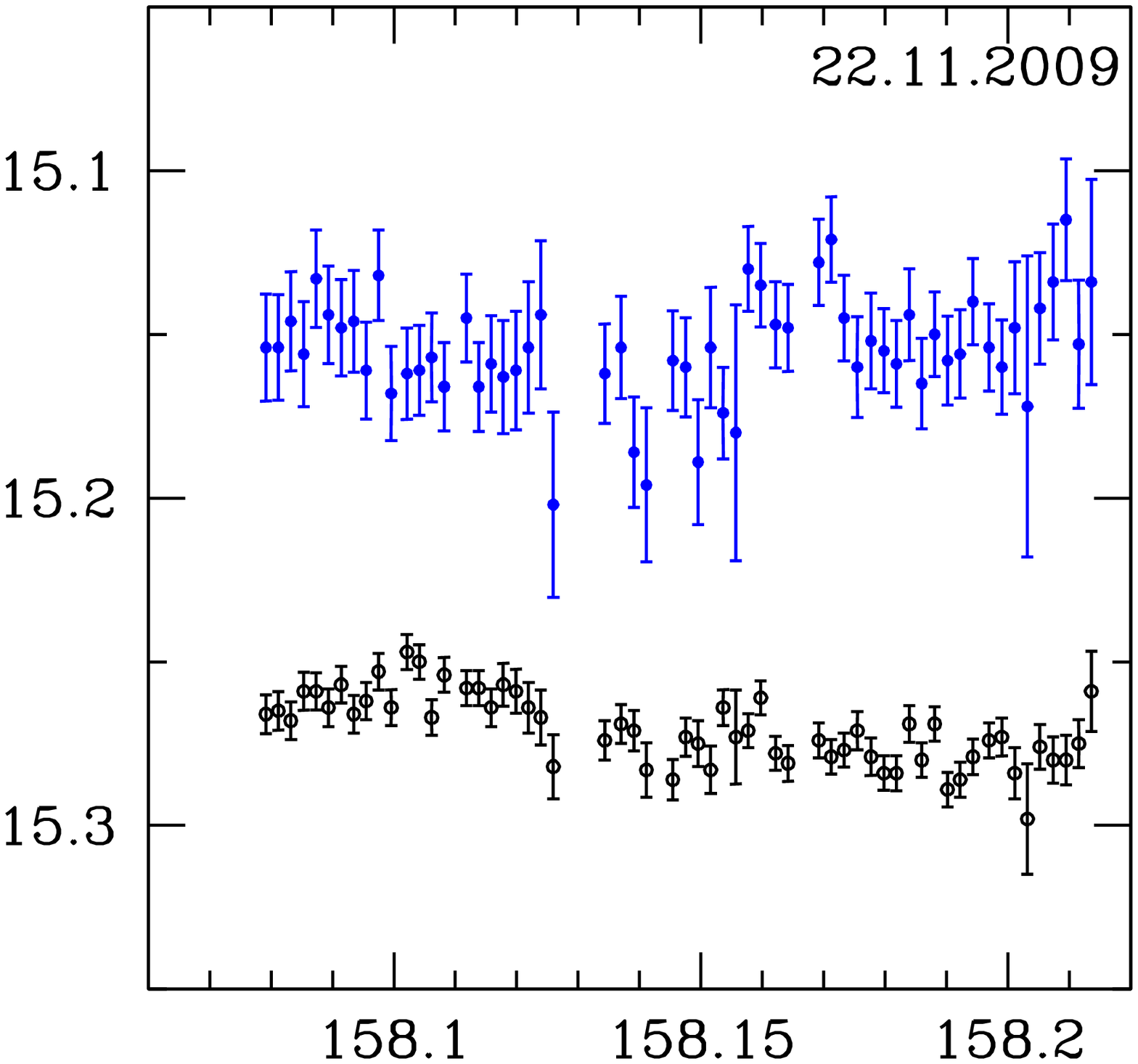}
\includegraphics[width=2.2in,height=2.5in]{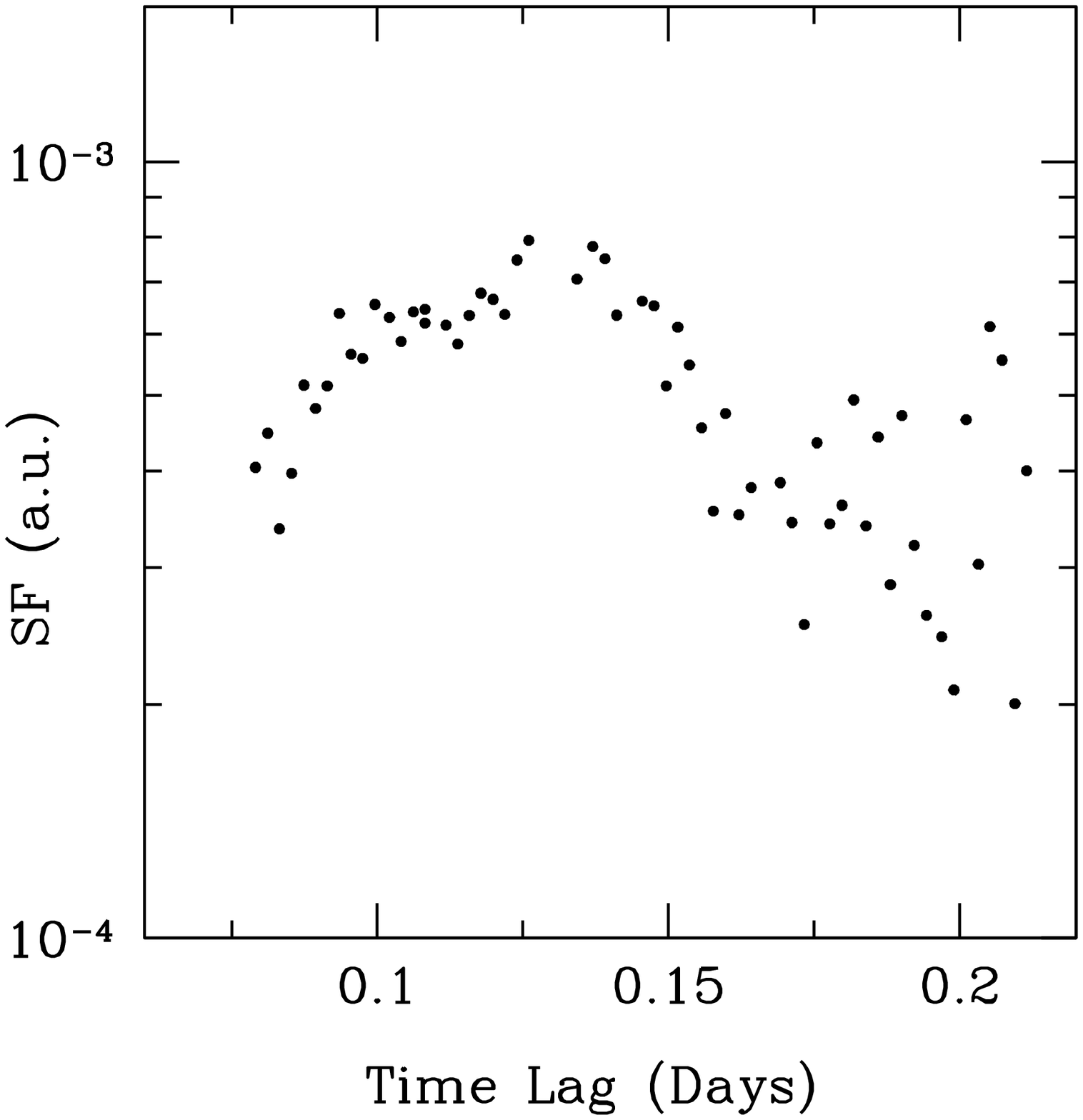}
\includegraphics[width=2.2in,height=2.5in]{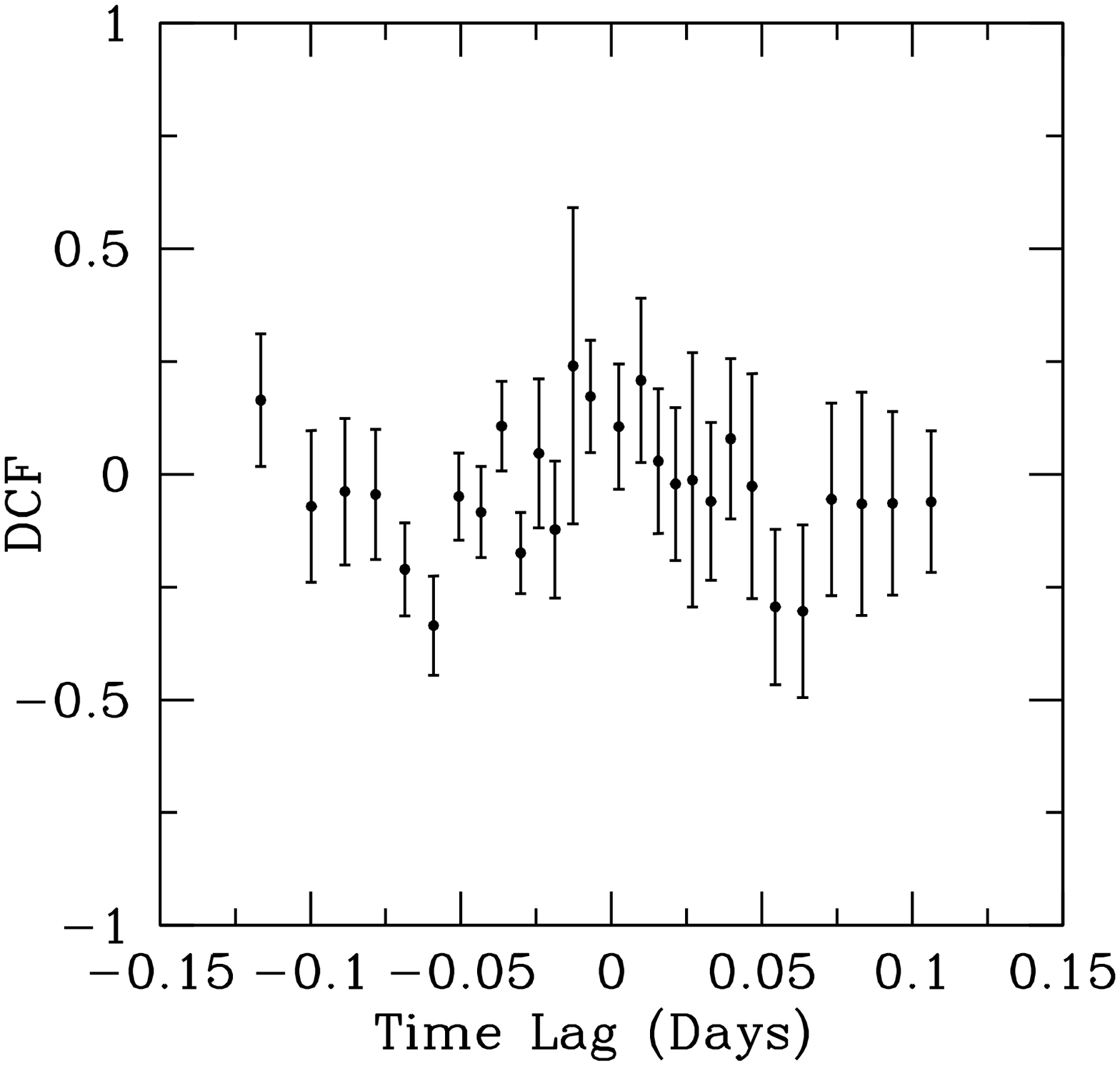}
\includegraphics[width=2.2in,height=2.5in]{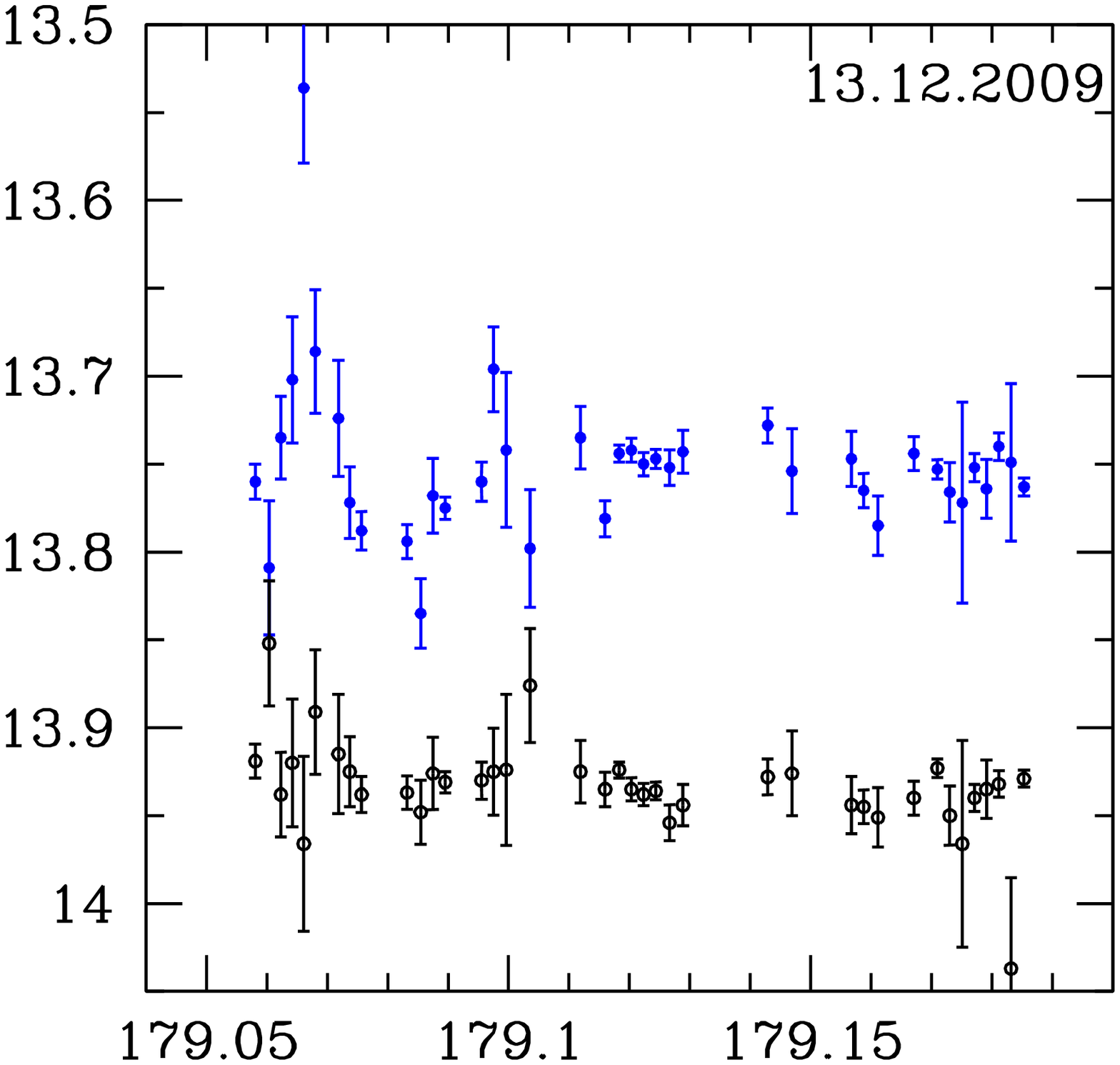}
\includegraphics[width=2.2in,height=2.5in]{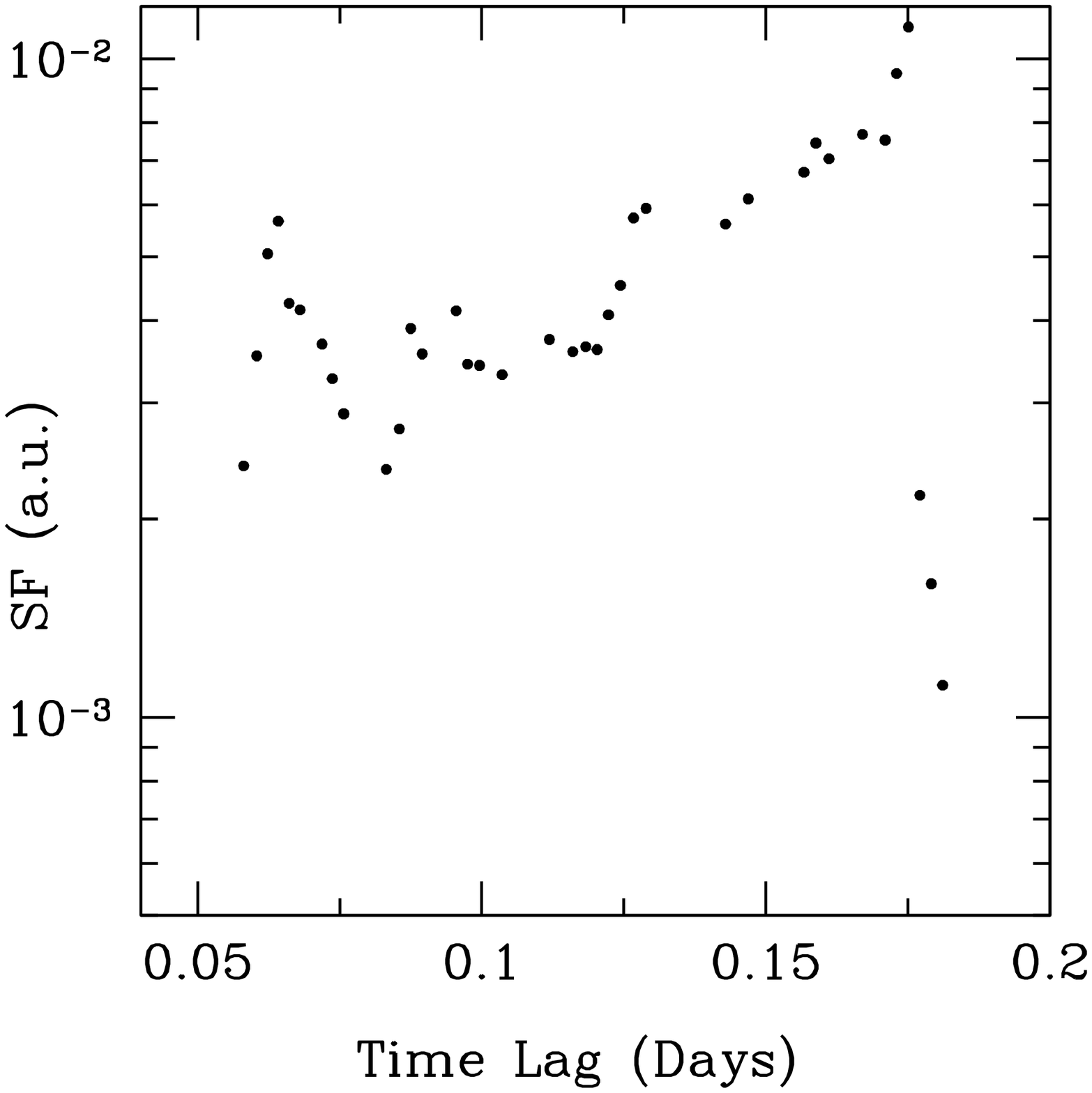}
\includegraphics[width=2.2in,height=2.5in]{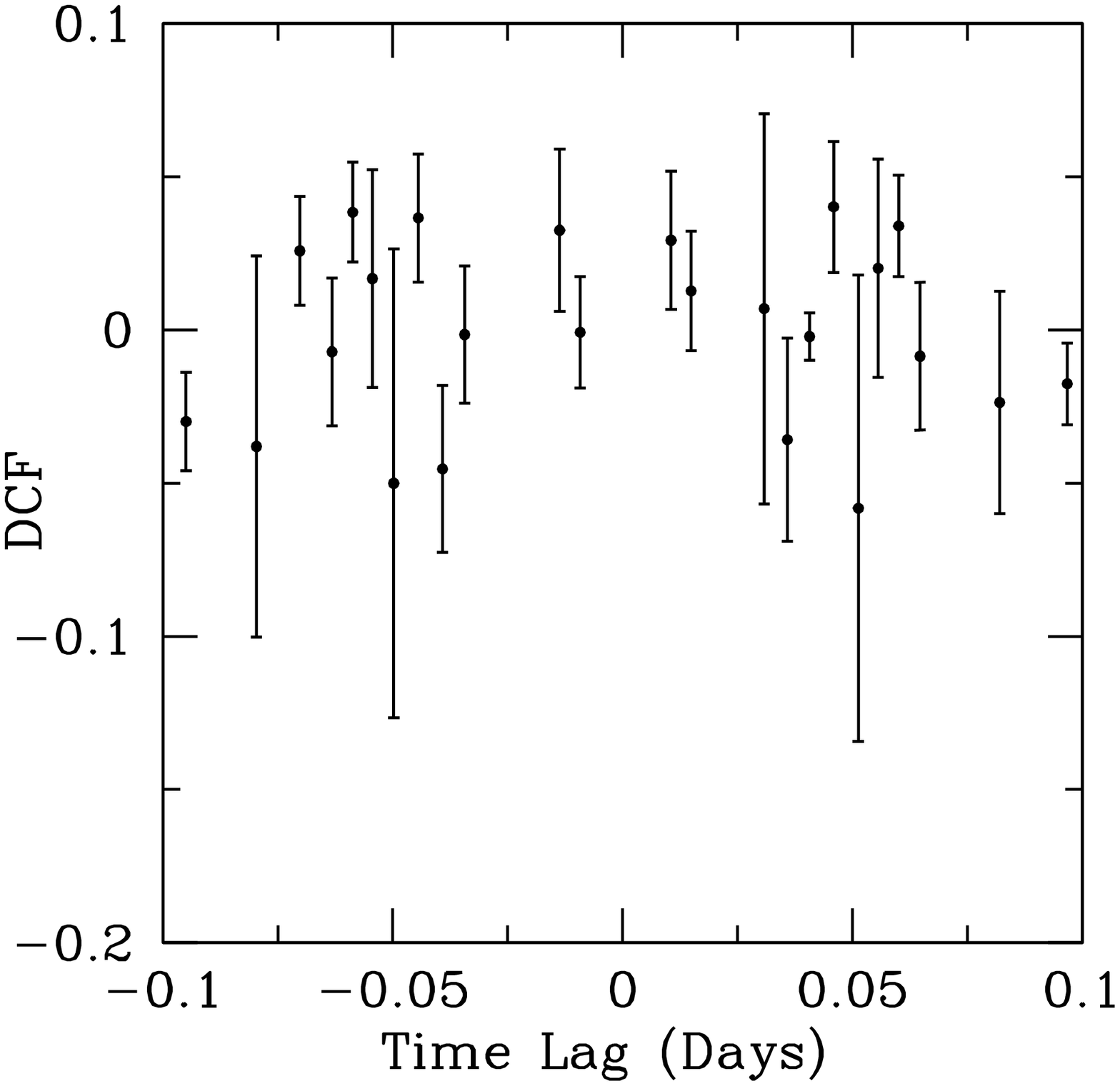}
\includegraphics[width=2.2in,height=2.5in]{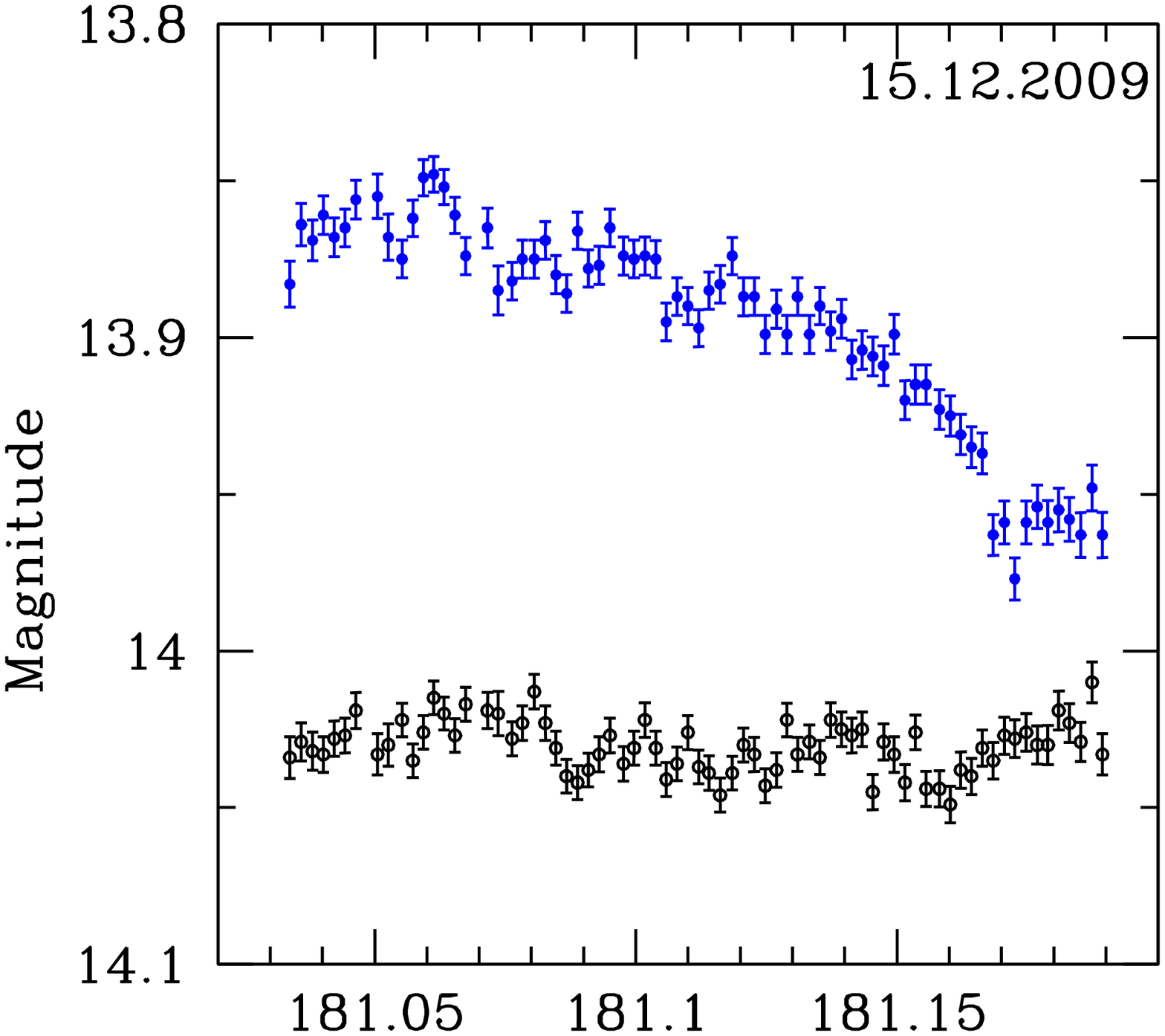}
\includegraphics[width=2.2in,height=2.5in]{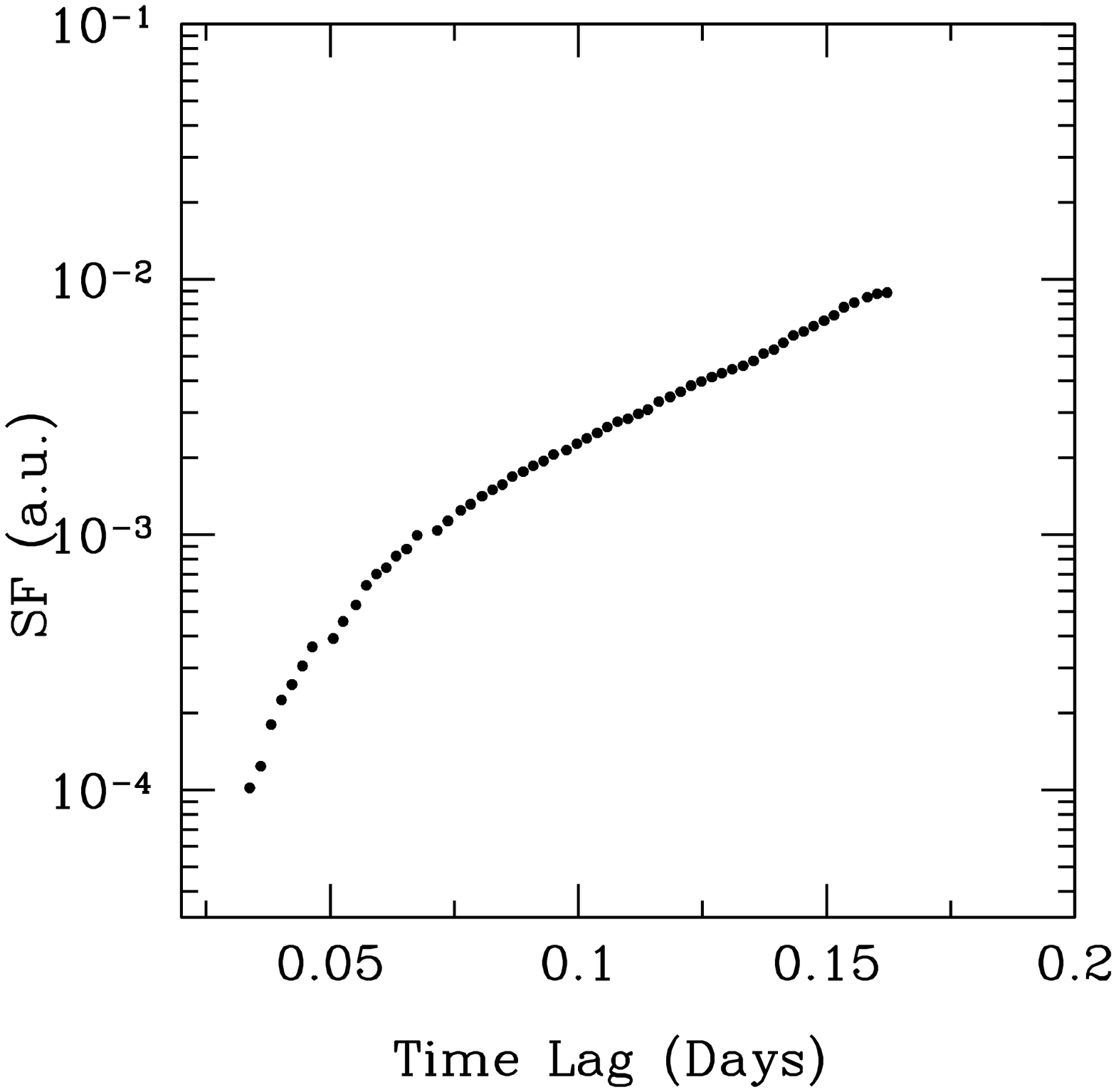}
\includegraphics[width=2.2in,height=2.5in]{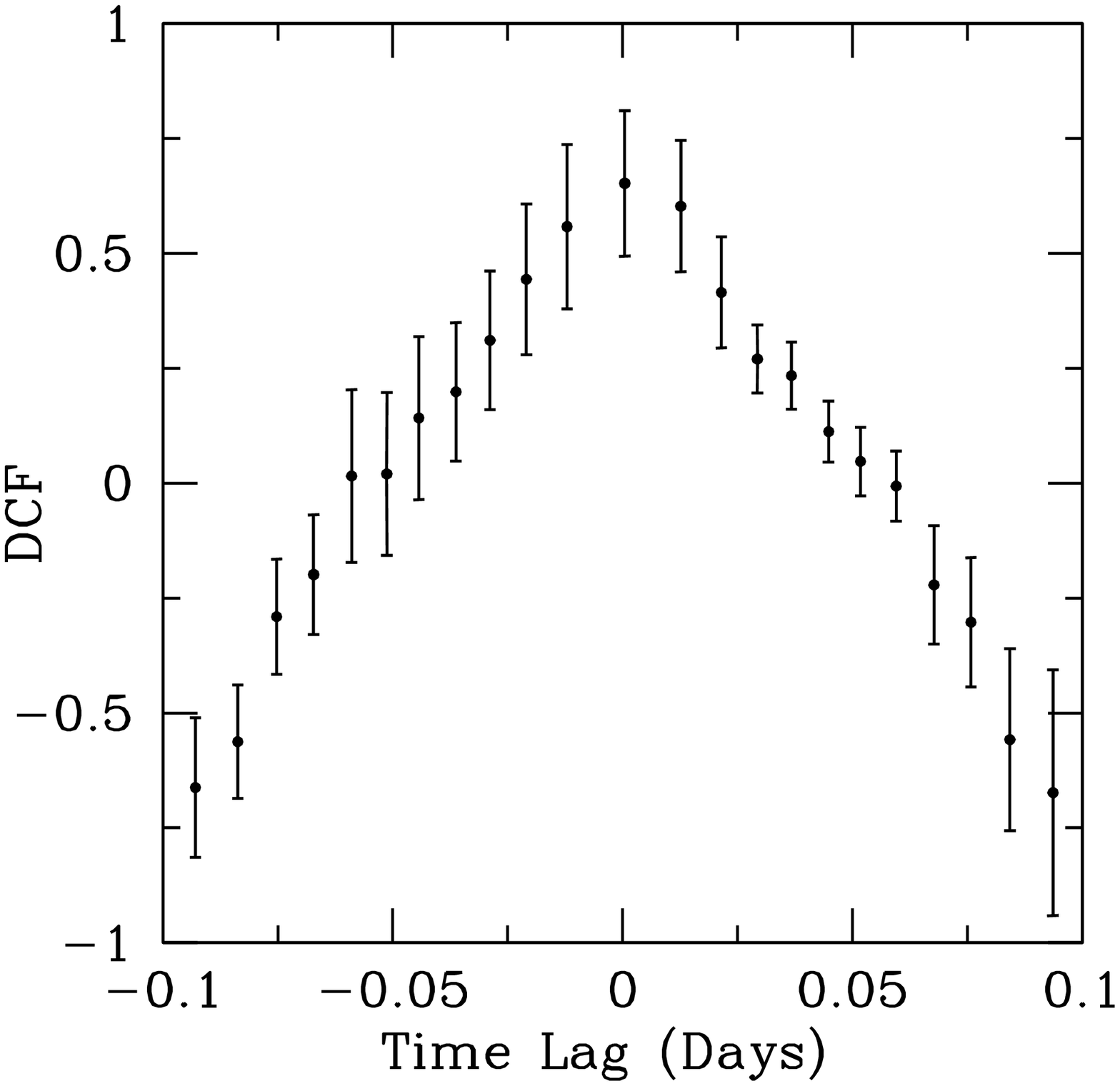}
\includegraphics[width=2.2in,height=2.5in]{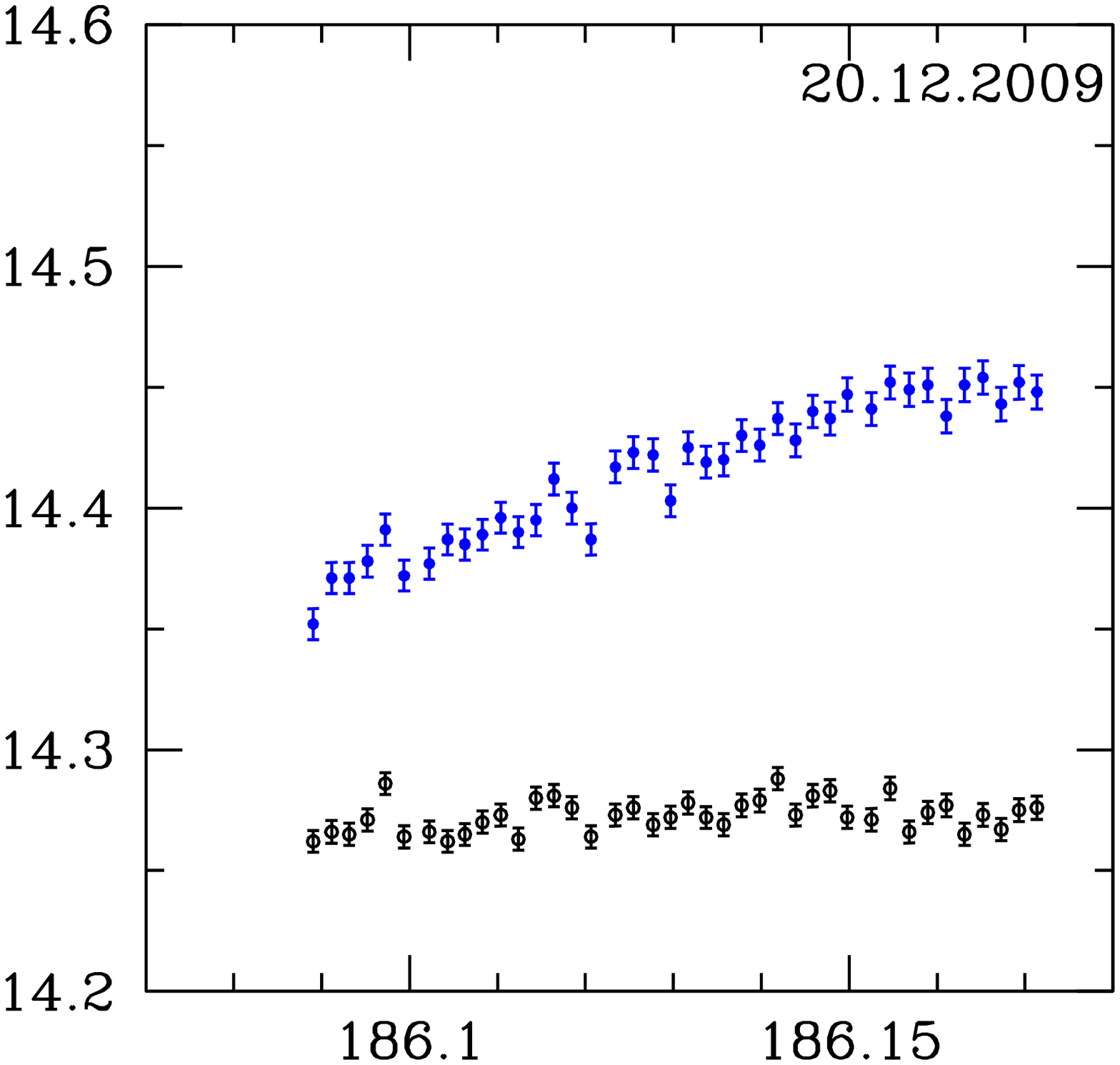}
\includegraphics[width=2.2in,height=2.5in]{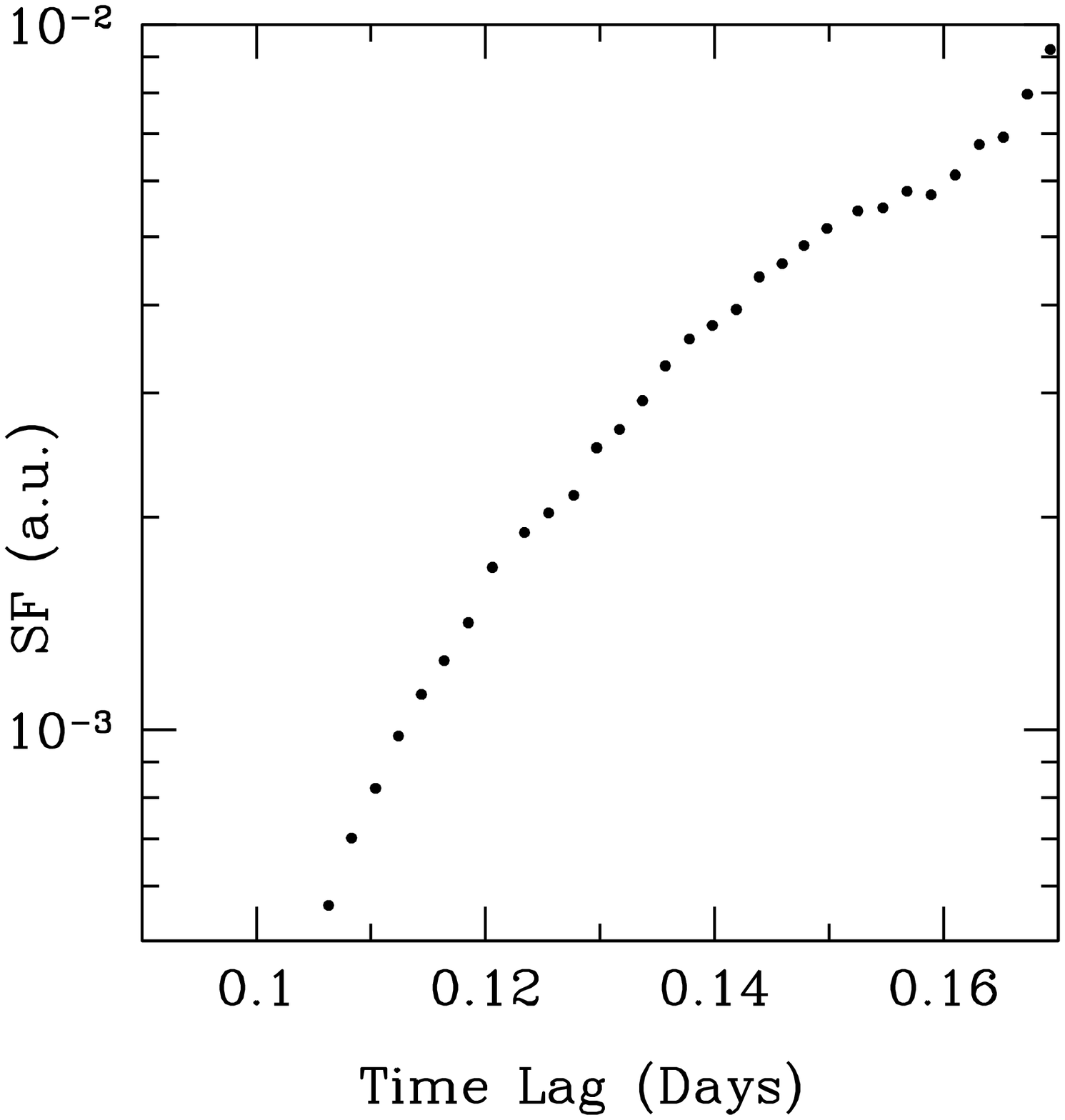}
\includegraphics[width=2.2in,height=2.5in]{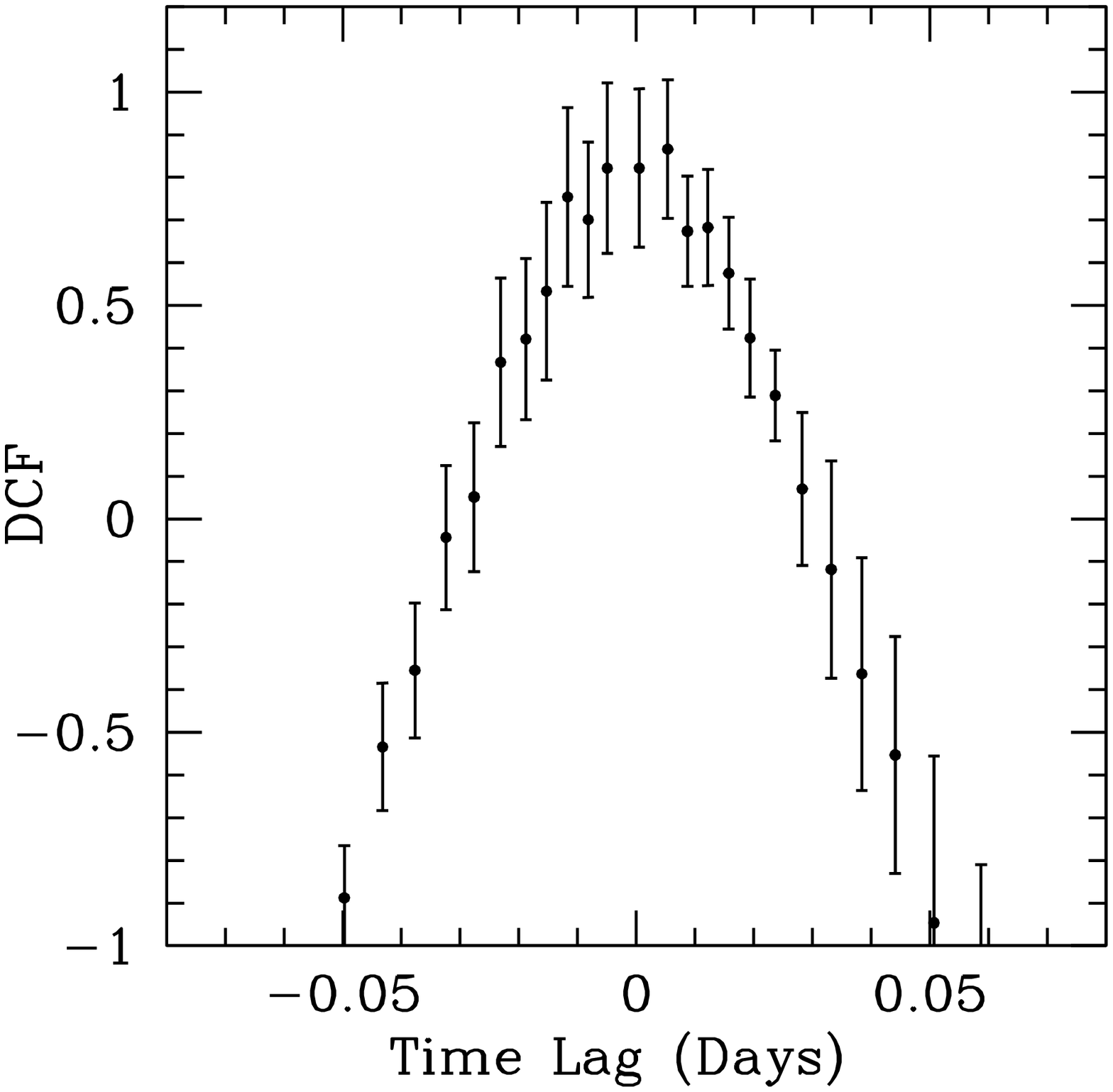}
\caption{ IDV LCs in the R-band for 3C 454.3 with their respective structure function (SF) 
and discrete correlation function (DCF) in each row from left to right, respectively.  }
\end{figure}

\clearpage
\begin{figure}
\includegraphics[width=3.0in,height=5.0in]{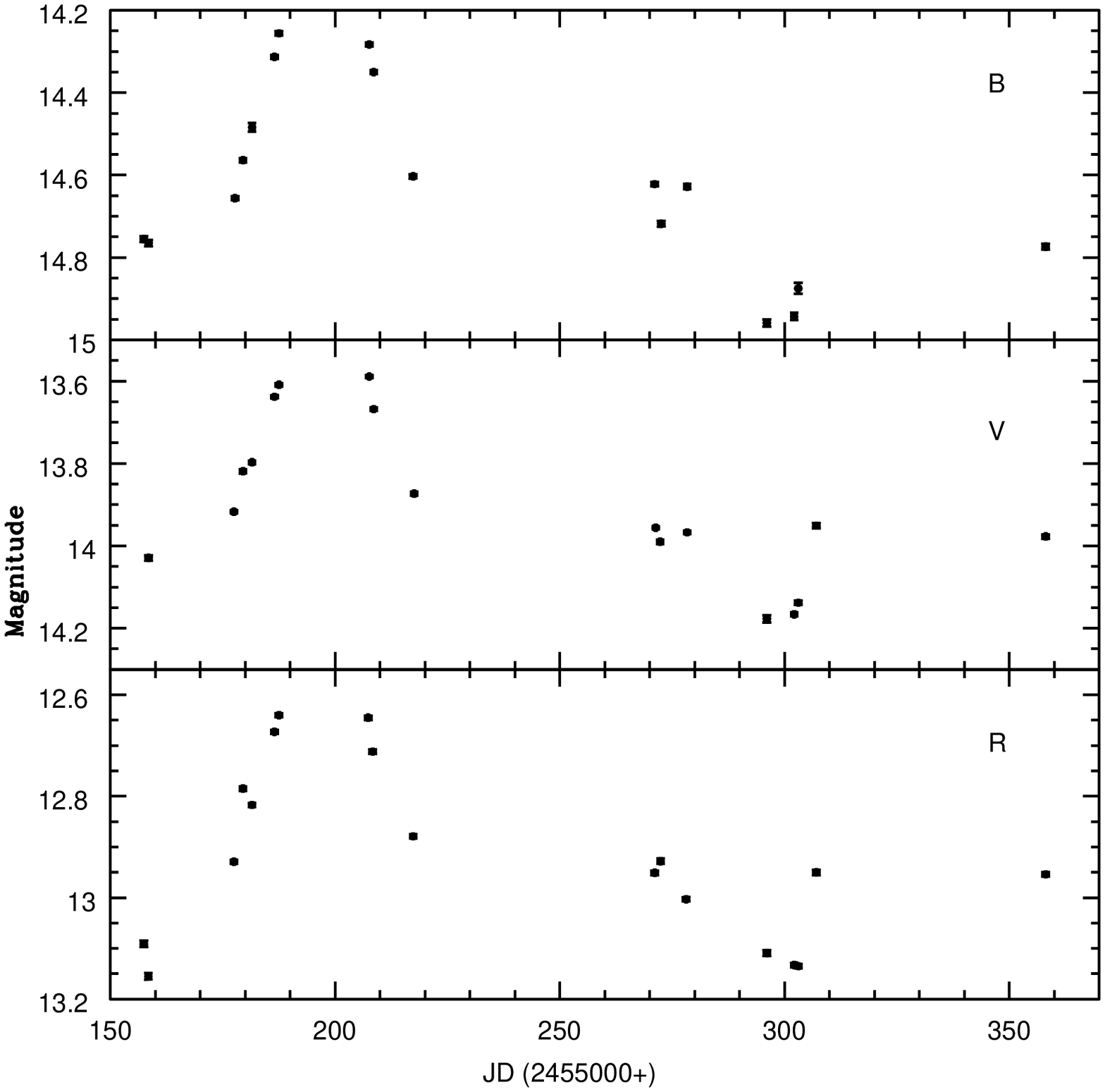}
\includegraphics[width=3.0in,height=5.0in]{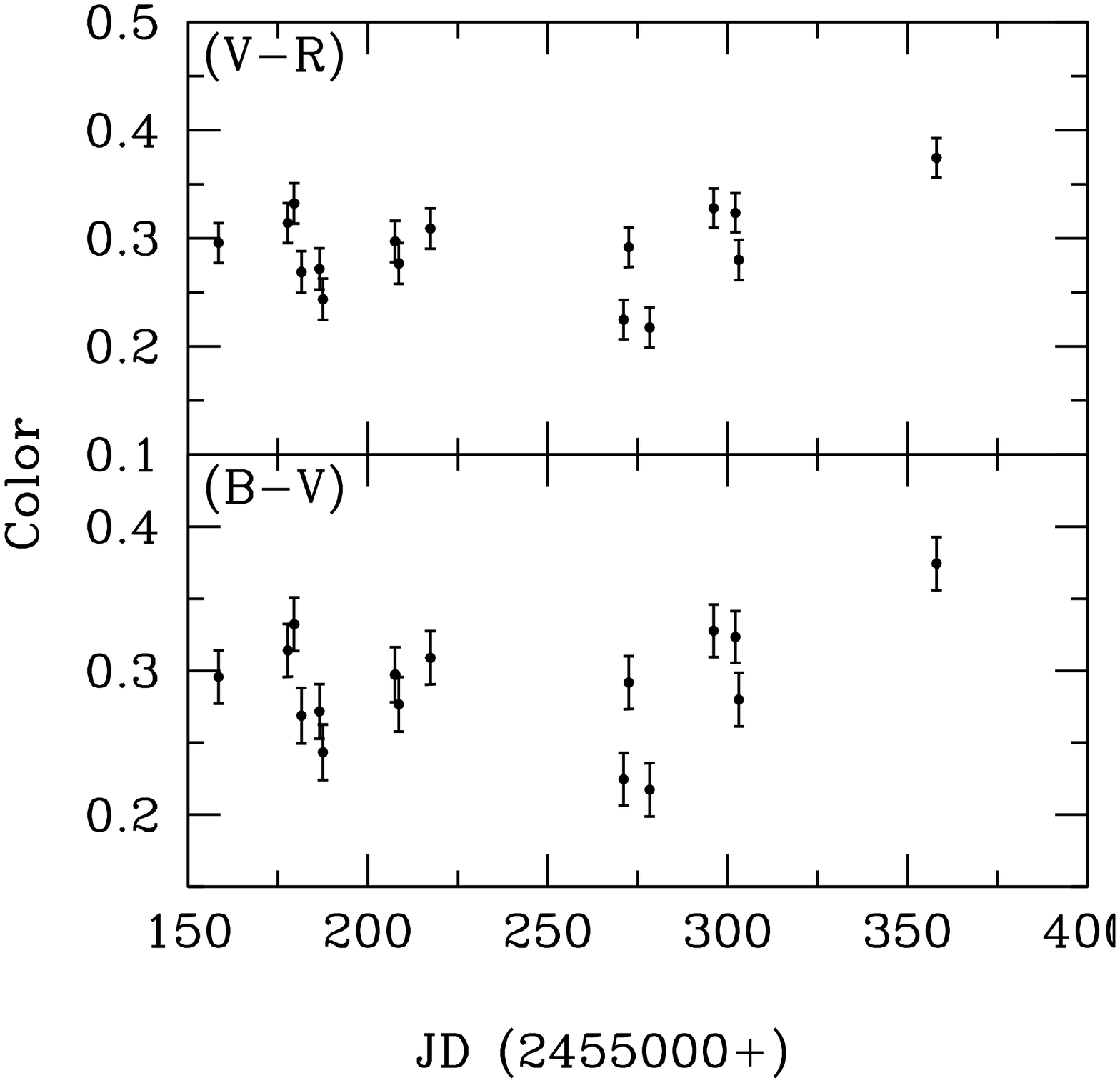}
\caption{Short-term variability LCs and color indices of Mrk 421 in the B, V and R bands
and (V-R) and (B-V) colors during the  2009--2010 season. }
\end{figure}

\clearpage
\begin{figure}
\centering
\includegraphics[width=7.0in,height=7.0in]{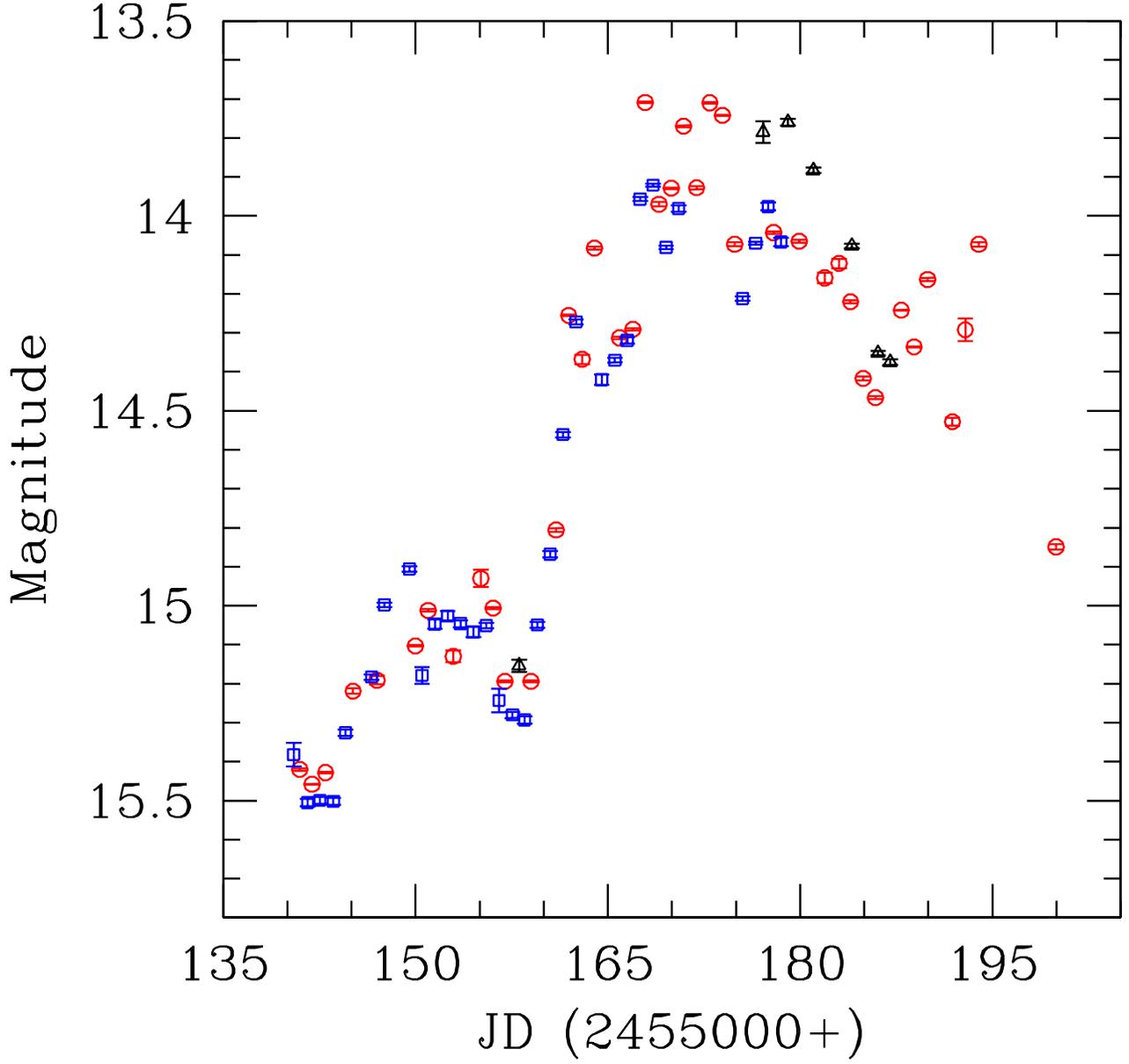}
\caption{Short-term variability LC of 3C 454.3 in R band. Open circles represent data from 
KANATA observatory, Japan; open triangles represent the data from ARIES, Nainital; open squares 
represents the data from SMARTS.}
\end{figure}

\clearpage
\begin{figure}
\centering
\includegraphics[width=7.0in,height=7.0in]{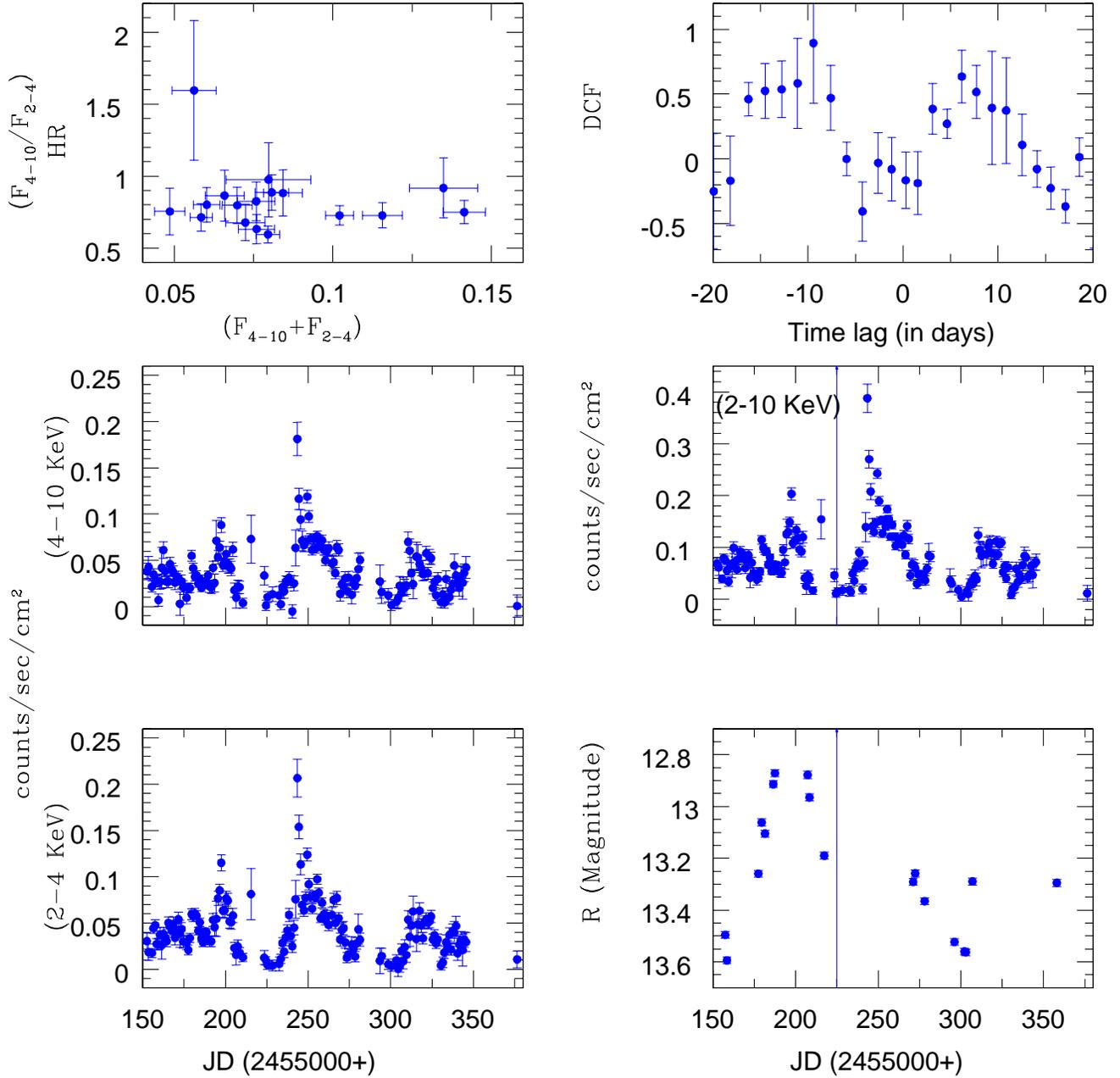}
\caption{X$-$ray LCs for Mrk 421: 2-4 KeV (lower left panel); 4-10 KeV 
(middle left panel) and 2-10 KeV (middle right panel).  Also R-band LC (lower right panel),
hardness intensity plot (upper left panel) and  DCF (upper right panel) performed on the optical vs X$-$ray 
data in the first flare region (lasting until the vertical line).  }
\end{figure}

\clearpage
\begin{figure}
 \centering
\includegraphics[width=2.2in,height=2.5in]{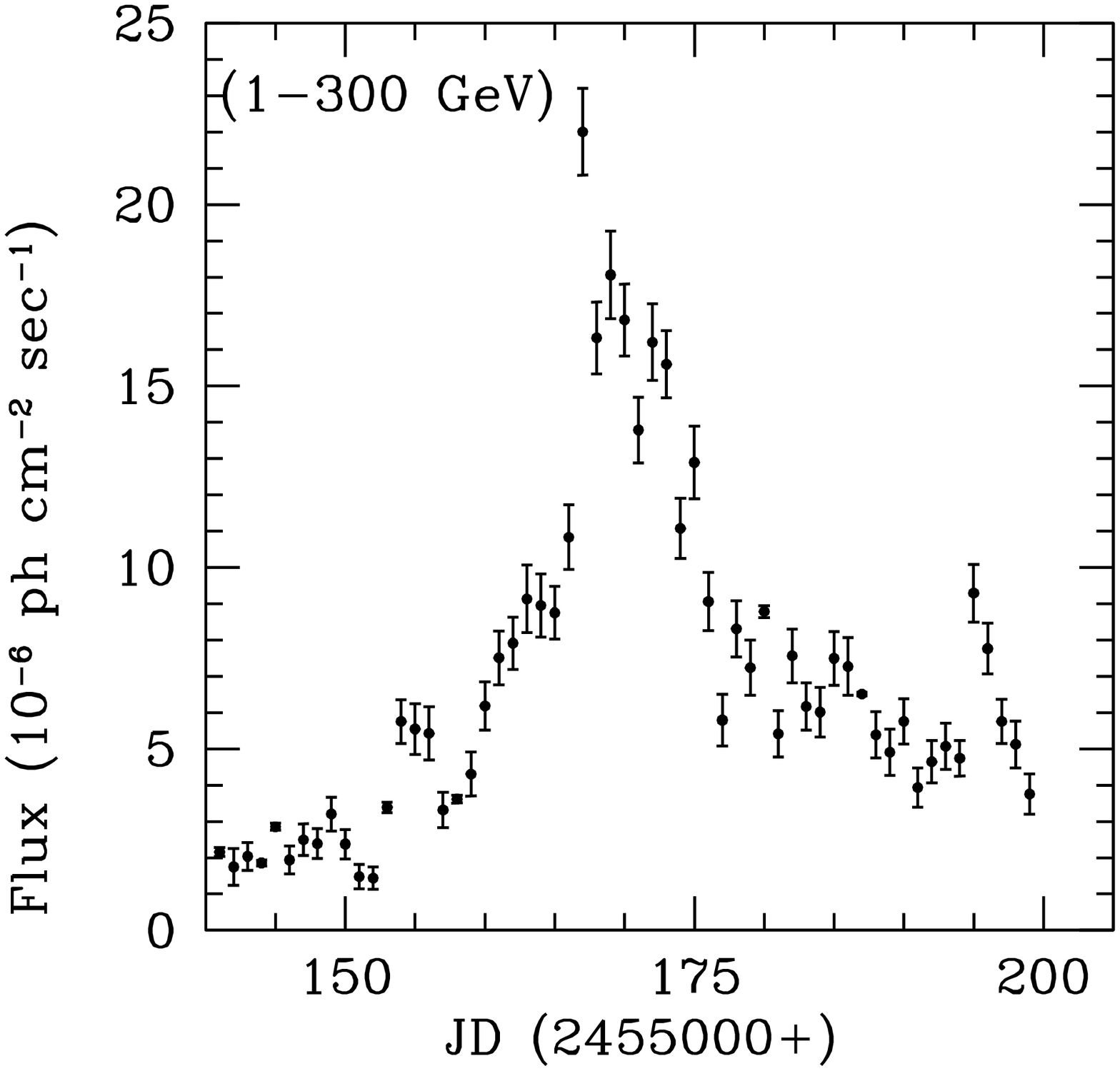}
\includegraphics[width=2.2in,height=2.5in]{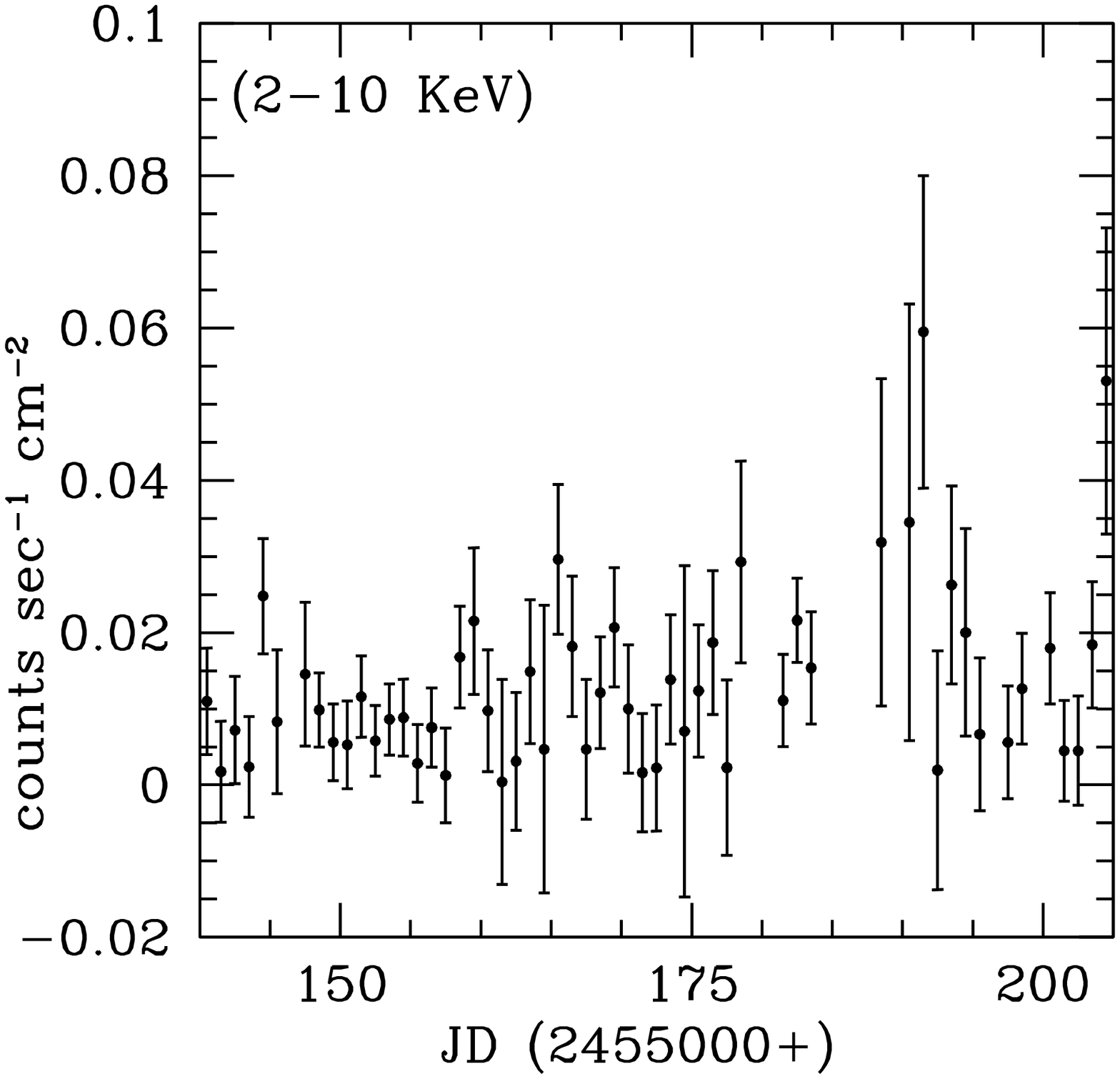}
\includegraphics[width=2.2in,height=2.5in]{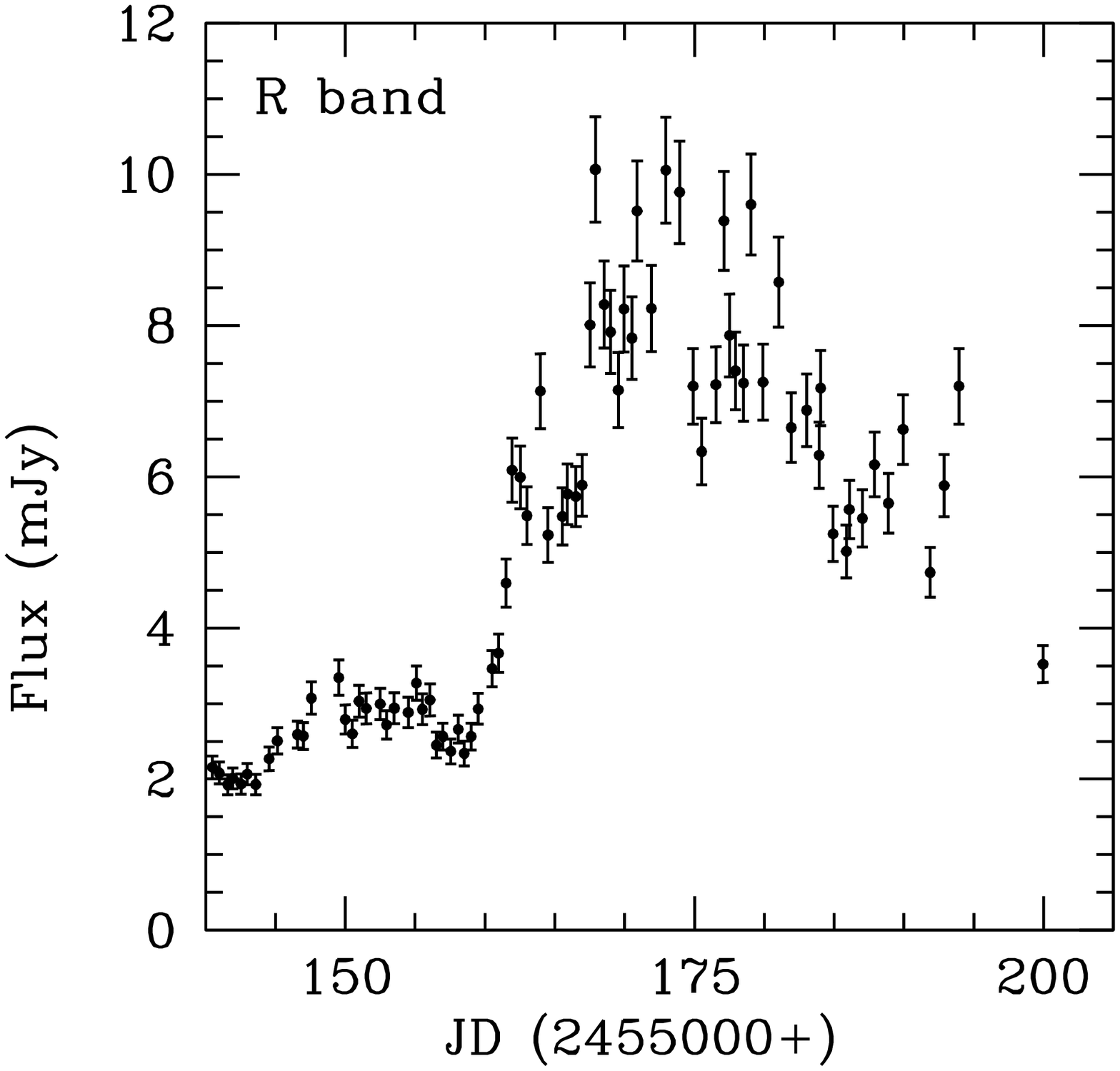}
\includegraphics[width=2.2in,height=2.5in]{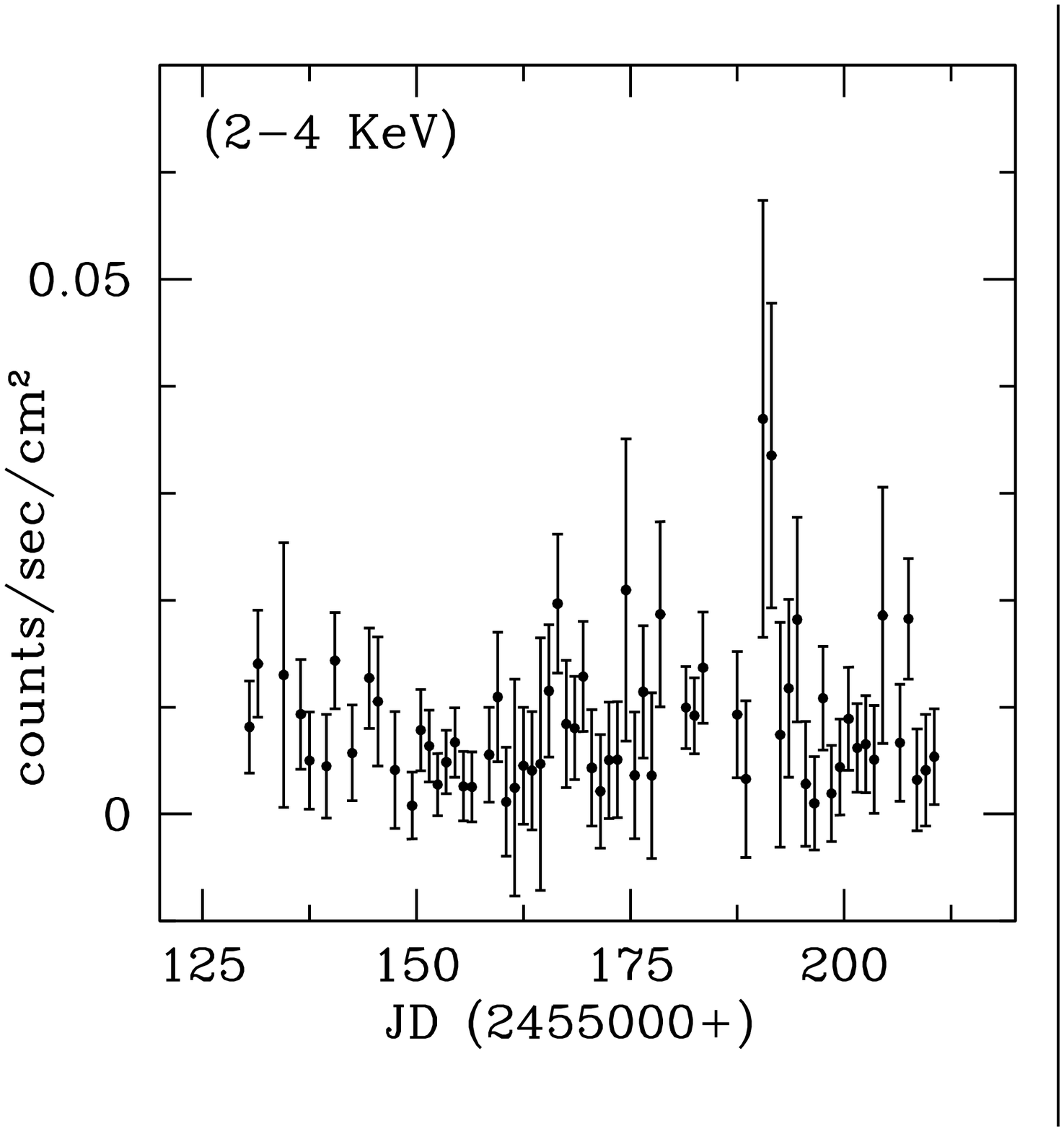}
\includegraphics[width=2.2in,height=2.5in]{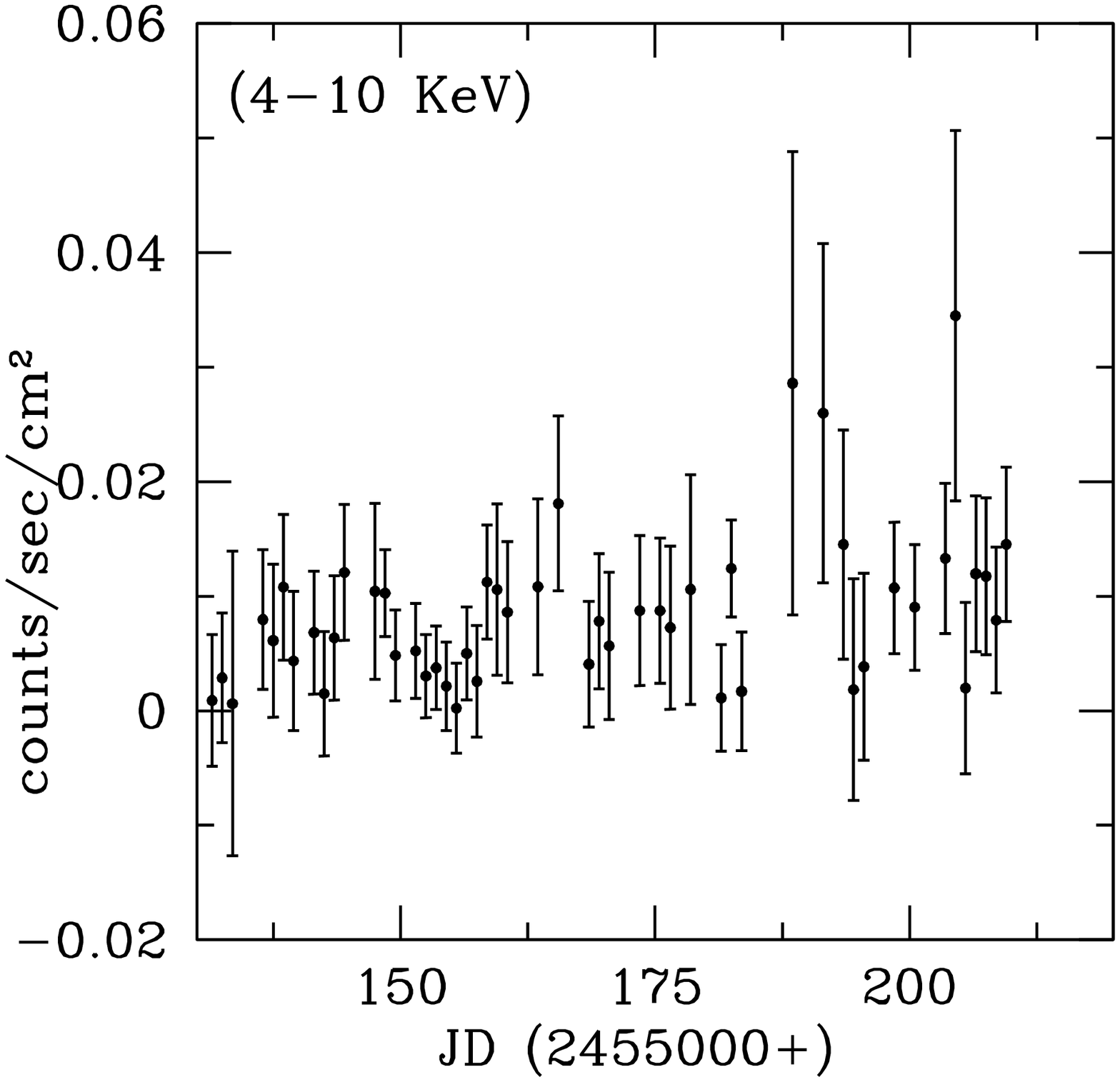}
\includegraphics[width=2.2in,height=2.5in]{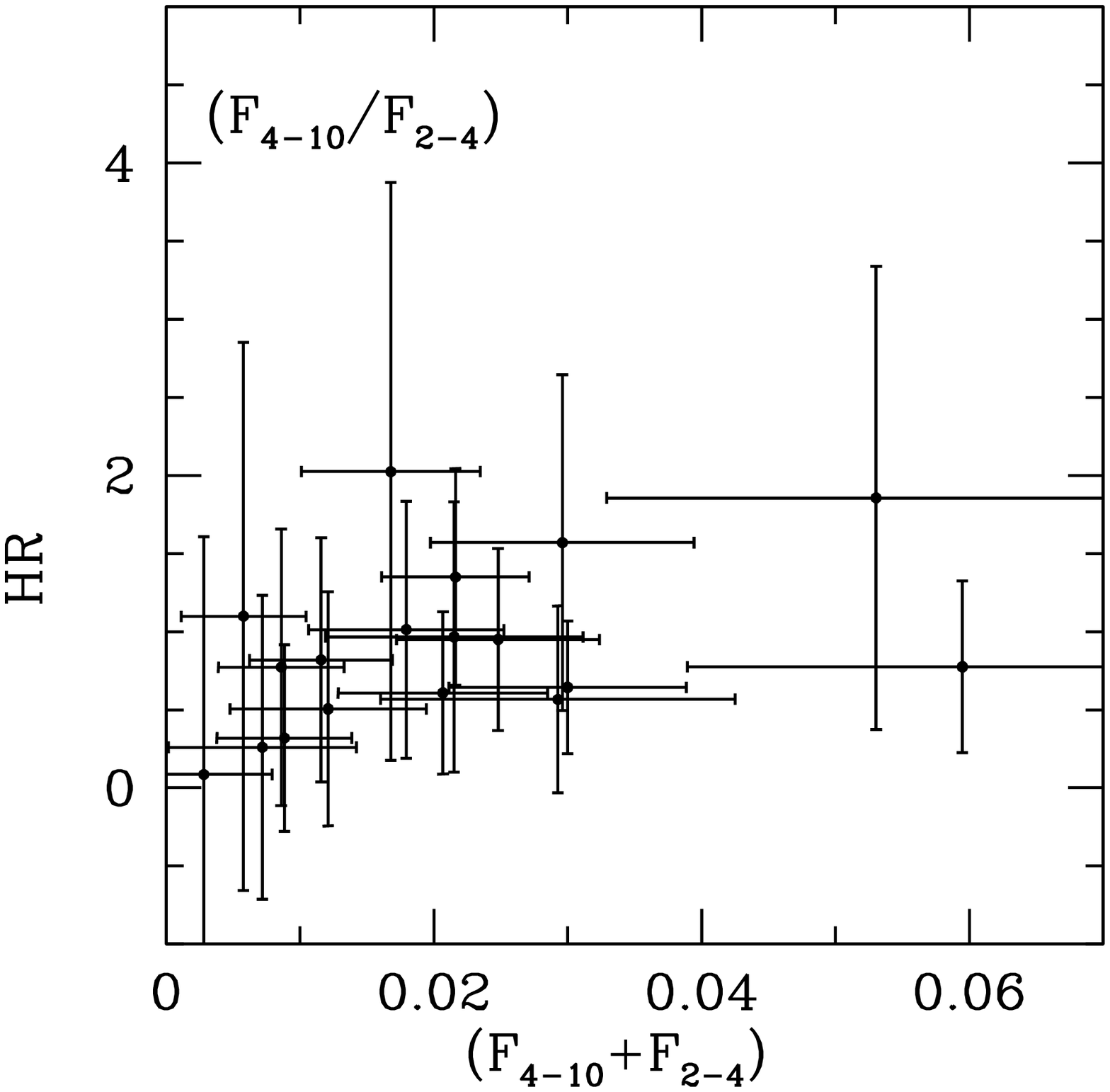}
\includegraphics[width=2.2in,height=2.5in]{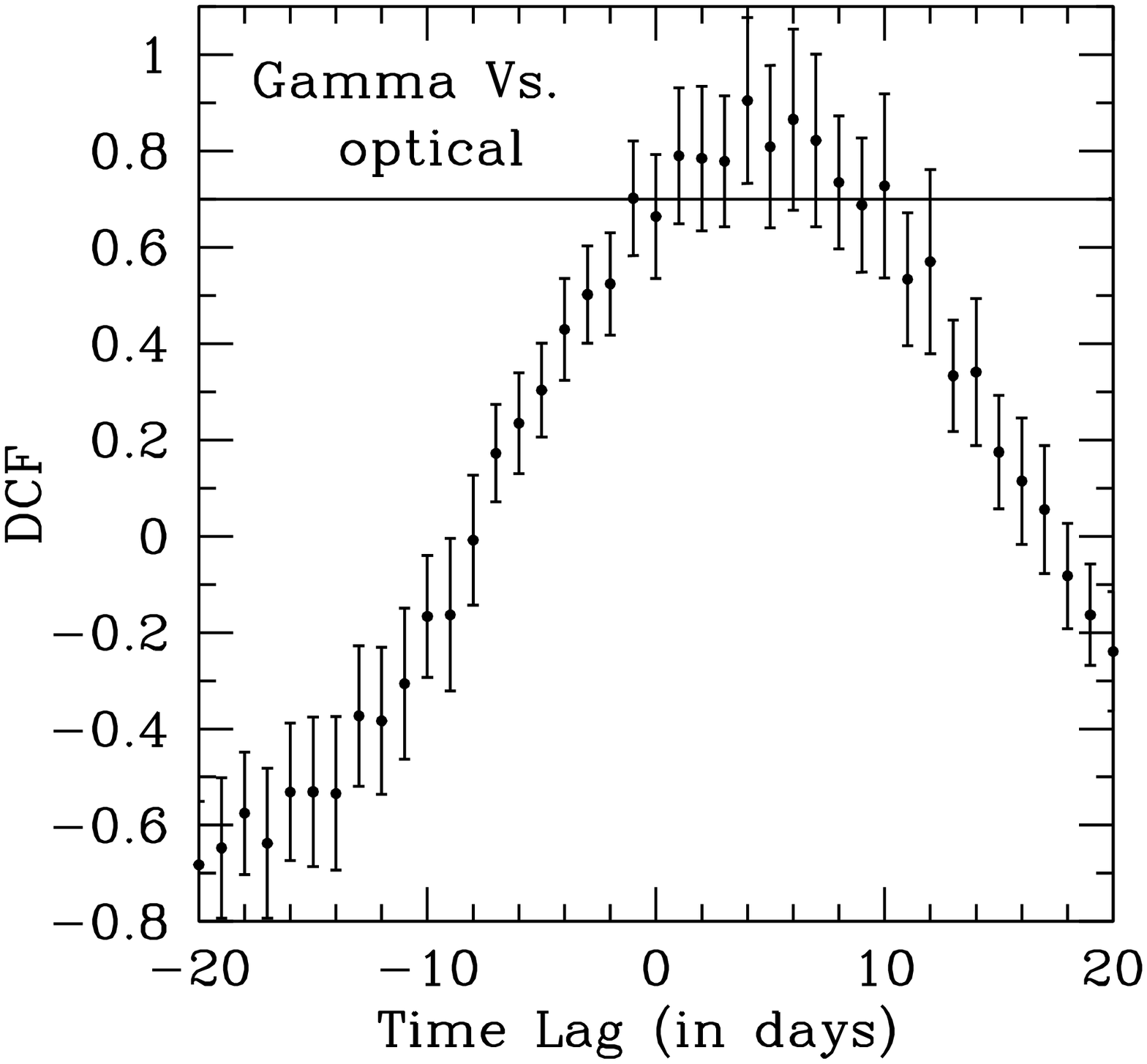}
\includegraphics[width=2.2in,height=2.5in]{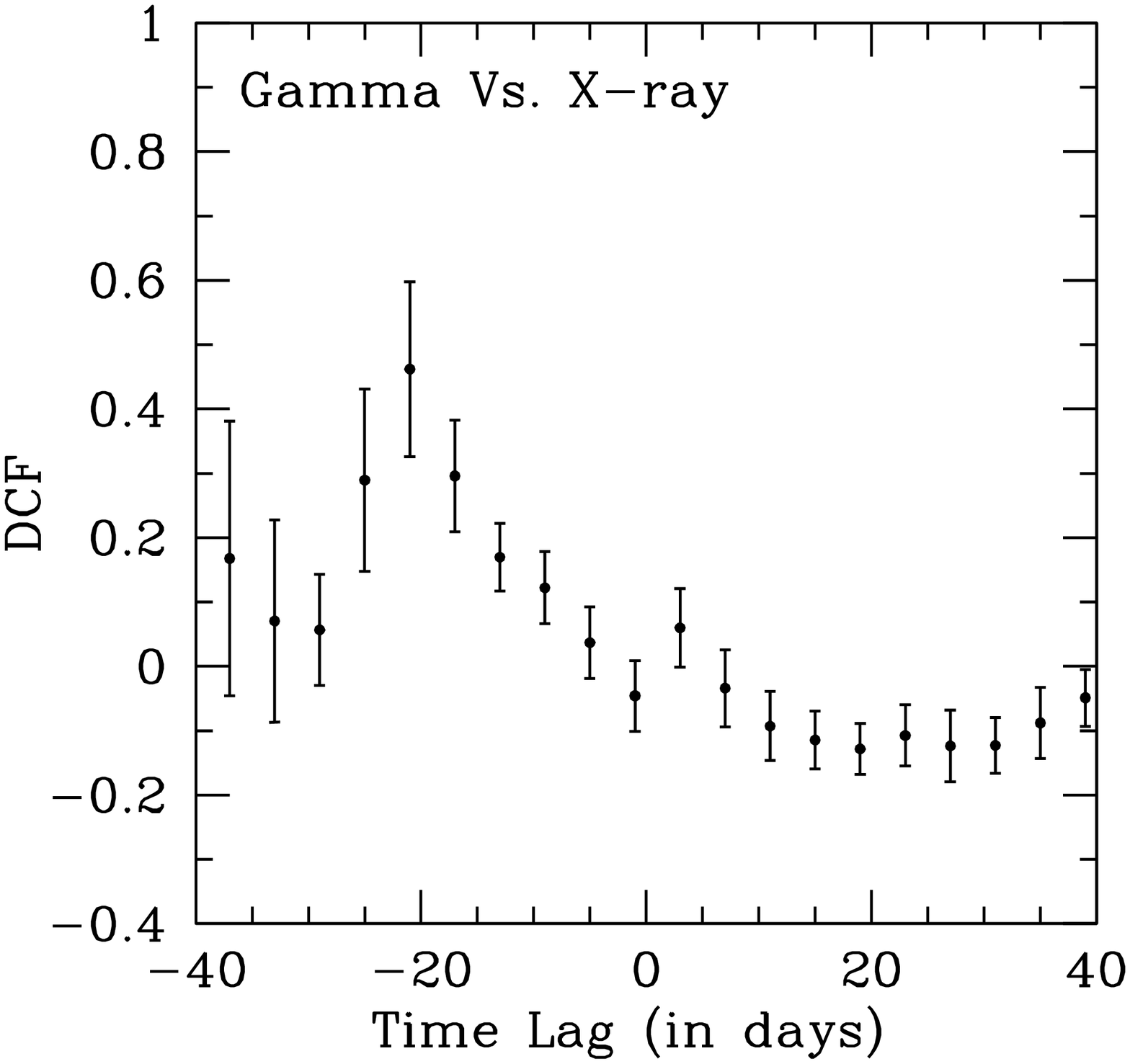}
\includegraphics[width=2.2in,height=2.5in]{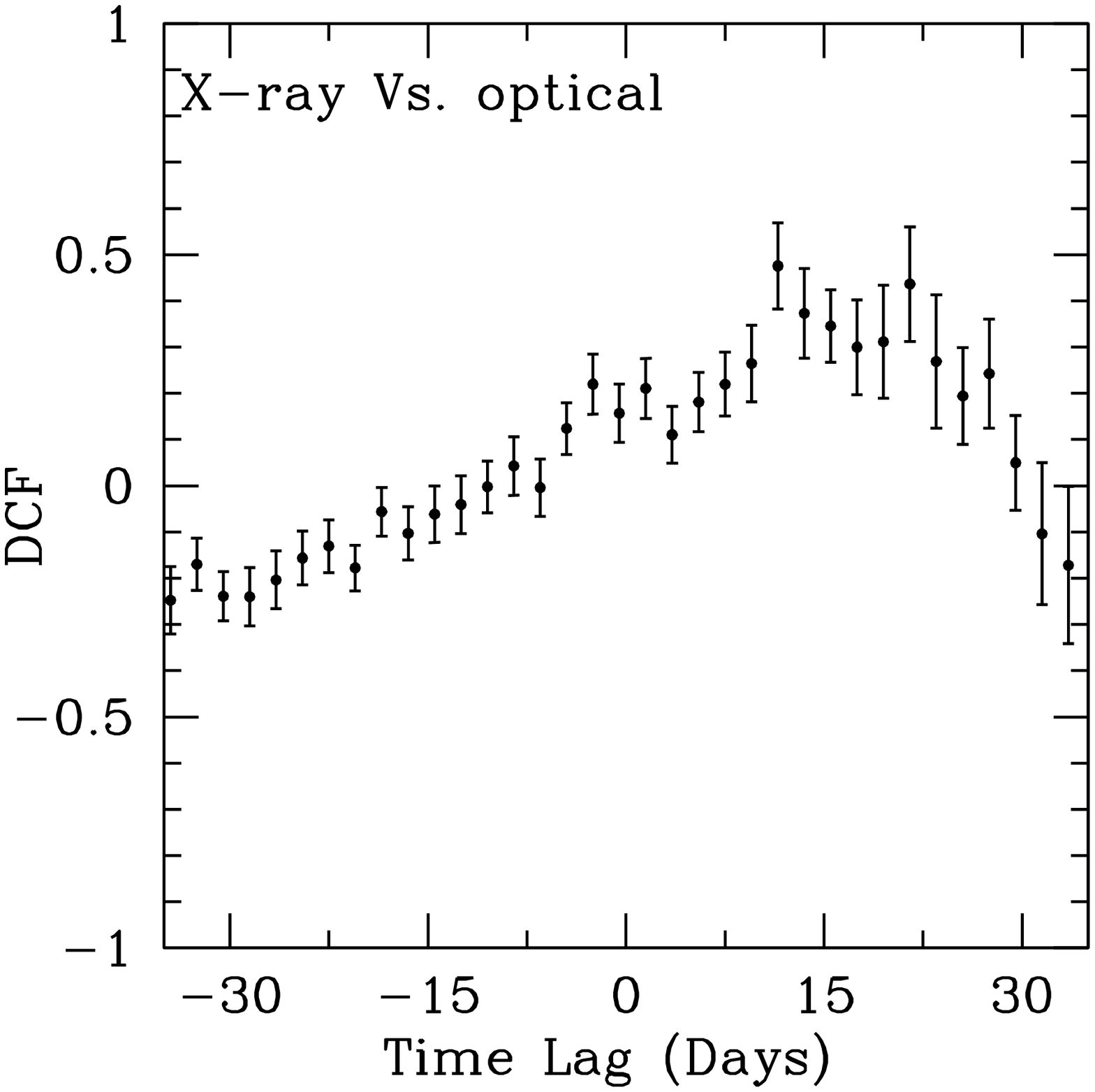}

\caption{Gamma, X-ray and optical LCs of 3C 454.3 (upper panels);
X$-$ray LCs  for 3C 454.3  in 2--4 keV, 4--10 keV and hardness intensity plot (middle panels);  
DCF between gamma vs.\ optical (horizontal line indicates 99\% significance level), 
$\gamma$-ray vs.\ X-ray and X-ray vs.\ optical (in lower panels).}
\end{figure}

\begin{table}
{\bf Table 1. Observation log of optical photometric observations of Mrk 421 and 3C 454.3 }

\vspace*{0.2in}
\begin{tabular}{lcccclcc} \hline \hline
Source                &Date of         & Filters      & Data          \\
name                &observation         &              & points          \\ \hline
Mrk421                &2009 Nov 21     &B,V,R         &1,1,36               \\
                      &2009 Nov 22     &B,V,R         &1,1,22       \\
                      &2009 Dec 11     &B,V,R         &1,1,2        \\
                      &2009 Dec 13     &B,V,R         &2,2,2       \\  
                     &2009 Dec 15     &B,V,R         &2,2,2          \\
                     &2009 Dec 20     &B,V,R         &2,2,2           \\
                       &2009 Dec 21     &B,V,R         &2,2,2          \\
                       &2010 Jan 10     &B,V,R         &1,1,84         \\
                       &2010 Jan 11     &B,V,R         &1,1,87        \\
                       &2010 Jan 20   &B,V,R           &1,1,80               \\
                        &2010 Mar 15     &B,V,R         &1,1,70              \\
                        &2010 Mar 16   &B,V,R             &1,1,70              \\
                       &2010 Mar 22   &B,V,R              &1,1,85               \\
                        &2010 Apr 09     &B,V,R             &1,1,80                \\
                         &2010 Apr 15   &B,V,R              &1,1,1               \\
                         &2010 Apr 16   &B,V,R              &1,1,1            \\
                         &2010 Apr 20   &B,V,R           &0,1,1                \\
                        &2010 Jun 10   &B,V,R             &1,1,1     \\
3C454.3                &2009 Nov 22     &B,V,R,I          &1,1,62,0    \\
                       &2009 Dec 11     &B,V,R,I          &1,1,46,1    \\
                       &2009 Dec 13     &B,V,R,I          &1,1,39,1    \\
                       &2009 Dec 15     &B,V,R,I          &1,1,46,1    \\
                       &2009 Dec 18     &B,V,R,I          &1,1,33,1    \\
                       &2009 Dec 20     &B,V,R,I          &1,0,41,1    \\
                       &2009 Dec 21     &B,V,R,I          &1,1,75,1    \\  \hline

\end{tabular}
\end{table}

%

\normalsize
\begin{table}
{\bf Table 2. IDV observations  of Mrk 421 and 3C 454.3}
\scriptsize

\noindent
\begin{flushleft} 
\begin{tabular}{lclcccc} \hline
Source Name    &Date   &N     &C-Test                    &F-Test                    &Variable &A (\%)\\
          &(dd.mm.yy)&   &$C_{1},C_{2}$  &$F_{1},F_{2},F_{c}(0.99),F_{c}(0.999)$  &          &  \\\hline  
Mrk 421     &21.11.09  &36    &0.71, 1.00        &0.52, 1.00, 2.23, 2.93            &NV       &-   \\
            &22.11.09  &22    &2.00, 1.73        &4.00, 2.98, 2.86, 4.13            &NV       &-   \\
            &10.01.10  &83    &1.52, 1.34        &2.31, 1.79, 1.68, 1.99            &NV       &-   \\
            &11.01.10  &85    &0.72, 1.09        &0.52, 1.18, 1.67, 1.98            &NV       &-   \\
            &20.01.10  &76    &0.49, 0.84        &0.24, 0.70, 1.72, 2.06            &NV       &-   \\
            &15.03.10  &66    &0.27, 1.06        &0.07, 1.12, 1.79, 2.18            &NV       &-   \\
            &16.03.10  &70    &0.52, 1.08        &0.27, 1.16, 1.76, 2.13            &NV       &-   \\
            &22.03.10  &83    &0.85, 1.37        &0.73, 1.87, 1.68, 1.99            &NV       &-   \\
            &09.04.10  &79    &0.45, 0.77        &0.20, 0.59, 1.70, 2.03            &NV       &-   \\
3C 454.3    &22.11.09  &62    &3.09, 2.91        &9.54, 8.48, 1.83, 2.24            &V        &8.66   \\
            &11.12.09  &46    &1.20, 1.20        &1.45, 1.45, 2.02, 2.57            &NV       &-   \\
            &13.12.09  &40    &5.24, 5.08        &27.49, 25.83, 2.14, 2.76          &V        &29.84   \\
            &15.12.09  &73    &4.10, 4.08        &16.85, 16.63, 1.74, 2.09          &V        &12.90   \\
            &18.12.09  &30    &1.41, 3.09        &1.98, 3.09, 2.42, 3.29            &NV       &-   \\
            &20.12.09  &40    &3.85, 4.20        &14.79, 17.63, 2.14, 2.76          &V       &9.34   \\
            &21.12.09  &75    &1.22, 1.12        &1.49, 1.26, 1.73, 2.07            &NV       &-   \\  \hline

\end{tabular} \end{flushleft}
V: Variable  NV: Non-variable
\end{table}

\normalsize
\begin{table}
{\bf Table 3. STV observations of Mrk 421 and 3C 454.3}
\scriptsize

\noindent
\begin{flushleft} 
\begin{tabular}{lclcccc} \hline
Source Name    &Band   &N     &C-Test                    &F-Test                    &Variable &A (\%)\\
               &       &   &$C_{1},C_{2}$  &$F_{1},F_{2},F_{c}(0.99),F_{c}(0.999)$  &          &  \\\hline
Mrk421         &B      &16   &6.16, 6.61   &38.00, 43.67, 3.52, 5.53            &V         &62.33  \\
               &V      &17   &9.27, 9.63   &85.95, 92.79, 3.37, 5.20            &V         &53.08  \\
               &R      &18   &5.98, 6.21   &35.72, 38.53, 3.24, 4.92            &V         &44.20 \\
               &(B-V)  &16   &0.12, 0.99   &0.01, 0.98, 3.52, 5.34              &NV        &-  \\
               &(V-R)  &15   &2.18, 2.20   &4.75, 4.83, 3.70, 5.93              &NV        &- \\
3C 454.3$^{*}$ &B      &7    &2.83, 2.63   &40.88, 60.96, 8.47, 20.03           &V         &125.51  \\
               &V      &6    &43.29, 43.02 &1874.33, 1850.54, 10.97, 29.75      &V         &124.98\\
               &R      &7    &45.48, 45.48 &2068.11, 2068.03, 8.47, 20.03       &V         &124.70\\
               &I      &6    &4.71, 3.92   &22.21, 15.40, 10.97, 29.75          &V        &59.10 \\ \hline            
\end{tabular} \end{flushleft}
*: C and F tests are performed only for ARIES data and amplitudes are calculated for the complete\\ 
light curve including data from SMARTS and KANATA.\\
V: Variable  NV: Non-variable\\
\end{table}


\begin{thebibliography}{}
\bibitem[Abdo et al.(2009)]{2009ApJ...699..817A} Abdo, A.~A., et al.\ 2009, 
\apj, 699, 817 



\bibitem[Abdo et al.(2010)]{2010ApJ...715..429A} Abdo, A.~A., et al.\ 2010,
\apj, 715, 429

\bibitem[Acciari et al.(2011)]{2011arXiv1106.1210A} Acciari, V.~A., et al.\ 
2011, arXiv:1106.1210 



\bibitem[Aharonian et al.(2005)]{2005A&A...437...95A} Aharonian, F., et al.\ 2005, \aap, 437, 95
\bibitem[Albert et al.(2007)]{2007ApJ...663..125A} Albert, J., et al.\ 2007, \apj, 663, 125
\bibitem[Aleksic et al.(2011)]{2011arXiv1106.1589A} Aleksic, J., et al.\ 
2011, arXiv:1106.1589 


\bibitem[Bailyn et al.(1999)]{1999AAS...195.8706B} Bailyn, C.~D., Depoy, 
D., Agostinho, R., Mendez, R., Espinoza, J., 
\& Gonzalez, D.\ 1999, BAAS, 31, 1502 
\bibitem[Bennett(1962)]{1962MmRAS..68..163B} Bennett, A.~S.\ 1962, \memras, 
68, 163 
\bibitem[Bennett et al.(2003)]{2003ApJS..148...97B} Bennett, C.~L., et al.\ 
2003, \apjs, 148, 97 
\bibitem[Blom et 
al.(1995)]{1995A&A...295..330B} Blom, J.~J., et al.\ 1995, \aap, 295, 330 
\bibitem[Bonnoli et al.(2011)]{2011MNRAS.410..368B} Bonnoli, G., 
Ghisellini, G., Foschini, L., Tavecchio, F., 
\& Ghirlanda, G.\ 2011, \mnras, 410, 368 
\bibitem[Bonning et al.(2009)]{2009ApJ...697L..81B} Bonning, E.~W., et al.\
2009, \apjl, 697, L81
\bibitem[Chakrabarti \& Wiita(1993)]{1993ApJ...411..602C} Chakrabarti, S.~K., \& Wiita, P.~J.\ 1993, \apj, 411, 602 
\bibitem[Chen et al.(2007)]{2007ATel.1278....1C} Chen, A., et al.\ 2007,
The Astronomer's Telegram, 1278, 1

\bibitem[Costa et al.(2008)]{2008ATel.1574....1C} Costa, E., et al.\ 2008, The Astronomer's Telegram, 1574, 1
\bibitem[de Diego(2010)]{2010AJ....139.1269D} de Diego, J.~A.\ 2010, \aj, 139, 1269 
\bibitem[Donnarumma et al.(2009)]{2009ApJ...707.1115D} Donnarumma, I., et 
al.\ 2009, \apj, 707, 1115 

\bibitem[Dermer 
\& Schlickeiser(1993)]{1993ApJ...416..458D} Dermer, C.~D., \& Schlickeiser, R.\ 1993, \apj, 416, 458 


\bibitem[Edelson \& Krolik(1988)]{1988ApJ...333..646E} Edelson, R.~A., \& Krolik, J.~H.\ 1988, \apj, 333, 646 
\bibitem[Emmanoulopoulos et al.(2010)]{2010MNRAS.404..931E} 
Emmanoulopoulos, D., McHardy, I.~M., \& Uttley, P.\ 2010, \mnras, 404, 931 

\bibitem[Escande
\& Tanaka(2009)]{2009ATel.2328....1E} Escande, L., \& Tanaka, Y.~T.\ 2009, The Astronomer's Telegram, 2328, 1

\bibitem[Fan \& Lin(1999)]{1999ApJS..121..131F} Fan, J.~H., \& Lin, R.~G.\ 1999, \apjs, 121, 131 

\bibitem[Foschini et al.(2010)]{2010arXiv1004.4518F} Foschini, L.,
Tagliaferri, G., Ghisellini, G., Ghirlanda, G., Tavecchio, F.,
\& Bonnoli, G.\ 2010, MNRAS, 408, 448

\bibitem[Fossati et al.(1998)]{1998MNRAS.299..433F} Fossati, G., Maraschi, L., Celotti, A., Comastri, A., \& 
Ghisellini, G.\ 1998, \mnras, 299, 433 
\bibitem[Fossati et al.(2008)]{2008ApJ...677..906F} Fossati, G., et al.\ 2008, \apj, 677, 906
\bibitem[Finke
\& Dermer(2010)]{2010ApJ...714L.303F} Finke, J.~D., \& Dermer, C.~D.\ 2010, \apjl, 714, L303


\bibitem[Fuhrmann et
al.(2006)]{2006A&A...445L...1F} Fuhrmann, L., et al.\ 2006, \aap, 445, L1
\bibitem[Fukugita et al.(1995)]{1995PASP..107..945F} Fukugita, M., 
Shimasaku, K., \& Ichikawa, T.\ 1995, \pasp, 107, 945 



\bibitem[Gaidos et al.(1996)]{1996Natur.383..319G} Gaidos, J.~A., et al.\ 1996, \nat, 383, 319
\bibitem[Gaur et al.(2010)]{2010ApJ...718..279G} Gaur, H., Gupta, A.~C., 
Lachowicz, P., \& Wiita, P.~J.\ 2010, \apj, 718, 279 


\bibitem[Ghisellini et al.(1998)]{1998MNRAS.301..451G} Ghisellini, G., Celotti, A., Fossati, G., Maraschi, L.,
\& Comastri, A.\ 1998, \mnras, 301, 451
\bibitem[Ghisellini \& Tavecchio(2008)]{2008MNRAS.386L..28G} Ghisellini, G., \& Tavecchio, F.\ 2008, \mnras, 386, L28
\bibitem[Giommi et al.(1995)]{1995A&AS..109..267G} Giommi, P., Ansari, S.~G., \& Micol, A.\ 1995, \aaps, 109, 267 
\bibitem[Giommi et
al.(2006)]{2006A&A...456..911G} Giommi, P., et al.\ 2006, \aap, 456, 911


\bibitem[Gupta et al.(2004)]{2004A&A...422..505G} Gupta, A.~C., Banerjee, D.~P.~K., Ashok, N.~M., \& Joshi, U.~C.\ 2004, \aap, 422, 505 
\bibitem[Gupta et al.(2008)]{2008ChJAA...8..395G} Gupta, A.~C., Acharya, B.~S., Bose, D., Chitnis, V.~R., \& Fan, J.-H.\ 2008, ChJAA, 8, 395 
\bibitem[Gupta et al.(2009)]{2009ATel.2352....1G} Gupta, A.~C., Gaur, H.,
\& Rani, B.\ 2009, The Astronomer's Telegram, 2352, 1


\bibitem[Hartman et al.(1993)]{1993ApJ...407L..41H} Hartman, R.~C., et al.\ 
1993, \apjl, 407, L41 

\bibitem[Hartman et al.(1999)]{1999ApJS..123...79H} Hartman, R.~C., et al.\ 
1999, \apjs, 123, 79 


\bibitem[Heidt \& Wagner(1996)]{1996A&A...305...42H} Heidt, J., \& Wagner, S.~J.\ 1996, \aap, 305, 42 
\bibitem[Hovatta et al.(2007)]{2007A&A...469..899H} Hovatta, T., Tornikoski, M., Lainela, M., Lehto, H.~J., Valtaoja, E., 
Torniainen, I., Aller, M.~F., \& Aller, H.~D.\ 2007, \aap, 469, 899 
\bibitem[Hufnagel \& Bregman(1992)]{1992ApJ...386..473H} Hufnagel, B.~R., \& Bregman, J.~N.\ 1992, \apj, 386, 473 
\bibitem[Isobe et al.(2010)]{2010arXiv1010.1003I} Isobe, N., et al.\ 2010, PASJ, 62, L55

\bibitem[Kerrick et al.(1995)]{1995ApJ...438L..59K} Kerrick, A.~D., et al.\ 1995, \apjl, 438, L59
\bibitem[Lichti et al.(2008)]{2008A&A...486..721L} Lichti, G.~G., et al.\ 2008, \aap, 486, 721
\bibitem[Lin et al.(1992)]{1992ApJ...401L..61L} Lin, Y.~C., et al.\ 1992, \apjl, 401, L61
\bibitem[Liu et al.(1997)]{1997A&AS..123..569L} Liu, F.~K., Liu, B.~F., \& Xie, G.~Z.\ 1997, \aaps, 123, 569
\bibitem[Mangalam \& Wiita(1993)]{1993ApJ...406..420M} Mangalam, A.~V., \& Wiita, P.~J.\ 1993, \apj, 406, 420 
\bibitem[Marscher \& Gear(1985)]{1985ApJ...298..114M} Marscher, A.~P., \& Gear, W.~K.\ 1985, \apj, 298, 114
\bibitem[Marscher(1996)]{1996ASPC..110..248M} Marscher, A.~P.\ 1996, in Blazar Continuum Variability ASP
Conference Series 110, eds. by. H. R. Miller, J. R. Webb \& J. C. Noble, p. 248.
\bibitem[Matsuoka et al.(2009)]{2009PASJ...61..999M} Matsuoka, M., et al.\ 2009, \pasj, 61, 999
\bibitem[Maraschi et al.(1999)]{1999ApJ...526L..81M} Maraschi, L., et al.\ 1999, \apjl, 526, L81
\bibitem[Michelson et al.(1992)]{1992IAUC.5470....2M} Michelson, P.~F., et al.\ 1992, \iaucirc, 5470, 2
\bibitem[Michelson(2007)]{2007AIPC..921....8M} Michelson, P.~F.\ 2007, in AIP Conf. Series, Vol. 921, 
The First GLAST Symposium, ed. S. Ritz, P. Michelson \& C. A. Meegan (Melville, NY; AIP), 8

\bibitem[Miller(1975)]{1975ApJ...201L.109M} Miller, H.~R.\ 1975, \apjl, 201, L109
\bibitem[Nilsson et 
al.(2007)]{2007A&A...475..199N} Nilsson, K., Pasanen, M., Takalo, L.~O., Lindfors, E., Berdyugin, A., Ciprini, S., \& Pforr, J.\ 2007, \aap, 475, 199 


\bibitem[Padovani \& Giommi(1995)]{1995MNRAS.277.1477P} Padovani, P., \& Giommi, P.\ 1995, \mnras, 277, 1477
\bibitem[Pacciani et al.(2010)]{2010ApJ...716L.170P} Pacciani, L., et al.\
2010, \apjl, 716, L170
\bibitem[Paltani et 
al.(1997)]{1997A&A...327..539P} Paltani, S., Courvoisier, T.~J.-L., Blecha, A., \& Bratschi, P.\ 1997, \aap, 327, 539 

\bibitem[Pian et
al.(2006)]{2006A&A...449L..21P} Pian, E., et al.\ 2006, \aap, 449, L21


\bibitem[Pittori et al.(2008)]{2008ATel.1583....1P} Pittori, C., et al.\ 2008, The Astronomer's Telegram, 1583, 1
\bibitem[Punch et al.(1992)]{1992Natur.358..477P} Punch, M., et al.\ 1992, \nat, 358, 477
\bibitem[Raiteri et 
al.(1998)]{1998A&AS..127..445R} Raiteri, C.~M., Ghisellini, G., Villata, M., de Francesco, G., Lanteri, L., Chiaberge, M., Peila, A., \& Antico, G.\ 1998, \aaps, 127, 445 

\bibitem[Raiteri et
al.(2007)]{2007A&A...473..819R} Raiteri, C.~M., et al.\ 2007, \aap, 473, 819


\bibitem[Raiteri et
al.(2008)]{2008A&A...485L..17R} Raiteri, C.~M., et al.\ 2008, \aap, 485, L17


\bibitem[Rebillot et al.(2006)]{2006ApJ...641..740R} Rebillot, P.~F., et al.\ 2006, \apj, 641, 740
\bibitem[Romero et al.(1999)]{1999A&AS..135..477R} Romero, G.~E., Cellone, S.~A., \& Combi, J.~A.\ 1999, \aaps, 135, 477 
\bibitem[Rutman(1978)]{1978IEEEP..66.1048R} Rutman, J.\ 1978, IEEE 
Proceedings, 66, 1048 


\bibitem[Sandage(1966)]{1966ApJ...144.1234S} Sandage, A.\ 1966, \apj, 144, 
1234 

\bibitem[Sasada et al.(2009)]{2009ATel.2333....1S} Sasada, M., et al.\
2009, The Astronomer's Telegram, 2333, 1

\bibitem[Schlegel et al.(1998)]{1998ApJ...500..525S} Schlegel, D.~J., 
Finkbeiner, D.~P., \& Davis, M.\ 1998, \apj, 500, 525 



\bibitem[Sikora et al.(1994)]{1994ApJ...421..153S} Sikora, M., Begelman, 
M.~C., \& Rees, M.~J.\ 1994, \apj, 421, 153 


\bibitem[Sikora et al.(2009)]{2009ApJ...704...38S} Sikora, M., Stawarz, 
{\L}., Moderski, R., Nalewajko, K., \& Madejski, G.~M.\ 2009, \apj, 704, 38 


\bibitem[Simonetti et al.(1985)]{1985ApJ...296...46S} Simonetti, J.~H., 
Cordes, J.~M., \& Heeschen, D.~S.\ 1985, \apj, 296, 46 


\bibitem[Stein et al.(1976)]{1976ARA&A..14..173S} Stein, W.~A., Odell, S.~L., \& Strittmatter, P.~A.\ 1976, \araa, 14, 173 
\bibitem[Stetson(1987)]{1987PASP...99..191S} Stetson, P.~B.\ 1987, \pasp, 99, 191 
\bibitem[Stetson(1992)]{1992JRASC..86...71S} Stetson, P.~B.\ 1992, \jrasc, 86, 71 
\bibitem[Striani et al.(2009)]{2009ATel.2322....1S} Striani, E., et al.\
2009, The Astronomer's Telegram, 2322, 1
\bibitem[Striani et al.(2010)]{2010ApJ...718..455S} Striani, E., et al.\
2010, \apj, 718, 455


\bibitem[Takahashi et al.(1994)]{1994IAUC.5993....2T} Takahashi, T., et al.\ 1994, \iaucirc, 5993, 2
\bibitem[Takahashi et al.(1995)]{1995IAUC.6167....1T} Takahashi, T., et al.\ 1995, \iaucirc, 6167, 1
\bibitem[Takahashi et al.(2000)]{2000ApJ...542L.105T} Takahashi, T., et al.\ 2000, \apjl, 542, L105

\bibitem[Tavecchio et al.(2002)]{2002ApJ...575..137T} Tavecchio, F., et 
al.\ 2002, \apj, 575, 137 


\bibitem[Ter{\"a}sranta et al.(2004)]{2004A&A...427..769T} Ter{\"a}sranta, H., et al.\ 2004, \aap, 427, 769
\bibitem[Ter{\"a}sranta et al.(2005)]{2005A&A...440..409T} Ter{\"a}sranta, H., Wiren, S., Koivisto, P., Saarinen, V., \& Hovatta, T.\ 2005, \aap, 440, 409
\bibitem[Tornikoski et al.(1994)]{1994A&A...289..673T} Tornikoski, M., Valtaoja, E., Terasranta, H., Smith, A.~G., Nair, A.~D., 
Clements, S.~D., \& Leacock, R.~J.\ 1994, \aap, 289, 673 
\bibitem[Tosti et al.(2008)]{2008ATel.1628....1T} Tosti, G., Chiang, J.,
Lott, B., Do Couto E Silva, E., Grove, J.~E.,
\& Thayer, J.~G.\ 2008, The Astronomer's Telegram, 1628, 1

\bibitem[Tramacere et al.(2009)]{2009A&A...501..879T} Tramacere, A., Giommi, P., Perri, M., Verrecchia, F., \& Tosti, G.\ 2009, \aap, 501, 879


\bibitem[Tsunemi et al.(2010)]{2010arXiv1009.6106T} Tsunemi, H., Ueda, S., Shigeyama, K., Mori, K., Aoyama, S., \& Takagi, S.\ 2010, arXiv:1009.6106
\bibitem[Ulrich et al.(1975)]{1975ApJ...198..261U} Ulrich, M.-H., Kinman, T.~D., Lynds, C.~R., Rieke, G.~H., \& Ekers, R.~D.\ 1975, \apj, 198, 261
\bibitem[Ulrich et al.(1997)]{1997ARA&A..35..445U} Ulrich, M.-H., Maraschi, L., \& Urry, C.~M.\ 1997, \araa, 35, 445 
\bibitem[Urry \& Padovani(1995)]{1995PASP..107..803U} Urry, C.~M., \& Padovani, P.\ 1995, \pasp, 107, 803 
\bibitem[Ushio et al.(2009)]{2009ApJ...699.1964U} Ushio, M., et al.\ 2009, 
\apj, 699, 1964 

\bibitem[Vercellone et al.(2008)]{2008ApJ...676L..13V} Vercellone, S., et
al.\ 2008, \apjl, 676, L13


 \bibitem[Villata et
al.(2006)]{2006A&A...453..817V} Villata, M., et al.\ 2006, \aap, 453, 817
\bibitem[Villata et
al.(2007)]{2007A&A...464L...5V} Villata, M., et al.\ 2007, \aap, 464, L5
\bibitem[Villata et al.(2008)]{2008ATel.1625....1V} Villata, M., et al.\ 
2008, The Astronomer's Telegram, 1625, 1 


\bibitem[Wagner \& Witzel(1995)]{1995ARA&A..33..163W} Wagner, S.~J., \& Witzel, A.\ 1995, \araa, 33, 163 
\bibitem[Worrall et al.(1987)]{1987ApJ...313..596W} Worrall, D.~M., 
Tananbaum, H., Giommi, P., \& Zamorani, G.\ 1987, \apj, 313, 596 

\bibitem[Xie et al.(1988)]{1988A&AS...72..163X} Xie, G.-Z., Lu, R.-W., Zhou, Y., Hao, P.-J., Zhang, Y., Li, X.-Y., Liu, X., \& Wu, J.-X.\ 1988, \aaps, 72, 163 
\bibitem[Zhang et 
al.(2005)]{2005A&A...444..767Z} Zhang, S., Collmar, W., \& Sch{\"o}nfelder, V.\ 2005, \aap, 444, 767 
\bibitem[Zhang et al.(2006)]{2006ApJ...637..699Z} Zhang, Y.~H., Treves, A., 
Maraschi, L., Bai, J.~M., \& Liu, F.~K.\ 2006, \apj, 637, 699 



\end{thebibliography}
\end{document}